\documentclass[fleqn,usenatbib]{mnras}

\usepackage{newtxtext,newtxmath}

\usepackage[T1]{fontenc}

\DeclareRobustCommand{\VAN}[3]{#2}
\let\VANthebibliography\thebibliography
\def\thebibliography{\DeclareRobustCommand{\VAN}[3]{##3}\VANthebibliography}

\usepackage{graphicx}	
\usepackage{amsmath}	
\usepackage{arydshln}
 
\usepackage{verbatim}
\usepackage{breqn}
\usepackage[export]{adjustbox}
\usepackage{booktabs}
\usepackage{pgfplotstable}

\newcommand{\lya}{Ly$\alpha$}
\newcommand{\lyb}{Ly$\beta$}

\newcommand{\kms}{$km s^{-1}$}

\newcommand{\HI}{\mbox{H\,{\sc i}}}

\newcommand{\CIV}{\mbox{C\,{\sc iv}}}

\newcommand{\SiIV}{\mbox{Si\,{\sc iv}}}

\newcommand{\MgII}{\mbox{Mg\,{\sc ii}}}

\newcommand{\NHI} {$N_{\rm HI}$}

\newcommand{\PG}[1]{{\color{magenta} { [Prakash: #1]}}}

\title[High-$z$ clustering of \lya\ absorbers]{Measurement of redshift space two- and three-point correlation of \lya\ absorbers at  $1.7<z<3.5$: Implications on evolution of the physical properties of IGM}

\author[Maitra et al.]{Soumak Maitra$^{1}$\thanks{E-mail: soumak93@gmail.com},
	Raghunathan Srianand$^{1}$, Prakash Gaikwad$^{2,3}$
	\\
	\\
	$^{1}$ IUCAA, Postbag 4, Ganeshkhind, Pune - 411007, India\\
	$^{2}$ Institute of Astronomy, University of Cambridge, Madingley Road, Cambridge, CB3 0HA, UK\\
	$^{3}$ Kavli Institute for Cosmology, University of Cambridge, Madingley Road, Cambridge, CB3 0HA, UK\\
	}

\date{Accepted XXX. Received YYY; in original form ZZZ}

\pubyear{2015}

\begin{document}
\label{firstpage}
\pagerange{\pageref{firstpage}--\pageref{lastpage}}
\maketitle

\begin{abstract}

We present redshift space two-point ($\xi$), three-point ($\zeta$) and reduced three-point (Q) correlations  of \lya\ absorbers (Voigt profile components having \HI\ column density, $N_{\rm HI}>10^{13.5}$cm$^{-2}$) over three redshift bins spanning $1.7< z<3.5$ using high-resolution spectra of 292 quasars.
We detect positive $\xi$ up to 8 $h^{-1}$ cMpc in all three redshift bins. 
The strongest detection of $\zeta =1.81\pm 0.59$ (with Q$=0.68\pm 0.23$), is in $z=1.7-2.3$ bin at $1-2h^{-1}$ cMpc.
The measured $\xi$ and $\zeta$ values show an increasing trend with \NHI, while Q remains relatively independent of \NHI. We find $\xi$ and $\zeta$ to evolve strongly with
redshift.
Using
simulations, we find that  $\xi$ and 
$\zeta$ seen in real space may be strongly amplified by peculiar velocities in redshift space. Simulations suggest that while feedback, thermal and pressure smoothing effects influence the clustering of \lya\ absorbers at small scales, i.e $<0.5h^{-1}$ cMpc, the \HI\ photo-ionization rate ($\Gamma_{\rm HI}$) has a strong influence at all scales.
The strong redshift evolution of $\xi$ and $\zeta$ (for a fixed \NHI-cutoff) is driven by the redshift evolution of the relationship between \NHI\ and baryon overdensity. 
Our simulation 
using best-fitted $\Gamma_{\rm HI}(z)$ measurements produces consistent clustering signals with observations at $z\sim 2$
but under-predicts the clustering at higher redshifts. One possible remedy is to have higher
values of $\Gamma_{\rm HI}$ at higher redshifts.
Alternatively the discrepancy could be related to non-equilibrium and inhomogeneous  conditions prevailing during He~{\sc ii} reionization not captured by our simulations.
\end{abstract}

\begin{keywords}
		Cosmology: large-scale structure of Universe - Cosmology: diffuse radiation - Galaxies: intergalactic medium - Galaxies: quasars : absorption lines
\end{keywords}



\section{Introduction}

    The Intergalactic Medium (IGM) manifests itself in the form of \lya\ forest absorption in the spectra of distant bright sources (quasars, galaxies and $\gamma-$ray bursts). The observed properties of these absorption features are governed by the matter distribution in the IGM along with its thermal and ionization state. The baryonic content of the IGM is known to trace the underlying dark matter density fluctuations at large scales; while at small scales, it experiences some additional smoothing due to the gas pressure \citep[see][]{bi1997}. Majority of the baryons exist in a photo-ionized low-density (i.e typical over-density, $\Delta$, in range 1 to 10 at $2<z<4$) state in the IGM at high redshifts. In the post \HI\ reionization era ($z<6$), the ionization state of the low-density \HI\ gas in the IGM and its evolution is determined by a uniform meta-galactic UV background \citep{haardt1996,haardt2012,khaire2015b,onorbe2017,khaire2019,puchwein2019,faucher2020}. The thermal state of this photo-ionized low-density gas (decided mainly by \HI\ and He~{\sc ii} reionization) and its evolution is well studied and the existence of a temperature-density (T$-\Delta$) relation is well-established \citep[see][]{hui1997}.

    The IGM studies are generally based on observational data in conjunction with hydrodynamical simulations. In particular, for a given set of cosmological and astrophysical parameters, the physical conditions are derived using the  observed properties of \lya\ forest such as,  neutral hydrogen column density ($N_{\rm HI}$) distribution, Doppler parameter ($b$) distribution, mean transmitted flux PDF, wavelet and curvature statistics and flux power spectrum \citep[see ][]{cen1994, Petitjean1995,zaldarriaga2002, springel2005, smith2011,becker2011, bolton2012,garzilli2012, rudie2012,boera2014,gaikwad2017a,gaikwad2017b,gaikwad2020,gaikwad2021}. Besides understanding the astrophysical properties of IGM, \lya\ forest spectra has been useful in constraining cosmological parameters \citep[][]{viel2004a,viel2004b,mcdonald2005} and placing bounds on masses of warm dark matter particles \citep[][]{viel2013w,irsic2017} and neutrino \citep[][]{palanque2015a,palanque2015b,yeche2017,walther2020}.

    A quintessential role in understanding the matter distribution in the IGM is played by the clustering studies of \lya\ forest. Most of these studies focused on redshift space clustering  using several quasar spectra and transmitted flux  \citep[longitudinal correlation or power spectrum in fourier space, see][]{mcdonald2000,mcdonald2006,croft2002,seljak2006}. One can also study clustering in the transverse direction between adjacent sightlines (transverse correlation) of closely spaced projected quasar pairs or gravitationally lensed quasars \citep{Smette1995,Rauch1995,petitjean1998,aracil2002, Rollinde2005, Coppolani2006, Dodorico2006, hennawi2010, maitra2019}. Transverse correlation statistics based on transmitted flux is found to be more sensitive to the 3D baryonic distribution in comparison to longitudinal correlation, which is affected primarily by thermal broadening effects \citep[][]{peeples2010a,peeples2010b}. However, due to sparsity of projected quasar pairs, the transverse studies are relatively scant in number. Nevertheless they have been helpful in constraining the pressure broadening scales in IGM which is known to be sensitive to the thermal history of the Universe \citep[][]{kulkarni2015,rorai2017}.

   Higher order clustering studies in \lya\ forest are useful in probing non-gaussianity in the matter density distribution caused by non-linear gravitational evolution. While three-point clustering (using three-point correlation, or bispectrum in fourier space) has been studied largely at low redshifts using galaxies \citep[][]{gaztanaga1994,gaztanaga2005,sefusatti2006,kulkarni2007,mcbride2011a,mcbride2011b,guo2016} or at higher redshifts for large scales using quasars \citep[][]{carvalho2020}, \lya\ forest as an observable would be able to probe the matter clustering at smaller scales as well as at higher redshifts. In addition to constraining the second order quadratic bias in clustering, three-point statistics can also act as an independent tool along with the two-point statistics in constraining cosmological parameters \citep[][]{fry1994,verde2002} as well as the physical state of IGM. Infact, \citet{mandelbaum2003} had shown that three-point statistics in \lya\ forest  along with two-point statistics can determine the amplitude, slope and curvature of the slope of the matter power spectrum with better precision. It has also been pointed out three-point statistics is helpful in removing degeneracies between different cosmological parameters like the bias and amplitude of matter power spectrum \citep[][]{verde2002,Bernardeau2002}.

Nevertheless, three-point statistics in \lya\ forest remains largely unexplored. {A few studies have been performed in exploring three-point statistics using cross-correlation between \lya\ forest power spectrum and matter density field probed by CMB weak-lensing signal \citep{doux2016,chiang2018}. } \citet{viel2004a} had used a large sample of Ultraviolet Echelle Spectrograph ({\sc UVES}) QSO absorption spectra ({\sc LUQAS}) and found that the errorbars in the observed bispectrum (redshift space) of  \lya\ forest (at $2.0\le z\le 2.4$) transmitted flux was too large to distinguish them from a set of absorption spectra having randomized absorber positions.
For the past three years we have been exploring using the Voigt profile components of the \lya\ absorption to study two- and three-point correlation function of the IGM \citep{maitra2019, maitra2020, maitra2020b}. We have shown that such an approach allows us to study the clustering properties as a function of \NHI~and~$b$-parameters.

\citet{maitra2020b} had explored low-$z$ \lya\ forest ($z<0.48$) using HST-COS data \citep[of][]{danforth2016} and had detected longitudinal three-point correlation in \lya\ absorbers for the first time with a $2.6\sigma$ significance. While detailed analysis of redshift space three-point correlation is possible, studies of transverse three-point correlation using close projected quasar triplets are still mostly at theoretical stages \citep{tie2019,maitra2020} due to scarcity of bright targets. There are still some attempts to study higher order clustering properties of \lya\ forest using few projected quasar triplets at high-$z$ \citep{cappetta2010,maitra2019}.

Here, we use a large sample of high resolution ($\sim 6$ \kms) QSO spectra from a publicly available {\sc KODIAQ-DR2} survey \citep{omeara2015,omeara2017} and {\sc SQUAD-DR1} survey \citep{murphy2019} to study the {longitudinal (redshift space) three-point correlation of \lya\ forest along high-z quasar sightlines}. 
Similar to \citet{maitra2020b}, we will be using the Voigt profile components (denoted as absorbers in this paper) based approach of decomposing the \lya\ forest absorption into distinct Voigt profile components and calculate the two- and three-point correlation for them.
The large size of this sample also allows us to probe the redshift evolution of these correlation functions.

This paper is organised as follows. In Section~\ref{sec:Dataset}, we describe the final sample of \lya\ forest used for our analysis. In section~\ref{Simulation}, we provide details of the hydrodynamical simulations used to interpret our observational results. In section~\ref{sec:HI_dist}, we present details of completeness of our sample as a function of \NHI\ and the \HI\ column density distribution separately for {\sc KODIAQ} and {\sc SQUAD} data. This exercise is important to establish the internal consistency between the two datasets. Then using the combined dataset we present the \NHI\ distribution for three different redshift bins of our interest. In section~\ref{Sec_Corr}, we present our measurements of two- and three-point correlation for the three redshift bins (i.e., $1.7<z<2.3$, $2.3<z<2.9$ and $2.9<z<3.5$) from observations. In this section, we also present the \HI\ column density dependence of the correlation functions and how metal line contamination affects our measurements. We also present our clustering measurements in a tabulated form for future use.
In section~\ref{sec:simulations}, using simulated spectra, we quantify how the measured clustering is affected by (1) \HI-photoionization rate ($\Gamma_{\rm HI}$), (2) pressure broadening effects (or the so called Jeans smoothing), (3) line of sight peculiar velocities (or redshift space distortions) and (4) feedback processes.
In section~\ref{Sec:evolution}, we discuss the redshift evolution of clustering. We show the strong evolution observed is mainly dominated by evolution in the physical state of the IGM and discuss the possible evolution in $\Gamma_{\rm HI}$ that can explain the observed evolution. Our results are summarized in Section~\ref{Sec:summary}.

\section{Our sample of high-z \lya\ absorbers:}
\label{sec:Dataset}
\begin{table*}
\caption{Details of our \lya\ sample used in this study}
\begin{tabular}{@{}cccccccc@{}}
\toprule
Redshift interval & \multicolumn{2}{c}{Redshift path-length} & \multicolumn{2}{c}{Number of LOS}&\multicolumn{3}{c}{Median SNR} \\ 
 & KODIAQ & SQUAD & KODIAQ & SQUAD &KODIAQ & SQUAD & KODIAQ+SQUAD \\
\midrule
\midrule
$1.7-2.3$           & 16.19 &    16.32 &  110   &   59      & 10.4  & 11.7 & 11.2    \\
$2.3-2.9$          & 19.69    & 7.28    &  143  &  26       & 12.0   & 20.7 & 13.1   \\
$2.9-3.5$          & 11.40  & 5.07    &  82  &  18     & 14.1 & 20.3 &  15.5    \\
\bottomrule
\end{tabular}\label{Table_KODIAQ}
\end{table*}

In this work, we use two publicly available compilations of high resolution (i.e $\sim$ 6 \kms) quasar spectra to study the clustering properties of high-$z$ (i.e $1.7\le z \le 3.5$) \lya\ absorbers. 
The first  sample is from the second data release (DR2) of {\sc KODIAQ} survey \citep{omeara2017}\footnote{\url{https://www2.keck.hawaii.edu/koa/public/koa.php}}. {\sc KODIAQ-DR2} sample consists of spectra of 300 quasars at $0.07<z_{em}<5.29$ obtained with {\sc KECK/HIRES} at high resolution ($\rm 30000\leq R \leq 103000$). Just for uniformity we consider only spectra having spectral resolution R$\sim$ 40000-45000.
In the case of {\sc KODIAQ}, individual spectra of a target observed multiple times are available.
In such cases we have coadded them following the procedure described in the Appendix~A in \citet{gaikwad2021}. 
Additionally, we also use quasar spectra from the publicly available first data release (DR1) of the UVES Spectral Quasar Absorption Database ({\sc SQUAD}) comprising of fully reduced and continuum fitted high-resolution spectra of 467 quasars at $0<z_{em}<5$ observed using {\sc VLT/UVES} \citep{murphy2019}\footnote{\url{https://data-portal.hpc.swin.edu.au/dataset/uves-squad-dr1}}.

As we are mostly interested in the IGM studies, we do not consider a spectrum that contains Damped \lya\ (DLA; systems with log~\NHI$\ge$ 20.3) or sub-DLA  (i.e systems with 19$\le$log~\NHI$\le$20.3) systems and associated Broad Absorption Line (BAL) systems.
For this we manually scanned all the spectra in the combined sample. We also cross-checked the emission redshifts quoted in these data sets. We check for duplicate quasar spectra between the two samples (i.e {\sc SQUAD} and {\sc KODIAQ}) and considered the spectrum having higher median SNR and better wavelength coverage for our study.  We also removed spectrum with a median spectral Signal-to-Noise Ratio (SNR) per pixel $<5$ in the \lya\ forest region.
The final list of number of quasar spectra used, redshift range and median SNR are summarized in Table~\ref{Table_KODIAQ}.
This yields 213 quasar spectra from {\sc KODIAQ} and 77 quasar spectra from {\sc SQUAD} covering \lya\ forest in the range of  $1.7<z<3.5$. For our study we consider the \lya\ absorption spectra between the \lya\ and \lyb\ emission lines of the quasar and excluding 10$h^{-1}$pMpc proximity region close to the quasar.
The details of these sightlines (quasar name, emission redshift, absorption redshift range considered for our study and the median SNR in this range) are given in Tables~\ref{tab:squad_sample} and \ref{tab:kodiaq_sample} in the appendix. 

In order to probe the redshift evolution of the clustering, we divide the data in to three  different sub-samples based on the redshift range covered by the \lya\ absorption. 
We consider redshift bins of width $\delta z = 0.6$ centered around $z=[2.0,2.6,3.2]$. 
Details of the sub-sample: total redshift path length without correcting for sensitivity, number of quasar sightlines considered and the median SNR of the subsample are summarized in Table~\ref{Table_KODIAQ} for the two data sets discussed above. 
It can be seen from this table that in the $1.7\le z\le 2.3$ bin the redshift path length for both samples are similar despite {\sc kodiaq} sample containing double the number of quasar sightlines. This is because redshift interval covered by individual quasar sightline in {\sc kodiaq} is typically less than that of {\sc squad}  in this redshift bin (see Tables~ \ref{tab:squad_sample} and \ref{tab:kodiaq_sample}).
Detailed description of the redshift pathlength corrected for detection sensitivity for a given \NHI\ is provided in section~\ref{sec:HI_dist}.

In our analysis, we do not remove the metal line contamination in the \lya\ forest. To quantify the effect of metal line contamination in the measured clustering we consider two sub-samples in the low-$z$ (37 quasar sightlines probing $1.9\le z\le 3.1$ \lya\ forest in the {\sc SQUAD} sample) and high-$z$ bins (18 quasar sightlines probing $2.9<z<3.5$ \lya\ forest in the {\sc SQUAD} sample). In both cases we manually checked the available spectra for the presence of strong metal absorption systems through \CIV, \SiIV, and \MgII\ doublets. After identifying these systems we systematically looked for associated metal line absorption that falls within the wavelength range of interest for our \lya\ absorption studies. We then masked these contaminating lines. We note that strongest contamination originates from strong metal lines having absorption components spread over few hundred \kms. We find the metal line contamination does not influence our clustering measurements and results of this exercise are presented in section~\ref{sec:metal_contamination}.

For our analysis, we decompose the observed \lya\ absorption spectrum into multiple Voigt profile components using the automated profile fitting routine {\sc viper} \citep[see,][for details]{gaikwad2017b}. We use only absorption lines that are detected above 3$\sigma$ significant levels. In what follows we call the individual Voigt profile component as "absorbers". In the series of studies we have highlighted the advantage of using "absorbers" for the clustering analysis 
\citep[refer to discussions in][]{maitra2019,maitra2020,maitra2020b} compared to most frequently used flux based statistics.

In this work we also consider different cosmological simulations to understand the measured clustering properties of the \lya\ absorbers. Before discussing our measurements we provide details of simulations used in our study in the following section.

\section{Simulation of \lya\ forest spectra}
\label{Simulation}

We have run hydrodynamical simulations using {\sc GADGET-3} code, which is a modified version of the publicly available code {\sc GADGET-2}\footnote{\url{http://wwwmpa.mpa-garching.mpg.de/gadget/}} \citep{springel2005} based on smoothed particle hydrodynamics.  The modified code self-consistently incorporates radiative heating and cooling processes in the thermal and ionization evolution equation, thereby simulating the physical properties of the IGM more accurately. The description of the simulation run and the procedure of forward modelling the simulated particle properties along sightlines to produce the transmitted flux is given in the following subsections \citep[see also,][]{maitra2020}.

\subsection{Description of simulation}
 We have generated a (100 $h^{-1}$ cMpc)$^3$ simulation box with $2\times 1024^3$ particles for a uniform metagalactic UV background \citep{khaire2019}. We use a standard flat $\Lambda$CDM cosmology with cosmological parameters based on \citet{planck2014} ( [$\Omega_{\Lambda}$, $\Omega_{m}$, $\Omega_{b}$, $h$, $n_s$, $\sigma_8$, $Y$] $\equiv$ [0.69, 0.31, 0.0486, 0.674, 0.96, 0.83, 0.24]). The initial condition for the simulations are generated at $z=99$ using {\sc 2lpt}\footnote{\url{https://cosmo.nyu.edu/roman/2LPT/}} \citep{scoccimarro2012} code, which evolves the density fields based on second-order lagrangian perturbation theory. We take the gravitational softening length to be $1/30^{th}$ of the mean inter-particle separation. We also turn on the {\sc quick\_lyalpha} flag in the simulation, which makes the simulation run faster by converting gas particles with overdensity $\Delta>10^3$ and temperature $T<10^5K$ to stars \citep[see][]{viel2004a}.  The simulations do not incorporate feedback processes sourcing from Active Galactic Nuclei (AGN) or Supernove driven galactic winds.
 
 \begin{figure}
	\includegraphics[width=8.05cm]{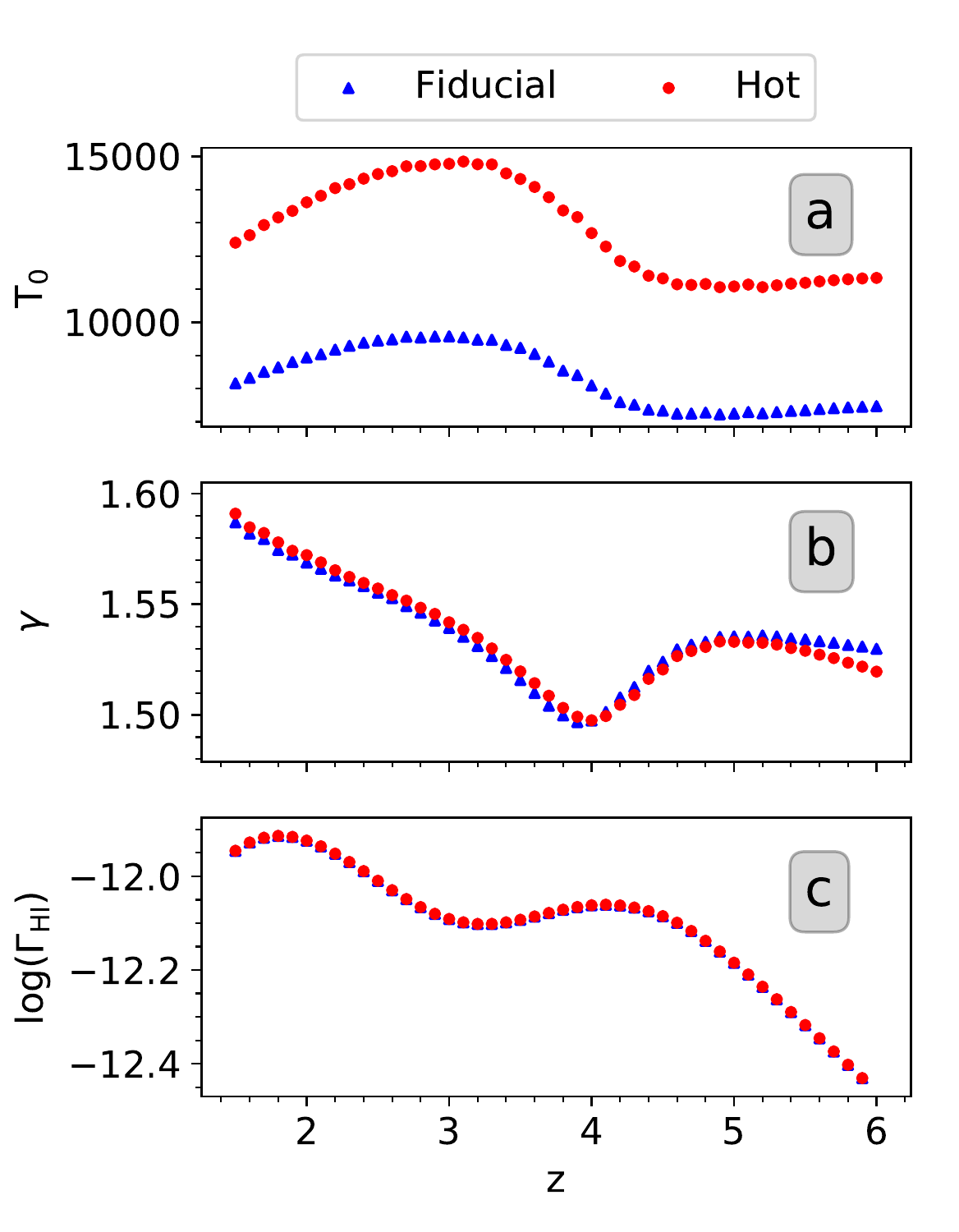}
	\includegraphics[width=8cm]{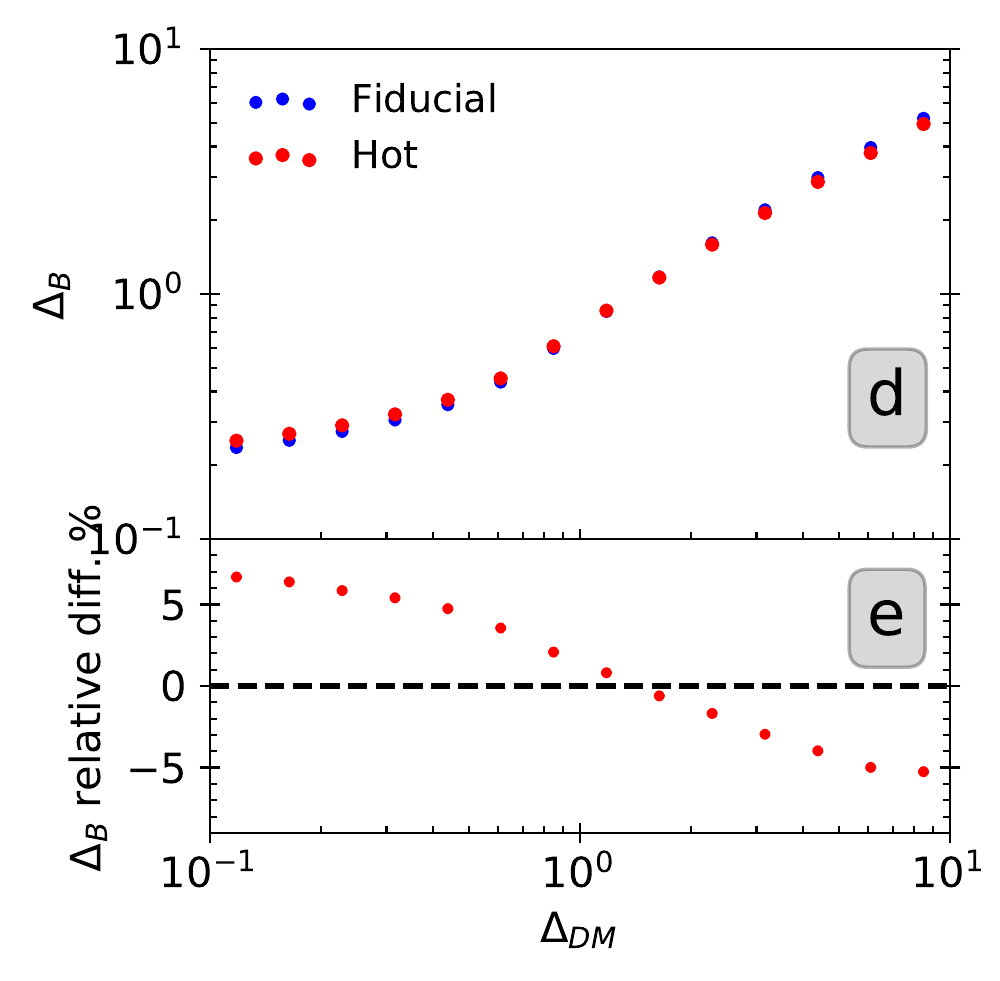}
	\caption{Panels (a) to (c) show the redshift evolution of 
	temperature at mean baryonic overdensity ($\rm T_0$), power-law index ($\gamma$) of the temperature-density relation and \HI\ photoionization rate ($\Gamma_{\rm HI}$), respectively for our "fiducial" 
	"hot" simulations (see section~\ref{Simulation} for details). Panel (d) shows the baryon-dark matter overdensity relation seen in the two simulation boxes at $z=2.0$. In panel (e) we plot the relative difference between the two distributions which quantifies the effect of additional pressure smoothing experienced by the baryons in the 'hot' simulations.}
\label{Sim_temp}
\end{figure}

We use two different simulation runs for this work. The fiducial model has been run with \citet{khaire2019}\footnote{\url{http://www.iucaa.in/projects/cobra/}} metagalactic UV background having a far UV quasar spectral index $\alpha$ ($f_{\nu}\propto \nu^{-\alpha}$) of 1.8. We also use a simulation run with an enhanced thermal history. In this case, we artificially double the background photo-heating rates with a uniform heat injection for all the particles independent of their densities, while keeping the photo-ionization rates same. In doing so, we not only increase the temperature of the baryon field at mean overdensity, but we also puff up the baryonic matter distribution due to increased gas pressure. The redshift evolution of the parameters related to thermal and ionization state of the gas [temperature at mean baryonic overdensity ($\rm T_0$), power-law index ($\gamma-1$) of the temperature-density relation (${\rm T=T_0} \Delta^{\gamma-1}$) and \HI\ photo-ionization rate ($\Gamma_{\rm HI}$)] for our ``fiducial" simulation ($\alpha=1.8$) and ``hot" simulation with enhanced UV background (Enhanced $\alpha=1.8$) has been plotted in panels (a) to (c) of Fig.~\ref{Sim_temp}. As can seen, $\Gamma_{\rm HI}$ is same for the two simulations and because of the uniform heat injection, $\gamma$ also remains more or less similar. The main difference between the ``fiducial'' and ``hot'' box is that the temperature is about $\sim 1.5$ times higher in the hot case.

To understand the effect of baryonic pressure smoothing between the two boxes, we shoot 4000 random sightlines through both the boxes for $z=2.0$ snapshot. We calculate the 1D dark matter and baryonic overdensities ($\Delta_{DM}$ and $\Delta_B$, respectively) along these sightlines using a standard Cloud-in-Cell algorithm for dark-matter field and SPH smoothing for baryonic field over 4096 grids. We then bin the $\Delta_{DM}$ values logarithmically in the range of 0.1 to 10 and find the median $\Delta_B$ corresponding to each $\Delta_{DM}$ bin for 4000 sightlines. We plot the $\Delta_{DM}$ vs $\Delta_B$ relation in panel (d) of Fig.~\ref{Sim_temp} for the "fiducial" and "hot" simulation box. In panel (e), we plot the relative percentage differences in $\Delta_B$ at each $\Delta_{DM}$ bin for the ``hot" model with respect to our ``fiducial" model. In the underdense regions ($\Delta_{DM}\sim0.1$), the $\Delta_B$ corresponding to "hot" model is higher than that for the "fiducial" model by 7\%.  in the overdense regions ($\Delta_{DM}\sim10$), the $\Delta_B$ corresponding to the "hot" model is less by 5\%. The fact that the overdense region becomes less overdense and underdense region becomes less underdense shows the  enhanced pressure smoothing effect in the baryonic density fields for the "hot" model. Thus, we will use both our simulation boxes to investigate the effect of the thermal state of the gas and pressure smoothing effects on the clustering properties of \lya\ absorbers. 

Note that in \citet{maitra2020}, we found the relative difference in $\Delta_B$ is very small when we consider UVB models with allowed range in power-law spectral index for the quasar SED (see their figure 14 for details). From the measurements presented in  \citet{gaikwad2021} we find the value of $T_0$ to in the range (0.95-1.45) $\times 10^4$ K for the redshift range of our interest. Our "hot" model typically produces higher $T_0$ values and hence the actual pressure smoothing effects may be smaller than that of our "hot" model.

Additionally, we will also use two hydrodynamical simulation boxes from publicly available Sherwood Simulation\footnote{\url{https://www.nottingham.ac.uk/astronomy/sherwood/}}  suite \citep{Bolton2017} to study the effect of wind and AGN feedback on the \lya\ clustering at $z\sim2.0$. Both these simulations were performed in a 80 h$^{-1}$ cMpc cubic box with $2\times512^3$ particle using P-Gadget-3 \citep{springel2005}. The ``80-512" model is run with {\sc quick\_lyalpha} flag \citep[as described in][]{viel2005}  without any stellar or wind  feedback. 
The second simulation ``80-512-ps13+agn" implements the star formation and energy driven wind model of \citet{Puchwein2013} and the AGN feedback. Both simulations have same initial seed density field, utilize \citet{haardt2012} UV background and use the same set of cosmological parameters from \citet{planck2014}, where $\Omega_m=0.308,\Omega_{\Lambda}=0.692, \Omega_b=0.0482,h=0.678,\sigma_8=0.829,n_S=0.961$.  As can be seen these are vary close to the cosmological parameter used for our simulations discussed above.

\subsection{Forward modelling of \lya\ forest spectra}\label{Forward_model}

We shoot random sightlines through the simulation box which contains particle information such as overdensity ($\Delta$), peculiar velocity ($v$) and temperature ($T$). We then use SPH smoothing to get 1D $\Delta$, $v$ and $T$ fields along these sightlines. We employ this SPH smoothing by sampling over 4096 uniform grids along box length. The choice of this gridding scheme is to keep the pixel size of the simulated spectra similar to the observations. To compute the \HI\ density ($n_{\rm HI}$) along a sightline, we consider two choices of $\Gamma_{\rm HI}$: 
\begin{enumerate}
    \item{} The $\Gamma_{\rm HI}$ from \citet{khaire2019} which is the default photo-ionizing background in our simulation boxes. We refer to this as $\Gamma_{\rm KS19}$. We use this while exploring the effects of various astrophysical parameters on clustering of \lya\ absorbers in simulations. 
    \item{} The $\Gamma_{\rm HI}$ obtained by matching the observed and simulated  mean transmitted fluxes, 
    which we will refer to as $\Gamma_{\rm obs}$. We use this while comparing simulated statistics with the observed ones.
\end{enumerate}

Using the $n_{\rm HI}$, $v$ and $T$ fields, we calculate the \lya\ optical depth $\tau$ as a function of wavelength along these sightlines \citep[see equation 30 of][]{choudhury2001}. The \lya\ transmitted flux is computed  as a negative exponential of $\tau$. In the case of Sherwood simulations, we sample over a sparser 2048 uniform grids, since the number of particles is lesser than our simulation box. We do not use these simulations for comparing the simulated statistics with observation, but rather to investigate the effect of wind and AGN feedback only. For the Sherwood simulations, we have rescaled optical depths with the $\Gamma_{\rm HI}$ from \citet{khaire2019} UVB model. 

To mimic the observed spectra, we first convolve the simulated transmitted flux with an instrumental Gaussian with an FWHM value of $\sim 6$ \kms\ (i.e., the typical instrumental resolution of {\sc KECK/HIRES} and {\sc VLT/UVES}). We then add a Gaussian random noise to the flux to mimic the observed noise properties. Similar to $\Gamma_{\rm HI}$, we make two choices for generating noise along the sightlines:
\begin{enumerate}
    \item Gaussian random noise corresponding to a uniform SNR of 50 per pixel. We will refer to this as $\rm SNR_{50}$ and will use it while exploring the effects of various astrophysical parameters on clustering of \lya\ absorbers in simulations. We will typically use this in conjunction with $\Gamma_{\rm KS19}$ photo-ionizing background.
    \item Gaussian random noise based on the observed SNR distribution across the sightlines in our {\sc KODIAQ} and {\sc SQUAD} sample. We randomly draw a value from the observed distribution of median SNR in a given redshift bin, and then assign that value uniformly across a simulated sightline. We repeat this procedure for all the sightlines. We will refer to this as $\rm SNR_{obs}$ and will use it while comparing simulated statistics with the observed ones, typically in conjunction with $\Gamma_{\rm obs}$. 
\end{enumerate}
Finally, we also decompose the simulated \lya\ forest transmitted flux into distinct Voigt profile components using {\sc viper} \citep[][]{gaikwad2017b}. We then use the "absorber" list created by {\sc viper} for our clustering analysis.

\section{ Observed \HI\ column density distribution}
\label{sec:HI_dist}
In this section, we estimate the $N_{\rm HI}$ distribution of the \lya\ absorbers in the {\sc KODIAQ} and {\sc SQUAD} sample and its redshift evolution.
We define the \NHI\ distribution as the number of \HI\ absorbers, N, within an absorption distance interval $X$ and $X+dX$ and having $N_{\rm HI}$ in the range $(N_{\rm HI}, N_{\rm HI}+dN_{\rm HI})$,
\begin{equation}
    f(N_{\rm HI},X)=\frac{d^2N}{d N_{\rm HI} dX},
\label{NHI_distribution_eqn}
\end{equation}
where the absorption distance $X$ is defined as $X(z)=\int_0^z dz \frac{H_0}{H(z)}(1+z)^2 $ \citep{bahcall1969}.
The $N_{\rm HI}$ is binned logarithmically in the range of log$N_{\rm HI}= (12.75,16.0)$ with bin width $dlog N_{\rm HI}=0.25$. We calculate the intrinsic $N_{\rm HI}$ distribution by estimating the incompleteness coming from regions having SNR lower than what is required for the detection of an absorber in a certain $N_{\rm HI}$ bin \citep[as explained in section 3.1 of][]{maitra2020b}.  In short,
for a given \NHI, we estimate the required spectral SNR so that the absorption line produced can be detected at $\ge 3\sigma$ level.
%
Only those regions for which the observed SNR is more than the estimated SNR are considered for the calculation of the redshift path length $dz$ \citep[see also][for details]{gaikwad2017b}. For this exercise we assume the line width to be the median b-value of our sample.

\begin{figure}
    \centering
    \includegraphics[width=8.5cm]{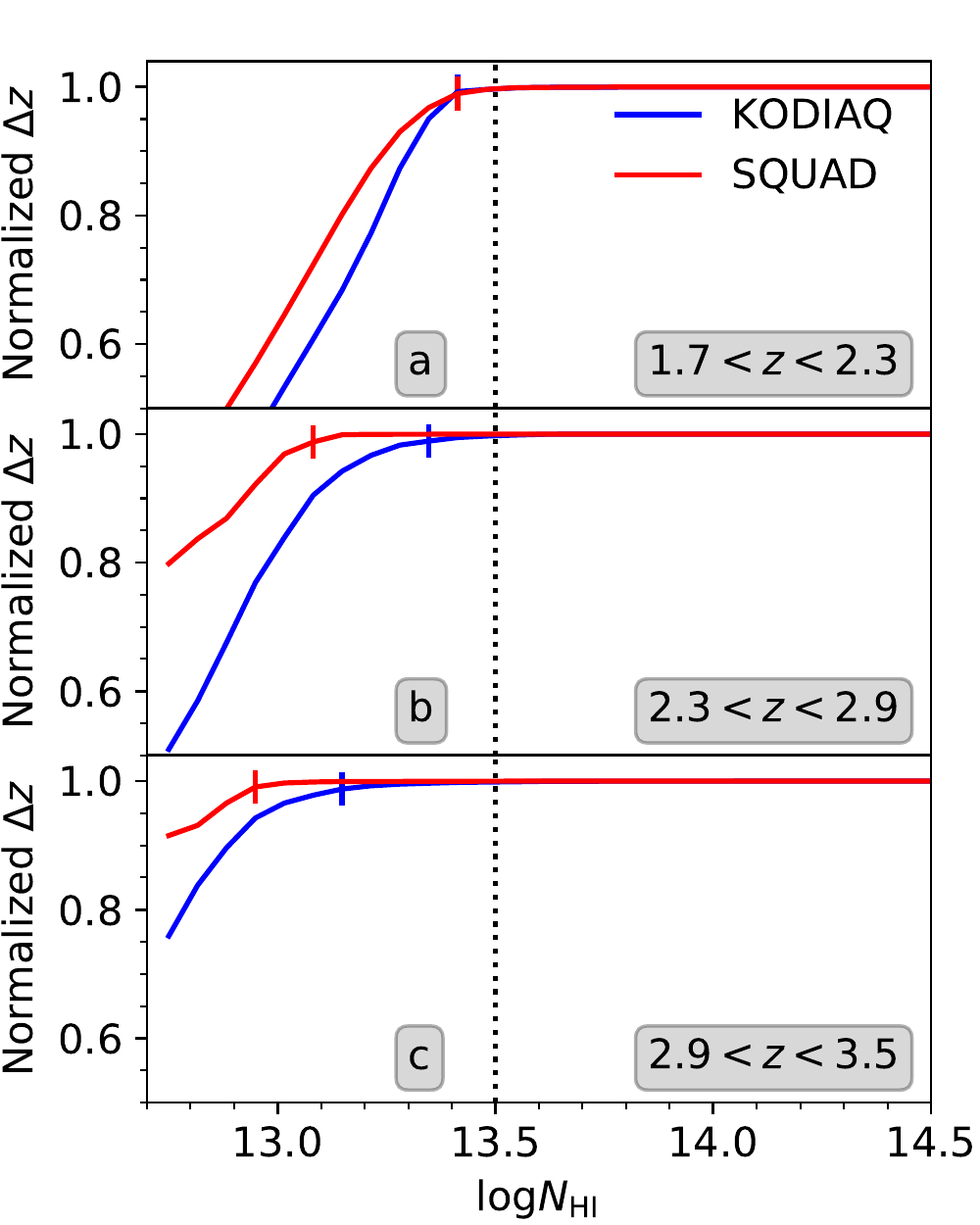}
    \caption{Sensitivity curve showing the total redshift path length ($\Delta z$) as a function of log$N_{\rm HI}$ for the {\sc KODIAQ} (blue curves) and {\sc SQUAD} (red curves) samples in the three redshift bins. The curve has been normalized to 1 for 100\% completeness limit. The coloured ticks on the sensitivity curve denotes the log$N_{\rm HI}$ value corresponding to completeness at the level of 99\%. We find that log$N_{\rm HI}=13.5$ is more than 99\% complete (vertial dotted line) in all the redshift bins. We use this value as a minimum $N_{\rm HI}$ threshold for studying clustering of \lya\ absorbers. Total redshift path length covered for each redshift for this \NHI\ threshold is provided in Table~\ref{Table_KODIAQ}. 
    }
    \label{Completeness}
\end{figure}

In Fig.~\ref{Completeness}, we plot the "sensitivity curve" showing the total redshift path length ($\Delta z$) over which the absorption produced by the gas having \HI\ column density \NHI\ can be detected at more than $3\sigma$ level for the {\sc KODIAQ} and {\sc SQUAD} samples in the three redshift bins. The coloured ticks on the sensitivity curve denote the log~$N_{\rm HI}$ value corresponding to completeness at the level of 99\%. We find that log$N_{\rm HI}=13.5$ is more than 99\% complete in all the redshift bins. So, for our study, we only consider \lya\ absorbers having $N_{\rm HI}>10^{13.5}$cm$^{-2}$. For this column density limit we get the redshift path lengths close to those summarized in Table~\ref{Table_KODIAQ}.

\begin{figure*}
	\includegraphics[width=18cm]{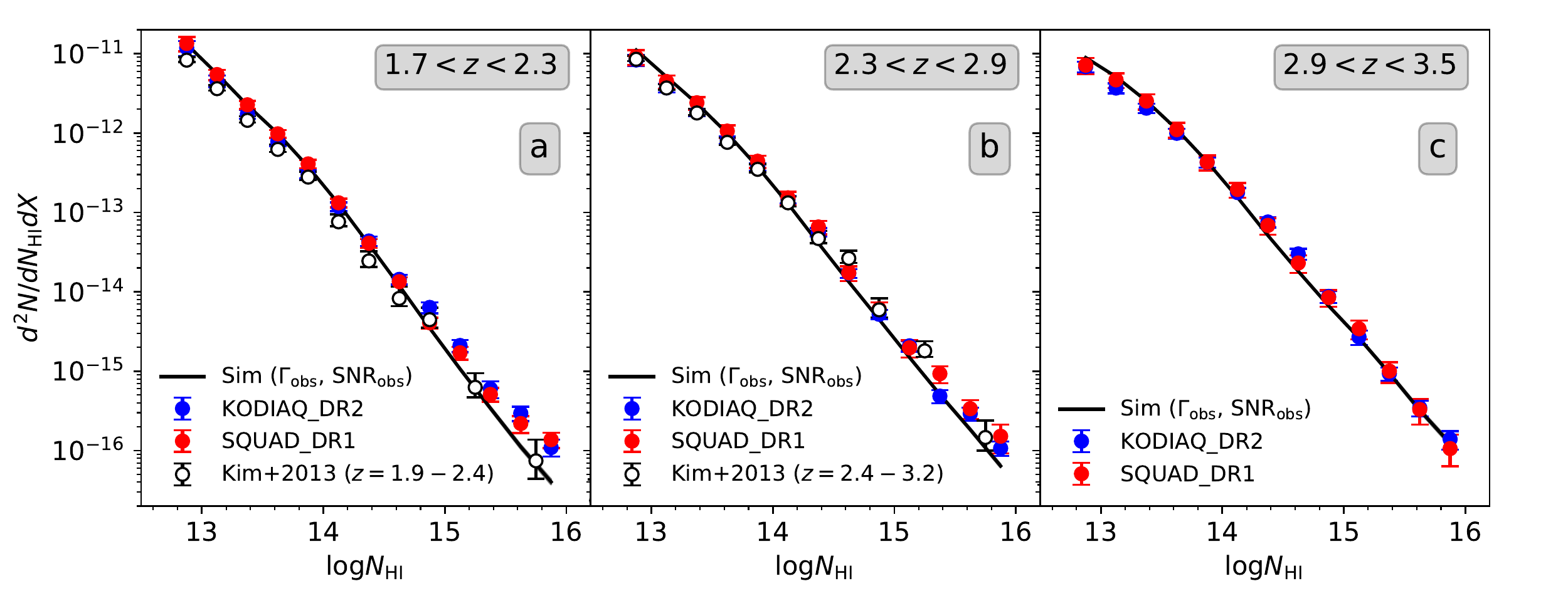}%

	\caption{ The \NHI\ distribution of \lya\ absorbers in the {\sc KODIAQ} and {\sc SQUAD} data sample at { three} different redshift bins.
	The errors correspond to larger of the two errors: one-sided poissonian uncertainty in the number of absorbers corresponding to $\pm 1\sigma$ or bootstrapping error about the mean distribution. The hollow dots represent { measurements by}
	 \citet{kim2013} in two redshift bins: $z=1.9-2.4$ and $2.4-3.2$. The black curves represent $N_{\rm HI}$ distribution from our fiducial simulation obtained by matching observed mean transmitted flux and having random Gaussian noise similar to the observed SNR distribution.}
\label{NHI_distribution_KODIAQ_vs_SQUAD}
\end{figure*}

In Fig.~\ref{NHI_distribution_KODIAQ_vs_SQUAD}, we plot the \NHI\ distribution for the {\sc KODIAQ} and {\sc SQUAD} samples separately for the 3 redshift bins of $z=[1.7-2.3,~2.3-2.9~{\rm and}~ 2.9-3.5]$. The error in the distribution is estimated using two methods: one-sided poissonian uncertainty in the number of absorbers corresponding to $\pm 1\sigma$ and bootstrapping error computed over all the sightlines for a given redshift bin. The larger of the two errors is taken to be the error associated with the distribution. The $N_{\rm HI}$ distribution is similar between the two samples within the errorbars. The figure also clearly reveals the redshift evolution in the distribution. The number of absorbers in the lower $N_{\rm HI}$ bin decreases and in the higher $N_{\rm HI}$ bins increases with increasing redshift. We compare our $N_{\rm HI}$ distributions with 
those obtained by \citet{kim2013} in two redshift bins: $z=1.9-2.4$ and $2.4-3.2$ using 18 high resolution, high SNR quasar spectra obtained from the ESO VLT/UVES archive. We find reasonable agreement between our $N_{\rm HI}$ distribution and that of \citet{kim2013} for the $2.3\le z\le 2.9$ bin. However, in the $1.7<z<2.3$ bin we do see our values are slightly higher than those of \citet{kim2013}.

As a sanity check for our simulations, we also plot the $N_{\rm HI}$ distribution obtained using the forward modelled sightlines for our fiducial simulation (with $\Gamma=\Gamma_{\rm obs}$ and $\rm SNR=SNR_{obs}$) at $z=2.0$, 2.6 and 3.2 in Fig.~\ref{NHI_distribution_KODIAQ_vs_SQUAD}. The simulated and observed distributions match well with each other in all the three redshift bins. However, we do find that the simulations to slightly underproduce the number of high column density(i.e log(\NHI)$\ge$15.0) absorbers at $z=2.0$ and 2.6 in comparison to the {\sc KODIAQ} and {\sc SQUAD} samples.

\begin{figure}
	\includegraphics[width=8cm]{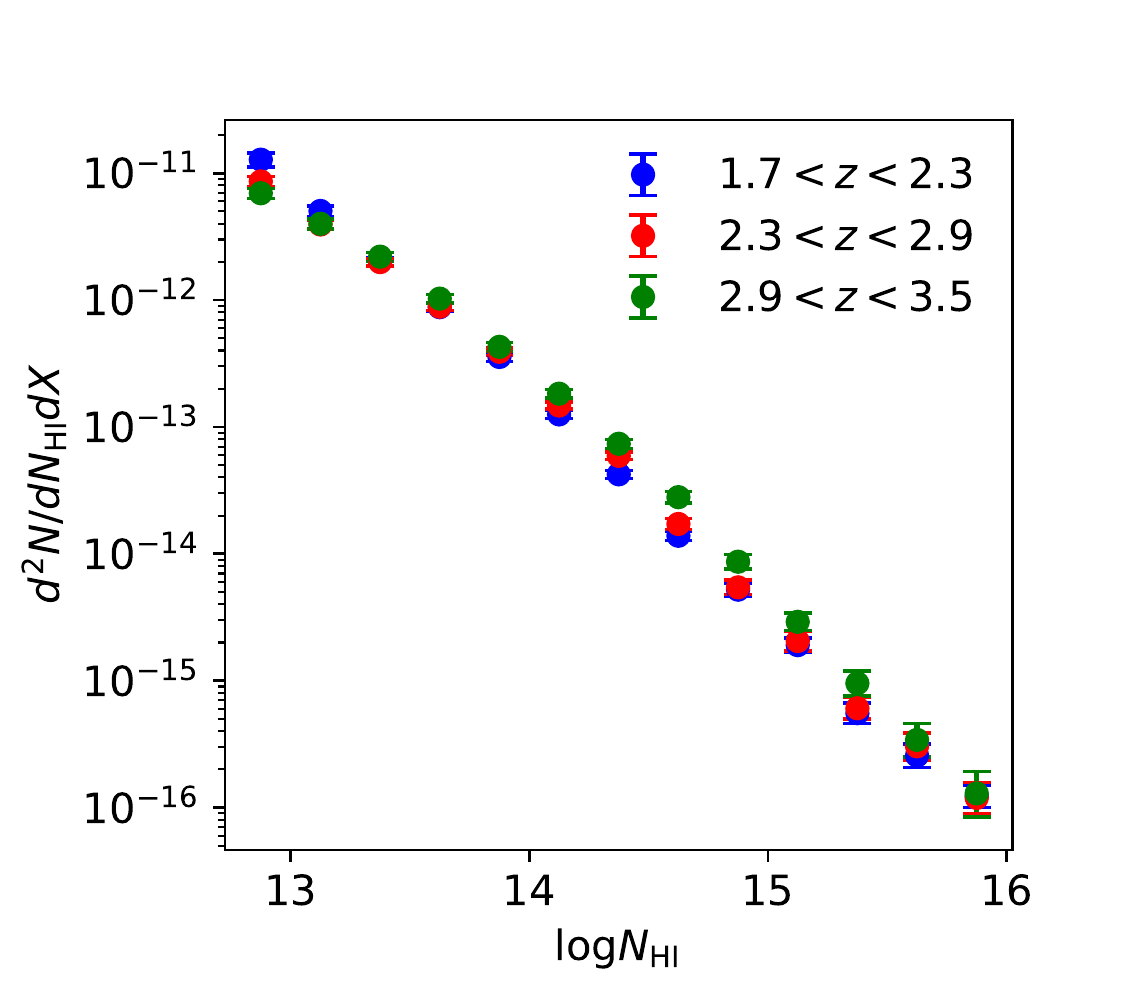}%

	\caption{ Measured \HI\ column density distribution of \lya\ absorbers in the combined {\sc KODIAQ+SQUAD} data sample at { three} redshift intervals.
	{ The errorbars are obtained as explained in Fig~\ref{NHI_distribution_KODIAQ_vs_SQUAD}}.}
\label{NHI_distribution_redshift}
\end{figure}

In Fig.~\ref{NHI_distribution_redshift}, we plot the $N_{\rm HI}$ distribution from the combined {\sc KODIAQ+SQUAD} sample for the three redshift bins. These values are tabulated in Table.~\ref{tab:NHI_dis} in the Appendix. We find indications for evolution of the $N_{\rm HI}$ distribution. The lowest redshift bin $1.7\le z\le 2.3$ has higher number of low $N_{\rm HI}$ absorbers ($N_{\rm HI}\sim 10^{13}$cm$^{-2}$) in comparison to the $2.9\le z \le 3.5$ bin. On the other hand, the highest redshift bin $2.9\le z \le 3.5$ has larger number of high $N_{\rm HI}$ absorbers ($N_{\rm HI}>10^{14}$cm$^{-2}$).

\section{Clustering of \lya\ absorbers}\label{Sec_Corr}
We study the clustering of \lya\ absorbers by decomposing the \lya\ forest into distinct Voigt profile components. This absorber-based approach allows one to study clustering of the \lya\ absorbers as a function of \NHI\ 
\citep[see][for early clustering studies based on voigt profile \lya\ components at high-$z$]{cristiani1997}.  In this section, we calculate the longitudinal (redshift space) two-point and {longitudinal} three-point correlation of absorbers in {\sc KODIAQ} and {\sc SQUAD} samples and { their dependence on $N_{\rm HI}$ and $z$}. 

\begin{figure*}
	\includegraphics[viewport=0 38 650 265,width=16cm, clip=true]{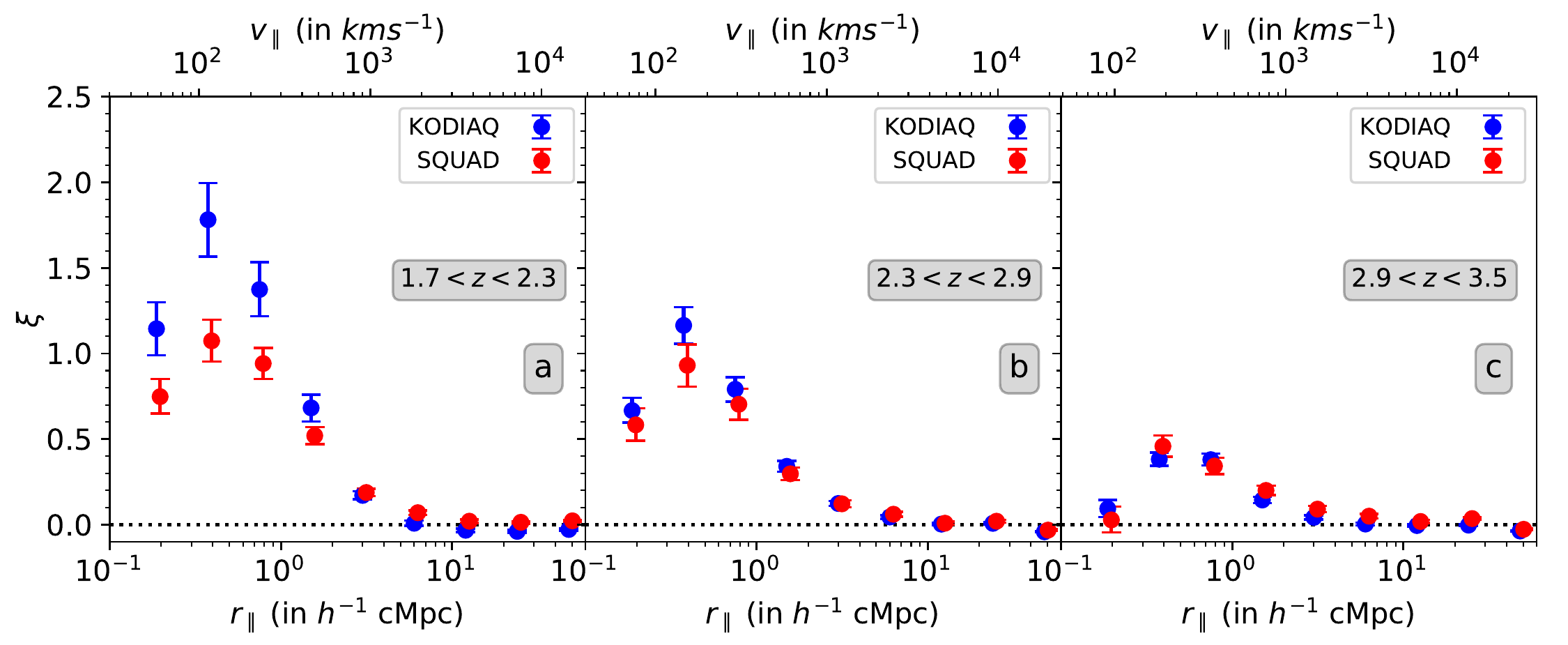}
	
	\includegraphics[viewport=0 0 650 234,width=16cm, clip=true]{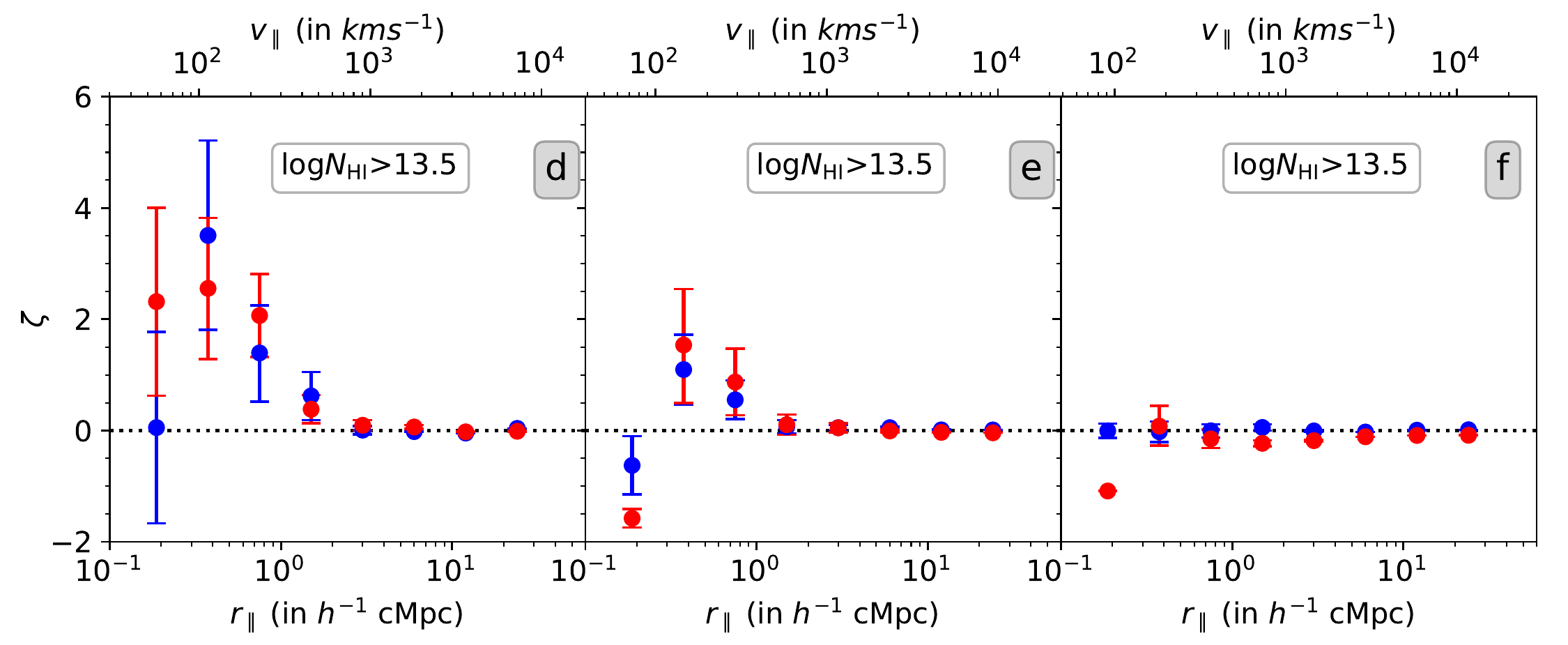}

	\caption{Absorber-based longitudinal two-point and three-point correlation (top to bottom) of \lya\ absorbers ($N_{\rm HI}>10^{13.5}$cm$^{-2}$) as a function of comoving longitudinal scale for three different redshift intervals of interest in this work. The velocity scale corresponding to the comoving longitudinal length scales are indicated in the top abscissa.
	The correlations are shown for {\sc KODIAQ} and {\sc SQUAD} sample separately. The errorbar shown corresponds to the larger of the two errors: one-sided poissonian uncertainty in the number of absorber pairs (or triplets) corresponding to $\pm 1\sigma$ or bootstrapping error. Both the sample show significant detections of two- and three-point correlation function.}
\label{Corr_KODIAQ+SQUAD}
\end{figure*}

\begin{figure*}
	\includegraphics[viewport=0 38 650 265,width=16cm, clip=true]{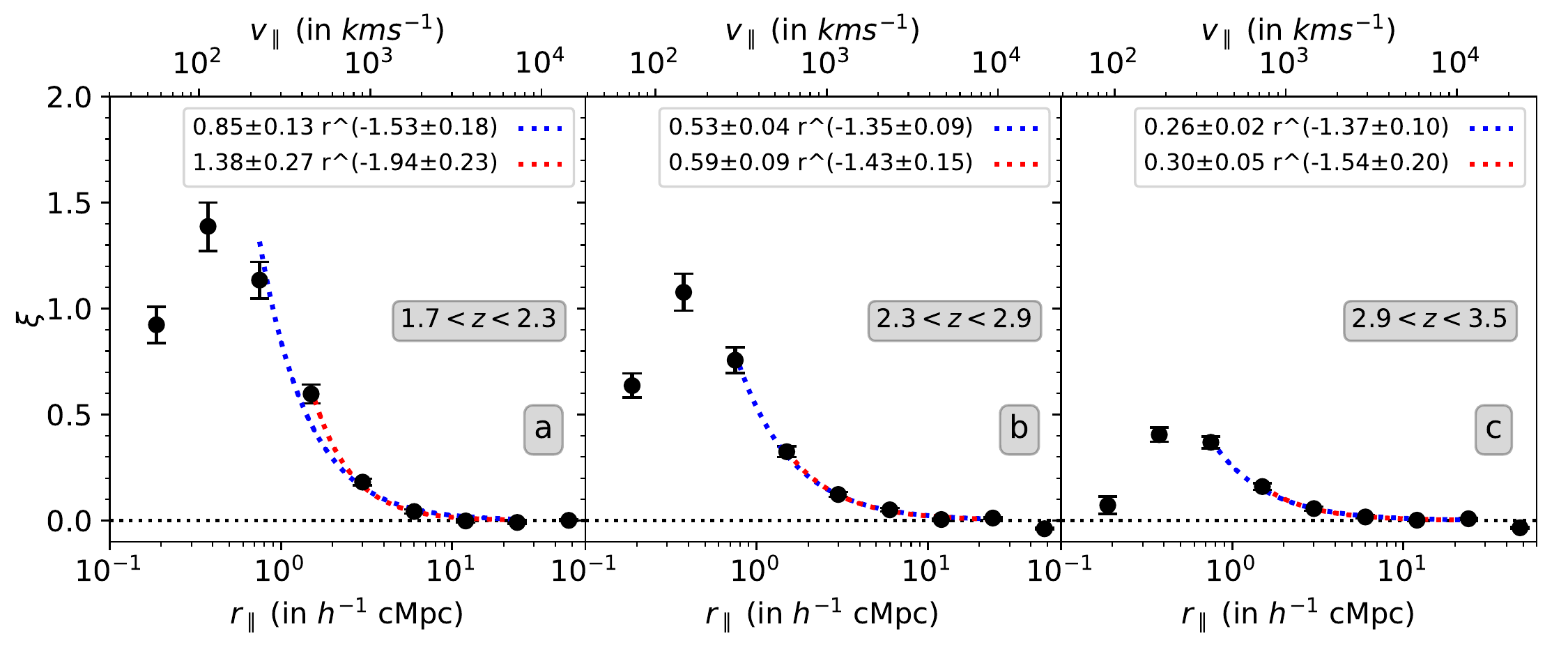}
	
	\includegraphics[viewport=0 38 650 234,width=16cm, clip=true]{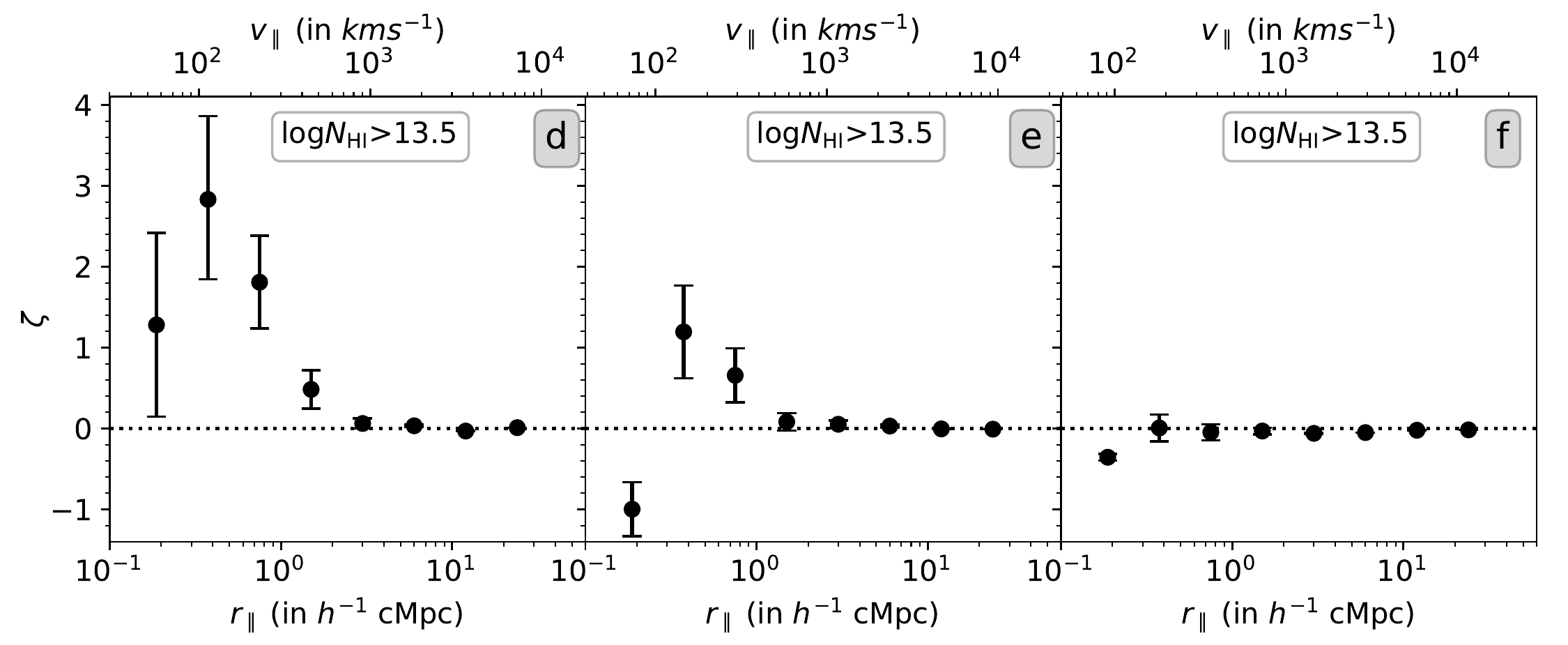}
	
	\includegraphics[viewport=0 0 650 234,width=16cm, clip=true]{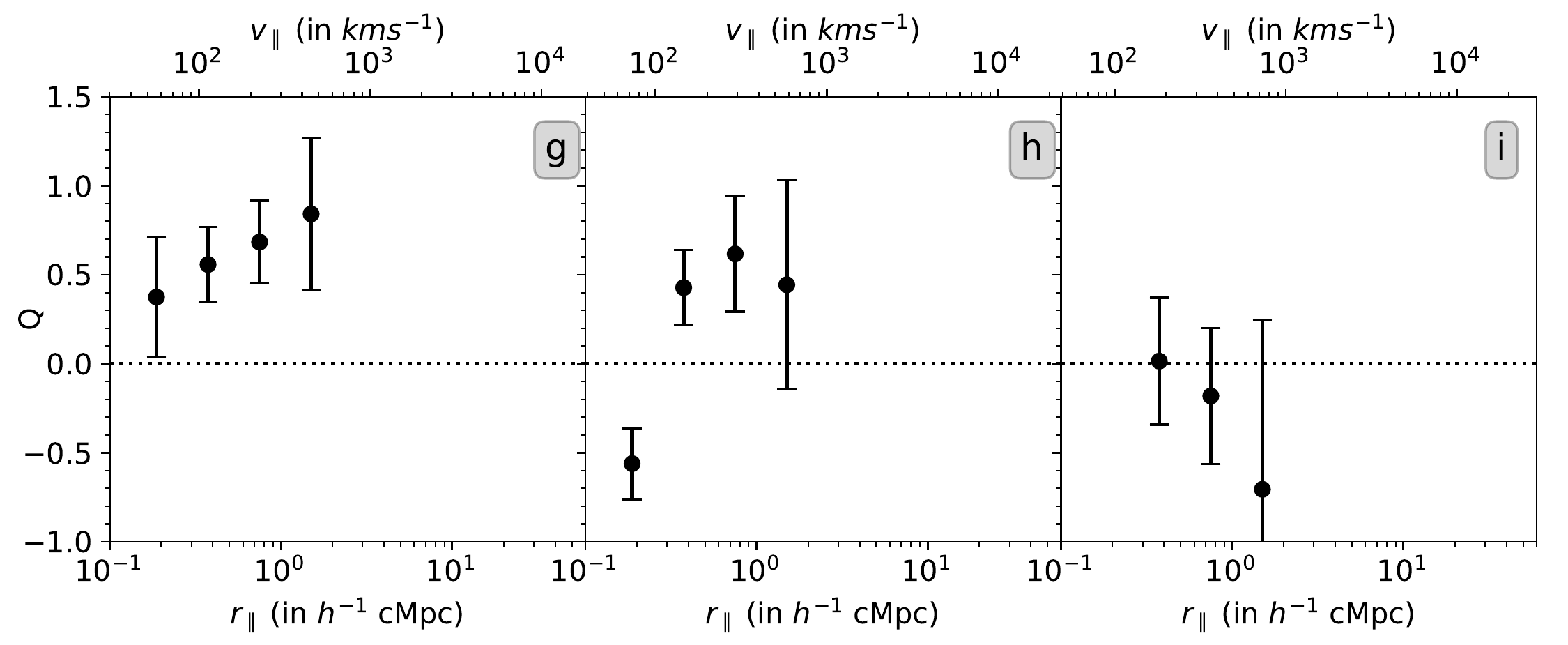}

	\caption{Absorber-based longitudinal two-point, three-point and reduced three-point correlation (top to bottom) of \lya\ absorbers ($N_{\rm HI}>10^{13.5}$cm$^{-2}$) as a function of longitudinal scale for three different redshift intervals. The corresponding velocity scales are indicated in the top abscissa.
	The correlations are shown for combined {\sc KODIAQ+SQUAD} sample. The errorbar shown corresponds to the larger of the two errors: one-sided poissonian uncertainty in the number of absorber pairs (or triplets) corresponding to $\pm 1\sigma$ or bootstrapping error. The coloured dashed curves in the top panels show the power-law best fit to the profile of two-point correlation at $r_{\parallel}>0.5h^{-1}$cMpc (blue) and $r_{\parallel}>1h^{-1}$cMpc. In each panel, we also provide the best fitted { power-law}.}
\label{Corr_KODIAQ+SQUAD1}
\end{figure*}

\subsection{Longitudinal two-point correlation}

We define longitudinal two-point correlation $\xi$ of \lya\ absorbers as the probability excess of detecting a pair of absorbers along the sightlines in redshift space. The estimator that we use is of the form
\begin{equation}
		\xi( r_{\parallel})=\frac{<DD>}{<RR>}-1 \ ,
\end{equation}	
where "DD" and "RR" are the counts of data-data and random-random pair of absorbers respectively at a separation of $r_{\parallel}$. We choose a normal estimator for the two-point correlation since the clustering amplitudes are relatively independent of the choice of estimators at small scales, as shown in \citet{kerscher2000}. We checked and found that it holds for \lya\ absorbers at the scales of interest in this study. Same holds for three-point correlation too (see Eq.~\ref{zeta}). The estimation of correlation is { carriedout} for absorbers above a certain $N_{\rm HI}$ threshold. The random distribution of absorbers for calculating RR is { computed} based on the observed $N_{\rm HI}$ distribution of the absorbers discussed in Section~\ref{sec:HI_dist}. The number of random absorbers to be populated along a sightline is decided by integrating Eq.~\ref{NHI_distribution_eqn} over a certain $N_{\rm HI}$ threshold and $dx$ corresponding to a redshift path length $dz$ along the sightline. 

Since the observed SNR does not vary appreciably along individual sightlines  \citep[unlike what we encounter at low-$z$ in][]{maitra2020b}, we assume a constant median SNR along each individual sightline while populating the random absorbers for our clustering analysis. The two-point correlation is estimated using 100 random sightlines for each of the data sightline and then by normalizing DD and RR with the total number of data-data ($n_D$) and random-random ($n_R$) pair combinations $n_D(n_D-1)/2$ and $n_R(n_R-1)/2$, respectively. We compute the two-point correlation in logarithmically spaced $r_{\parallel}$ bins of [0.125-0.25, 0.25-0.5, 0.5-1, 1-2, 2-4, 4-8, 8-16, 16-32, 32-64] $h^{-1}$cMpc.

Firstly, the correlations are calculated separately for $N_{\rm HI}>10^{13.5}$cm$^{-2}$ absorbers in the {\sc KODIAQ} and {\sc SQUAD} samples and compared in panel (a)-(c) in Fig.~\ref{Corr_KODIAQ+SQUAD} in the three redshift intervals as a function of longitudinal separation $r_{\parallel}$. We calculate the associated errors using two methods: one-sided poissonian uncertainty in the number of absorber pairs corresponding to $\pm 1\sigma$ and bootstrapping error. We take the larger of the two errors to be the error associated with the correlation. 
It is evident from the figure that the two-point correlation function measured for the two samples match very well for the two highest redshift bins. However in the low-$z$ bin we see differences in the measurements from the two sample. Note that in this redshift range the data are collected in the blue part of the CCD where the data are typically noisy. Also while the redshift path length probed for the two samples are nearly same (see Table~\ref{Table_KODIAQ}) typical SNR achieved and redshift coverage per sightline in the {\sc SQUAD} spectra are much higher than that of {\sc KODIAQ} as can be seen from the appreciable difference in the number of quasar sightline used. We also notice that within the narrow redshift bin the distribution of redshift range covered is also different. The {\sc SQUAD} sightlines typically probe lower-$z$ than the {\sc KODIAQ} sightlines.

The two-point correlation has a non-zero amplitude within $r_{\parallel}=8h^{-1}$cMpc at $1.7\le z\le 2.3$ and $2.3\le z\le 2.9$ redshift bins and within $r_{\parallel}=4h^{-1}$cMpc at $2.9\le z\le 3.5$ redshift bin. At scales below 0.25$h^{-1}$cMpc ($\sim$75 \kms\ at $z=2.0$, $\sim$97\kms\ at $z=2.6$ and $\sim$122\kms\ at $z=3.2$), we find a suppression in the correlation amplitudes at all the redshift bins.
This characteristic suppression can be attributed to a lower limit on the scale for identification of multiple \lya\ absorption lines set by a finite line-width of the thermally broadened absorption along with instrumental resolution and other systematic effects.  In section~\ref{sec:simulations}, we explore various physical parameters that can influence the measured two-point correlations at different scales (including the suppression seen at small scales) in detail using simulated data.

Next, we combine the {\sc KODIAQ} and {\sc SQUAD} data samples and plot the combined two-point correlation in panels (a)-(c) in Fig.~\ref{Corr_KODIAQ+SQUAD1}. The values are tabulated in Table.~\ref{tab:Corr} in the Appendix. We detect positive two-point correlation upto a length scale of $r_{\parallel}=8h^{-1}$cMpc in all the redshift bins. The strongest two-point correlation is seen at $r_{\parallel}=0.25-0.5h^{-1}$cMpc bin at $1.7\le z\le 2.3$ with an amplitude of $\xi=1.39\pm 0.11$. Strongest correlation is seen in the same $r_{\parallel}$ bin for $z=2.3\le z\le 2.9$ and $2.9\le z\le 3.5$ too. Their amplitude decreases with increasing redshift ($\xi=1.08\pm 0.09$ at $z=2.3-2.9$ and $0.41\pm 0.04$ at $z=2.9-3.5$).
Our measurements are consistent with  the early results of \citet{cristiani1997} albeit with much better sensitivity and over smaller redshift bins as expected due to the large sample used.

In the top panels of Fig.~\ref{Corr_KODIAQ+SQUAD1}, we fit the observed radial profile of clustering using a power-law. As $\xi$ measured at small scales are possibly affected by systematic effects we consider the fit for $r>0.5$ and 1.0 $h^{-1}$cMpc.
The best fit values are also shown in each panel.
For $z$ = 2.3-2.9 and 2.9-3.5 the best fitted values of power-law parameters are consistent within errors on the cutoff value of $r_{\parallel}$. Based on these fits we find the $\xi$($r_{\parallel}$=1$h^{-1}$cMpc) = 0.53$\pm$0.04 and 0.26$\pm$0.02 respectively for these two redshifts. In the case of $z=1.7-2.3$ bin, it appears that the single power-law will not provide adequate fit to all the data above $r>0.5h^{-1}$ cMpc. We find $\xi$($r_{\parallel}$= 1$h^{-1}$cMpc) = $0.85\pm0.13$ based on fits to data points at $r>0.5h^{-1}$ cMpc. Thus the observed clustering of \lya\ absorbers with 
log~\NHI $>$ 13.5 show strong evolution in amplitude over the redshift range considered. If we approximate the redshift evolution to a power law then we find roughly $\xi \propto (1+z)^{-3.5}$. This evolution is much steeper than what is expected purely based on dark matter evolution \citep[i.e $(1+z)^{-5/3}$, see][]{cristiani1997}. We present a detailed  discussion on the redshift evolution in section~\ref{Sec:evolution}.

The powerlaw coefficient of radial profile of the two-point function ranges between $-1.35$ and $-1.94$.  
In \citet{maitra2020}, we found that the radial profile of the { simulated} two-point function in the longitudinal direction are well approximated by a powerlaw with the index of $\sim -0.8$. It will be important to confirm such a difference in real data using transverse correlations. In section~\ref{sec:simulations}, we discuss the effect of of peculiar velocities on the radial profile.

\subsection{Longitudinal three-point correlation}

For calculating the longitudinal three-point correlation function, we use a scheme similar to \citet{maitra2020b}. We define triplets as three collinear points {along the sightline} where the longitudinal separation between the first and second point is $r_{1,\parallel}$ and between the second and third point is $r_{2,\parallel}$. For this work, we consider triplets having $r_{1,\parallel}=r_{2,\parallel}=r_{\parallel}$ \footnote{ This configuration is equivalent to the squeezed triplet configurations ($\theta=0^{\circ}$) with 1:2 arm length ratio.}. The probability excess of detecting a \lya\ triplet along the sightlines in redshift space having this separation $r_{\parallel}$, is given as 
\begin{equation}
		{\rm PE_3}=\frac{<DDD>}{<RRR>}-1 \ ,
\end{equation}	
where "DDD" and "RRR" are the counts of data-data-data and random-random-random triplets of absorbers respectively. The three-point correlation $\zeta$ is computed by subtracting the two-point correlation contributions to this probability excess;
\begin{equation}\label{zeta}
		\zeta( r_{\parallel})={\rm PE_3}-\xi(r_1)-\xi(r_2)-\xi(r_3) \ ,
\end{equation}
where 
 $\xi(r_1),\xi(r_2)$ and $\xi(r_3)$ are the two-point correlations for arm lengths joining the three points considered for three-point correlation \citep[see Eq. 18 and 20 of][]{peebles1975}. The $\rm PE_3$ and $\zeta$ are estimated using 1000 random sightlines for each of the data sightline and then by normalizing DDD and RRR with the total number of data-data and random-random pair combinations $n_D(n_D-1)(n_D-2)/6$ and $n_R(n_R-1)(n_R-2)/6$, respectively. 
 { We compute the three-point correlation in the same set of logarithmically spaced bins used above for two-point correlation.}
 
 Similar to two-point correlation, we also plot the three-point correlation separately for the {\sc KODIAQ} and {\sc SQUAD} samples first as a function of longitudinal separation $r_{\parallel}$ in panel (d)-(f) of Fig.~\ref{Corr_KODIAQ+SQUAD}. The errorbars are calculated similar to two-point correlation. Both the samples give similar three-point correlation function in all the redshift bins. Only in the first $r_{\parallel}$ bin of $z=2.3-2.9$ and $2.9-3.5$ redshift intervals we do find some discrepancy which can come from difference in systematics in the two samples. We combine the two samples and plot the resulting three-point correlation as a function of $r_{\parallel}$ in panels (d)-(f) of Fig.~\ref{Corr_KODIAQ+SQUAD1}.
 We clearly detect three-point correlation at a much higher significant level.
 
 Similar to two-point correlation, the three-point correlation is also largely suppressed in the $r_{\parallel}=0.125-0.25h^{-1}$cMpc bin. At the lowest redshift bin, we detect positive three-point correlation upto a scale of 2$h^{-1}$cMpc. The strongest detections in three-point correlation comes at $r_{\parallel}=0.25-0.5$ and $0.5-1$ $h^{-1}$cMpc bins ($\zeta=2.83\pm 1.02$ and $\zeta=1.81\pm 0.59$, respectively) with a significance of 2.8 and 3.1 $\sigma$ level. In the intermediate redshift bin (i.e., $2.3\le z\le 2.9$), positive three-point correlation is seen up to a scale of 1$h^{-1}$cMpc only. The three-point correlation values at $r_{\parallel}=0.25-0.5$ and $0.5-1$ $h^{-1}$cMpc bins are $\zeta=1.19\pm 0.59$ and $\zeta=0.66\pm 0.35$, respectively. In the highest redshift bin (i.e $2.9\le z \le 3.5$), we detect no positive three-point correlation. 
 
 In case of galaxies, it is common to express the three-point correlation by normalizing it with a cyclic combination of two-point correlations associated with the three arms of the triplets. This quantity is called the reduced three-point correlation, Q, and is defined as,
 \begin{equation}
    Q=\frac{\overline{\zeta(r_1,r_2,r_3)}}{\overline{\xi(r_1)}\times\overline{\xi(r_2)}+\overline{\xi(r_2)}\times\overline{\xi(r_3)}+\overline{\xi(r_1)}\times\overline{\xi(r_3)}} \ ,
\label{Q_eqn}
\end{equation}
where $r_1,r_2$ and $r_3$ are the arm lengths joining the three points considered for three-point correlation \citep[see][]{groth1977,peebles1980}. In this study, we have taken $r_1=r_2= r_{\parallel}$ and $r_3=2\times r_{\parallel}$. We plot Q as a function of $r_{\parallel}$ in panels (g)-(i) of Fig.~\ref{Corr_KODIAQ+SQUAD1} for the three redshift bins. We plot Q up to a scale $r_{\parallel}=2h^{-1}$ cMpc where we have non-zero three-point correlation detection. In the lowest redshift bin, the Q values corresponding to the strongest three-point correlation detections at $r_{\parallel}=0.25-0.5$ and $0.5-1$ $h^{-1}$cMpc are $0.56\pm 0.21$ and $0.68\pm 0.23$, respectively. We do not see any indication of scale dependence in Q within the errorbars for the length scale probed. In the intermediate redshift bin, the Q values at the similar length scales are $0.43\pm 0.22$ and $0.62\pm 0.33$, respectively. As mentioned before, no positive three-point correlation is detected in the largest redshift bin. All the values of three-point and reduced three-point correlation are tabulated in Table.~\ref{tab:Corr} in the Appendix.

\subsection{Dependence of \NHI\ threshold}

\begin{figure*}
	\includegraphics[viewport=0 32 650 250,width=16cm, clip=true]{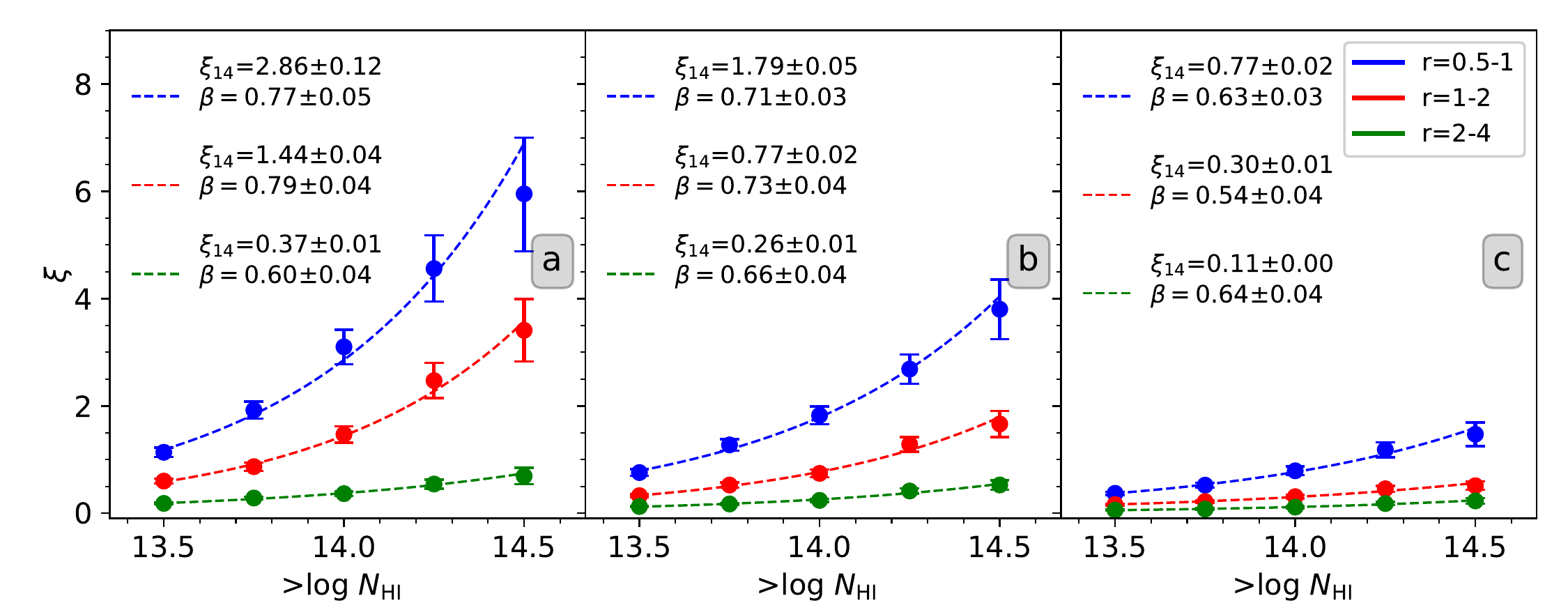}
	
	\includegraphics[viewport=0 32 650 250,width=16cm, clip=true]{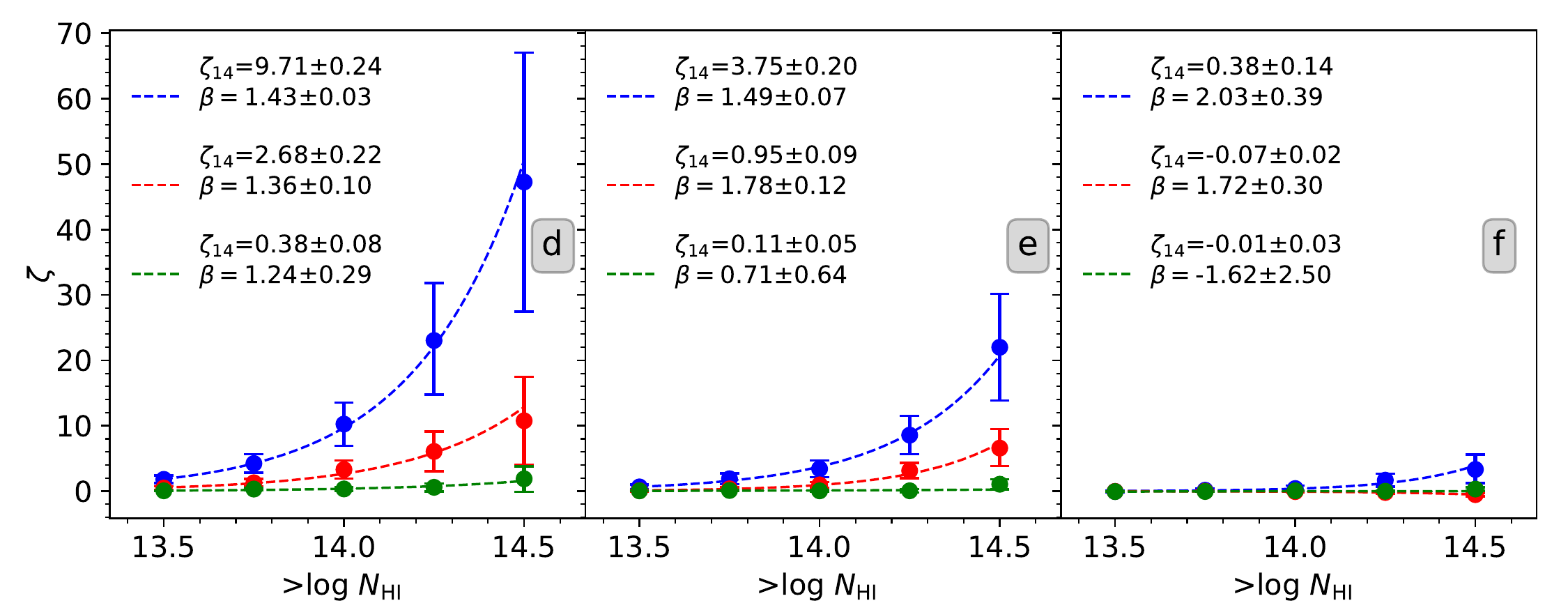}
	
	\includegraphics[viewport=0 0 650 250,width=16cm, clip=true]{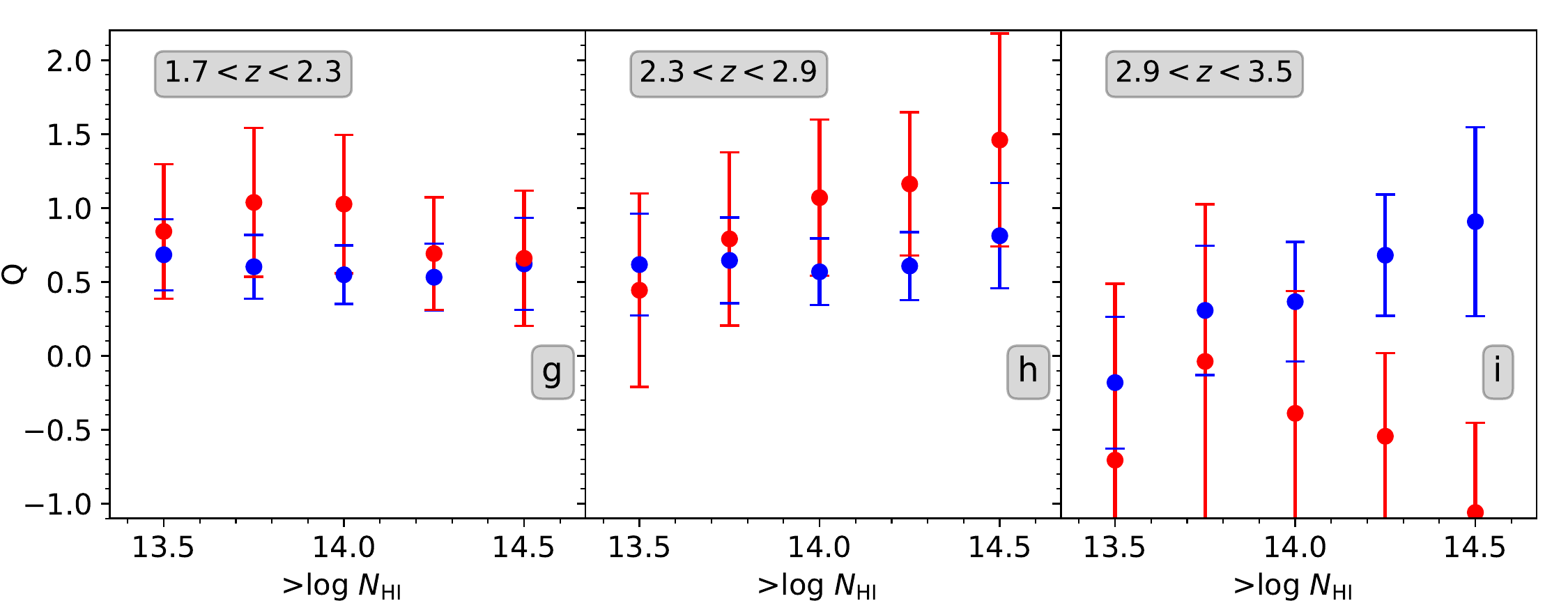}

\caption{Absorber-based longitudinal two-point, three-point and reduced three-point correlation (top to bottom) of \lya\ absorbers as a function of $N_{\rm HI}$ thresholds in different redshift intervals (left to right). The plots are made for scales $r_{\parallel}=$[0.5-1 ({ blue}), 1-2 ({ red}) and 2-4 ({ green})]$h^{-1}$cMpc. The dashed curves represent the power law-fit for the two-point ($\xi=\xi_{14}\mathcal{N}_{14}^{\beta}$, where $\mathcal{N}_{14}$ is the limiting ${N}_{\rm HI}$ in units of $10^{14}$cm$^{-2}$) and three-point correlation ($\zeta=\zeta_{14}N_{14}^{\beta}$) as a function of $N_{\rm HI}$ thresholds. The errorbars shown correspond to the larger of the two errors: one-sided poissonian uncertainty in the number of absorber pairs (or triplets) corresponding to $\pm 1\sigma$ or bootstrapping error. }
\label{Corr_NHI}
\end{figure*}

It is well known that the two-point correlation function of both low and high $z$ \lya\ absorbers strongly depends on \NHI \citep[][]{cristiani1997, Penton2002,  danforth2016}.
In our past studies involving redshift space clustering of low-$z$ \lya\ absorbers in observations \citep{maitra2020b} or transverse clustering of $z\sim 2$ absorbers in simulations \citep{maitra2020}, we had found that both two-point as well as three-point correlation depends strongly on the \NHI\ of the absorber. Here, we will investigate the \NHI\ dependence of redshift space two-point, three-point and reduced three-point correlation in the three redshift bins: $z = 1.7-2.3,~ 2.3-2.9$ and $2.9-3.5$. We do so for three longitudinal scales: $r_{\parallel}=0.5-1,\ 1-2$ and $2-4h^{-1}$cMpc. In case of reduced three-point correlation, we omit the $2-4h^{-1}$cMpc scale since the signals are very noisy. The results are plotted in Fig.~\ref{Corr_NHI}.

We find a strong \NHI\ dependence of both two-point and three-point correlation at all scales and in all the redshift bins. To quantify this trend, we fit the \NHI\ dependence of two-point correlation with a power-law fit of the form $\xi=\xi_{14}\mathcal{N}_{14}^{\beta}$, where $\mathcal{N}_{14}$ is limiting $N_{\rm HI}$ in units of $10^{14}$cm$^{-2}$. We do this for three-point correlation too ($\zeta=\zeta_{14}\mathcal{N}_{14}^{\beta}$). At $z=1.7-2.3$, we find strong \NHI\ dependence on two-point correlation. This dependence is slightly stronger at smaller scales with a power-law index $\beta>0.7$ in comparison to $r_{\parallel}=2-4h^{-1}$cMpc where $\beta\sim 0.6$. At the smallest comoving scale (0.5-1$h^{-1}$cMpc), the amplitude of the power-law fit for two-point correlation evolves strongly from $\xi_{14}=0.77\pm 0.02$ at $2.9\le z \le 3.5$ to $\xi_{14}=2.86\pm 0.12$ at $1.7\le z \le 2.3$. This trend is true for the two larger scales also. There is also a slight indication of evolution in the power-law index $\beta$ at the scales $r_{\parallel}=0.5-1$ and $1-2h^{-1}$cMpc for the two-point correlation.

In the case of three-point correlation, the \NHI\ dependence is stronger. In fact, the power-law index $\beta$ for three-point correlation is roughly double of the $\beta$ for two-point correlation. However, due to large errorbars on the fits, much of the trends seen in two-point correlation cannot be confirmed for three-point correlation. Interestingly, unlike two-point and three-point correlation, the reduced three-point correlation Q is relatively independent of $N_{\rm HI}$ thresholds and this remains true for all the redshift bins. This trend is similar to what has been found in our previous studies too.

\subsection{Effect of metal line contamination}
\label{sec:metal_contamination}

\begin{figure}
	\includegraphics[viewport = 29 39 500 290, clip =true, width=0.5\textwidth]{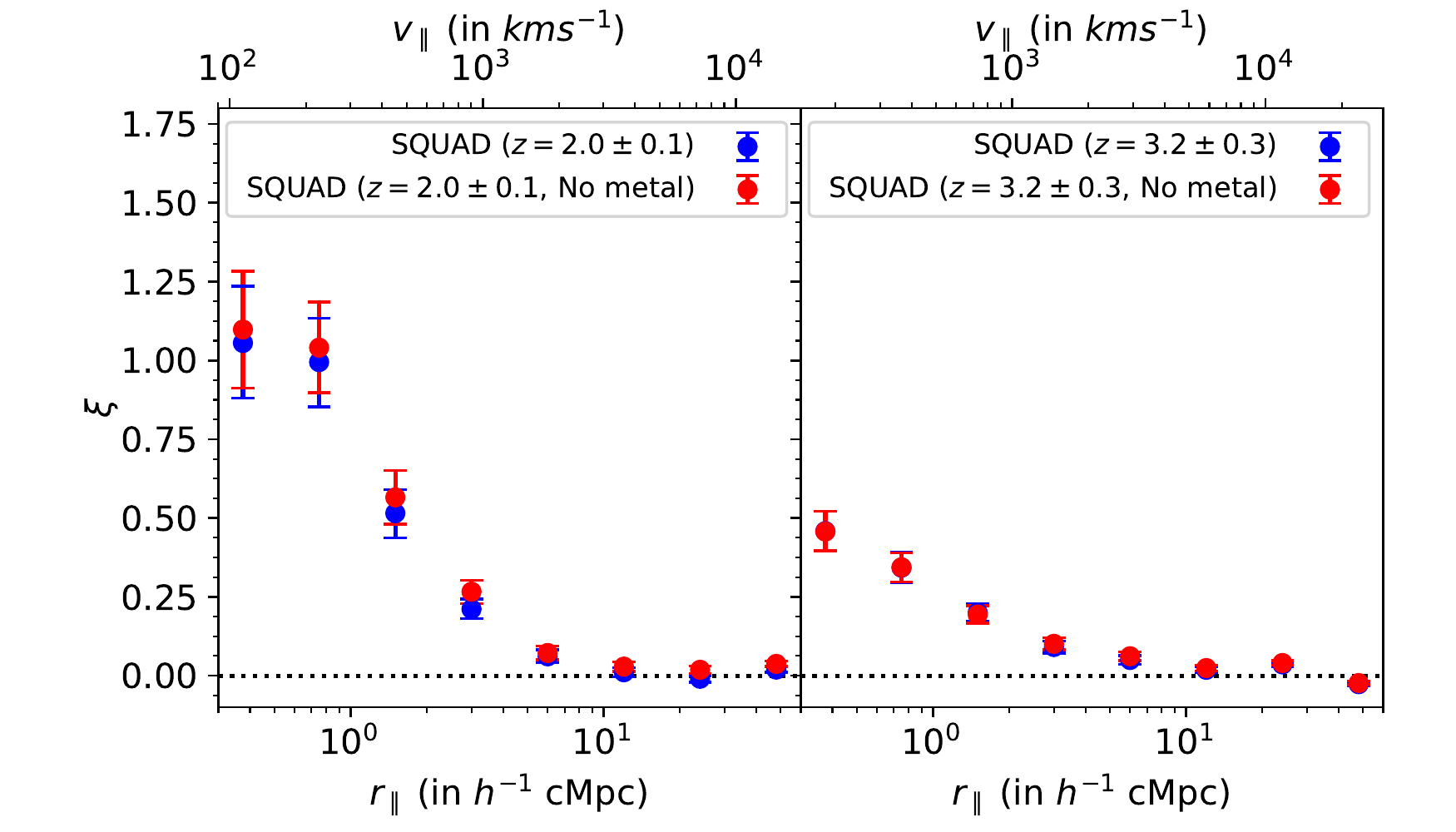}
	
	\includegraphics[viewport = 29 5 500 256, clip =true,width=0.5\textwidth]{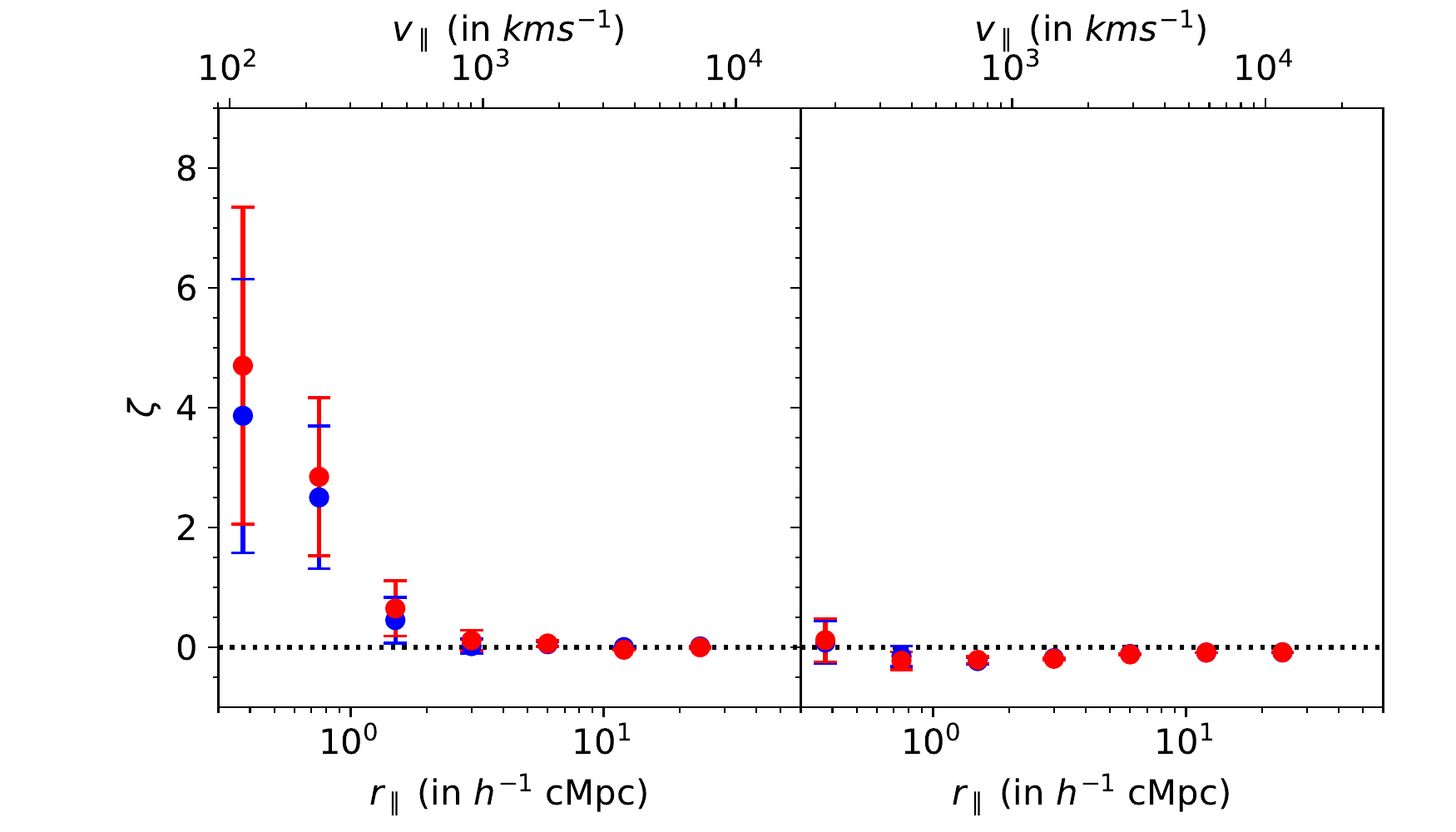}

	\caption{Effect of metal contamination on absorber-based longitudinal two-point and three-point correlation (top to bottom) of \lya\ absorbers ($N_{\rm HI}>10^{13.5}$cm$^{-2}$) as a function of longitudinal scale for different redshift intervals. { Measurements with and without metal line contamination are shown in blue and red points respectively}. The errorbars shown correspond to the larger of the two errors: one-sided poissonian uncertainty in the number of absorber pairs (or triplets) corresponding to $\pm 1\sigma$ or bootstrapping error.}
\label{Corr_metal}
\end{figure}

Clustering studies in \lya\ forest can be affected by contamination coming from absorption lines of metal ions. As a first step, we consider two sub-samples of \lya\ forest data from the full sample: one in the redshift range of $z=1.9\le z\le2.1$  and other for $2.9\le z\le 3.5$ in the {\sc squad} survey.
These two sub-samples  belong to two redshift extremities (described in section~\ref{sec:Dataset}) in our full sample and can represent the trend in redshift evolution of metal contamination, if any. Next, we identify metal line systems in the two sub-sample manually and mask the \lya\ forest affected by the contamination from these systems. 
First we find that the $N_{\rm HI}$ distribution of the absorbers in the two redshift sub-samples with and without metal contamination are nearly identical.
Same is true for two-point and three-point correlation, as can be seen in Fig.~\ref{Corr_metal}. The effect of metal contamination is much less significant as compared to the errorbars presented Fig.~\ref{Corr_metal} or even for our full sample in Fig.~\ref{Corr_KODIAQ+SQUAD}. So, we will neglect the effect of metal contamination for this study.

\section{Effect of various parameters on the clustering of \lya\ absorbers}\label{Astro_effects}
\label{sec:simulations}

\begin{figure}
    \centering
     \includegraphics[viewport=0 38 315 290,width=8cm, clip=true]{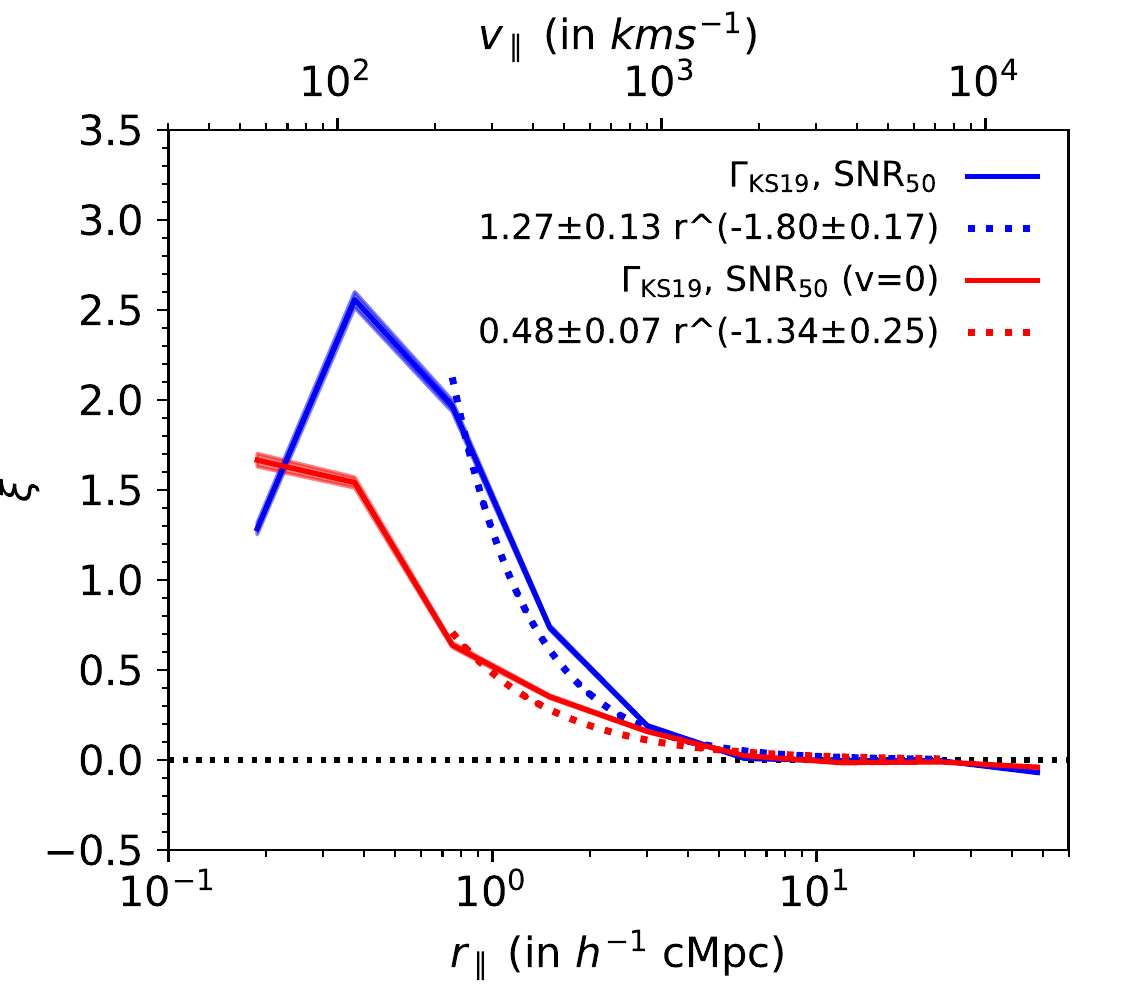}
     
	\includegraphics[viewport=0 0 315 260,width=8cm, clip=true]{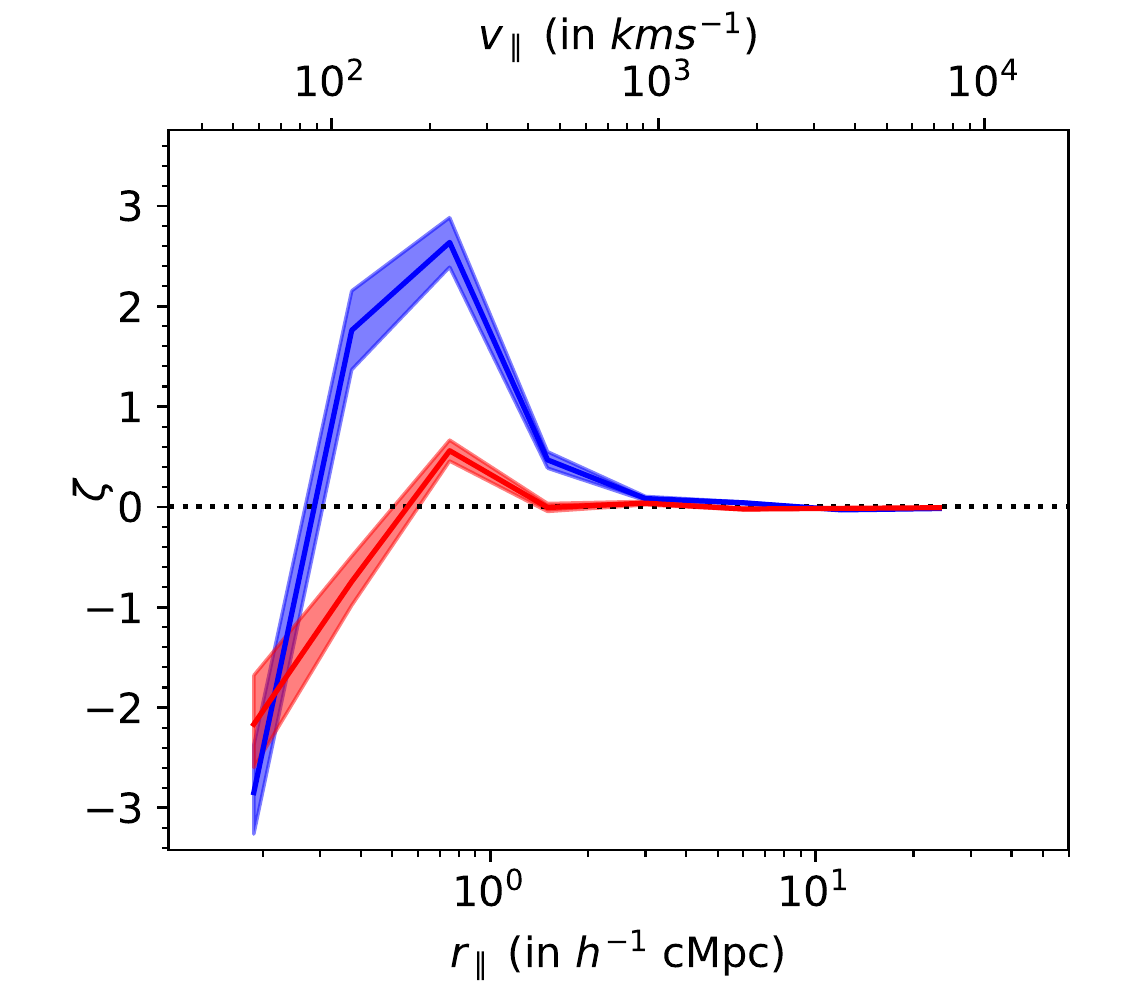}%

    \caption{ Effect of peculiar velocity on two-point and three-point correlation at $z=2.0$. { Results for models with (blue) and without (red) the effects of peculiar velocities are shown}. The dashed curves in the top panel show the power-law best fit models for the two-point correlation at $r_{\parallel}>0.5h^{-1}$cMpc.
    }
    \label{Corr_v}
\end{figure}

\begin{figure*}
    \centering
     \includegraphics[width=18cm]{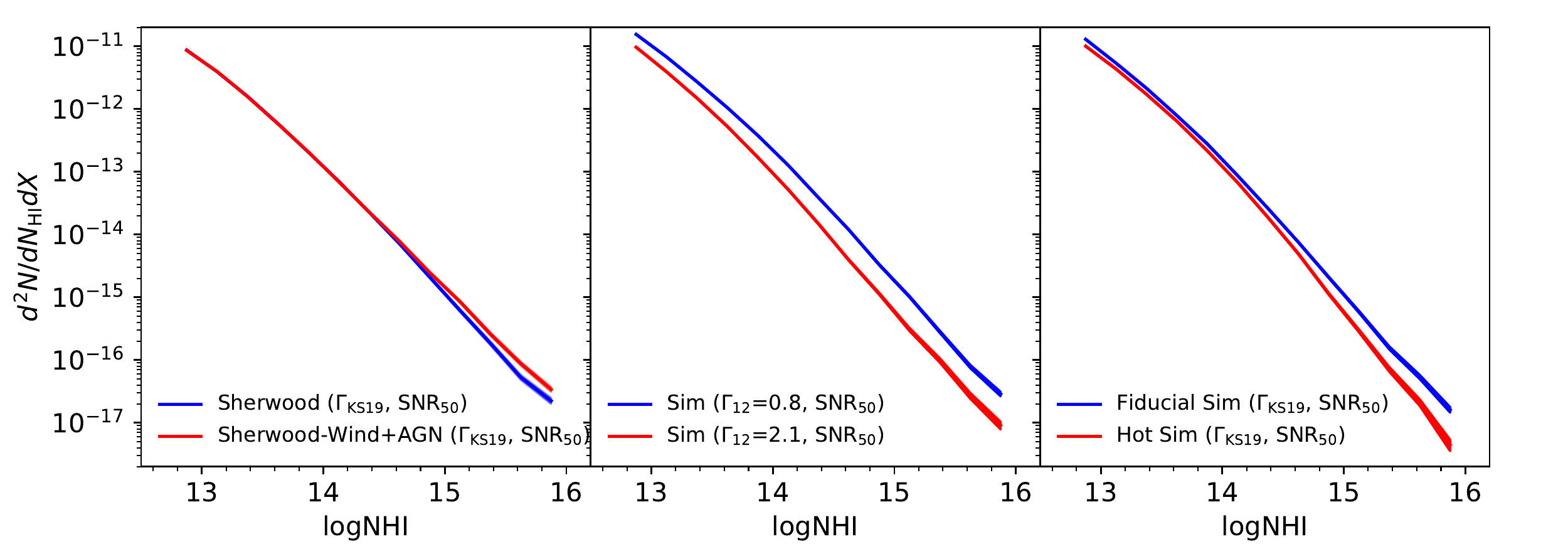}%
    \caption{Effect of feedback process (left), \HI\ photo-ionization rates ($\Gamma_{\rm HI}$; middl) and for thermal history (right) on the \HI\ column density distribution function at $z\sim 2$. { Various simulations used are indicated in each panel}.}
    \label{NHI_dis_effects}
\end{figure*}

In this section, we investigate the effect of peculiar velocity and astrophysical parameters of the IGM on the clustering properties of \lya\ absorbers using simulation box at $z=2$. Firstly, we check the effect of peculiar velocities using spectra generated by switching off the velocity field (other than thermal and natural broadenning) while generating the spectrum. Next we check how changing the UV background changes clustering by simply varying the $\Gamma_{\rm HI}$ parameter during the simulated flux generation stage. 
We will then study the effect of the  thermal state and thermal history of the IGM using the two simulation boxes mentioned in Section~\ref{Simulation}: the "fiducial" box with KS19 UVB (i.e $\alpha =1.8$) and the "hot" box with "Enhanced $\alpha=1.8$" UVB. We will also investigate the impact of wind and AGN feedback on the clustering signal using Sherwood simulations. 

\subsection{Effect of peculiar velocities}

We check the effect of peculiar velocities on the clustering of \lya\ absorbers using our fiducial simulation box. We turn off the peculiar velocities of the SPH particles in our simulation during the forward modelling process and generate simulated spectra. We compute the two- and three-point correlations of the \lya\ absorbers obtained by Voigt profile fitting of these spectra. We then compare this two- and three-point correlations obtained with the ones computed with peculiar velocities turned on. The results are plotted in Fig.~\ref{Corr_v}. In case of two-point correlation, we find that correlation amplitude is amplified at all scales greater than 0.25$h^{-1}$cMpc when the peculiar velocity is turned on. The amplification is strongest at $r_{\parallel}=0.5-1h^{-1}$cMpc where peculiar velocity increases the correlation amplitude by a factor $\sim 3$. The amplification becomes weaker at larger scales with an amplification of only $\sim 1.2$ at $r_{\parallel}=2-4h^{-1}$cMpc. It is also evident from the figure that the radial profile of $\xi(r)$ is steepened appreciably by the presence of peculiar velocity. The $\xi(r)$ obtained when the peculiar velocity is off is still steeper than $\xi(r)$ profile seen { along the transverse direction} in our simulations \citep[see figure 7  of][]{maitra2020}. It will be interesting to see whether such a trend is also seen in observed transverse correlations. Simultaneous measurement of transverse and longitudinal correlations will allow us to probe the parameters of the background cosmology \citep[the so called Alcock Paczy\'nski test,][]{Alcock1979}.

The effect of peculiar velocity is much more drastic in the case of three-point correlation. We find positive three-point correlation only at the scale of $r_{\parallel}=0.5-1h^{-1}$cMpc { when the peculiar velocities are switched-off}. At this scale, the three-point correlation is amplified by a factor of $\sim 5$ by the presence of peculiar velocity.
The trends seen here are consistent with what we found for low-$z$ absorbers in \citet{maitra2020b}. In that case, we notice the presence of peculiar velocities increasing the two- and three-point correlations by  factors of $\sim$ 1.7 and 2.1 respectively at scales of 1-2 pMpc ($z$ = 0.1 simulations).
Thus it appears that \lya\ absorbers observed over the full redshift range (i.e $0\le z\le 3.5$) are part of converging flows. { It also 
emphasises the importance of accurately modelling the peculiar velocity field (i.e redshift space distortions) in order to measure the correct spatial clustering (and bias parameters) from the redshift space clustering.}

\subsection{$\Gamma_{\rm HI}$ dependence}

\begin{figure*}
    \centering
     \includegraphics[viewport=0 0 315 290,width=6cm, clip=true]{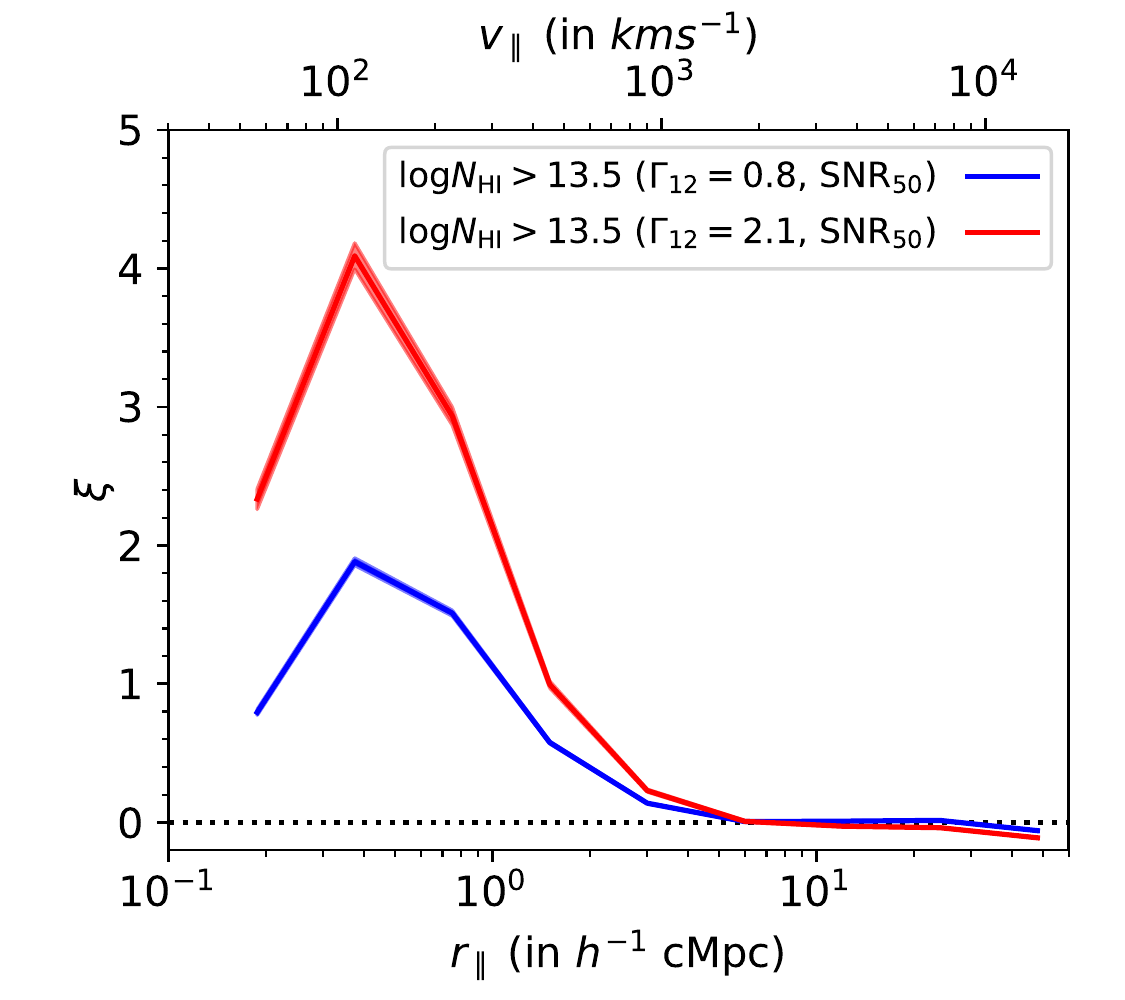}%
	\includegraphics[viewport=0 0 315 290,width=6cm, clip=true]{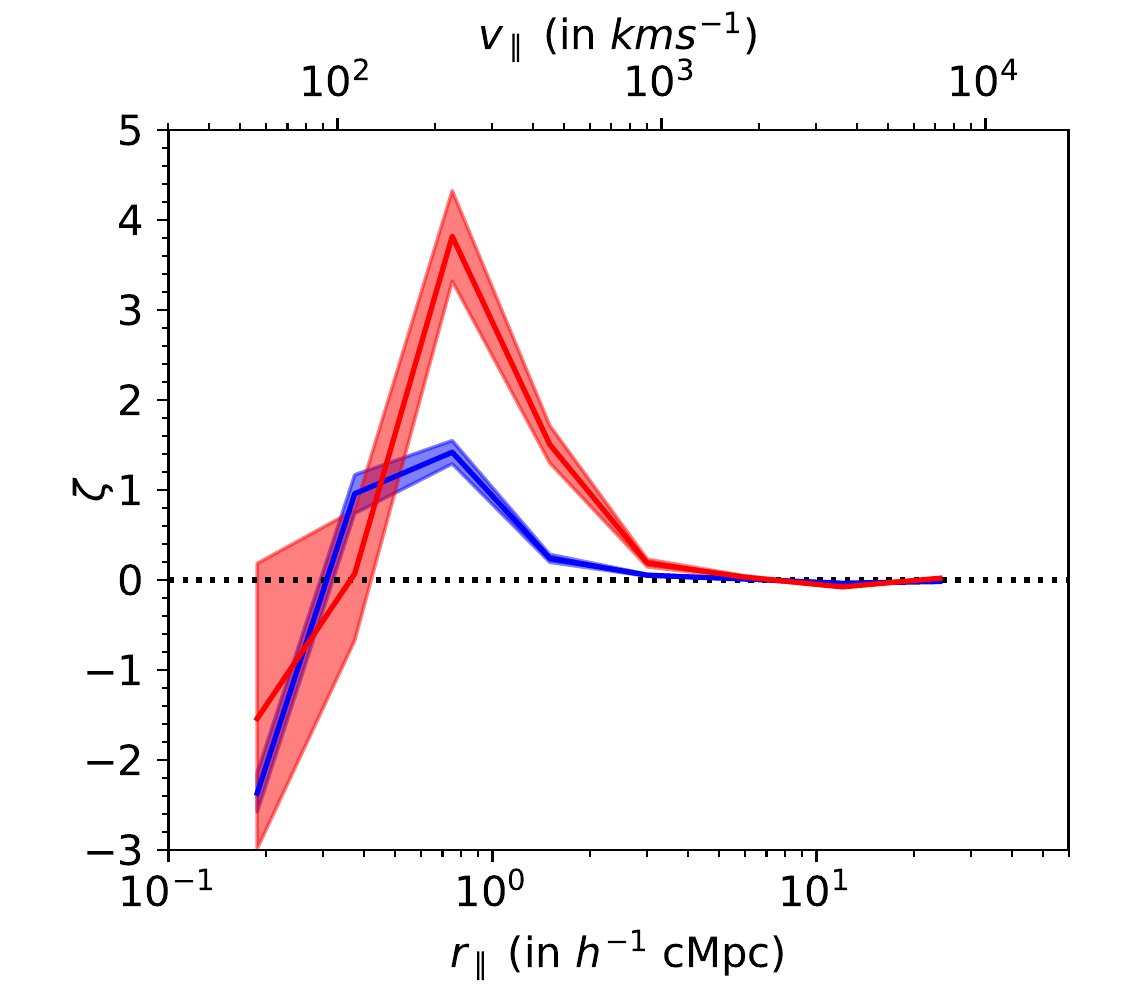}%
	\includegraphics[viewport=0 0 315 290,width=6cm, clip=true]{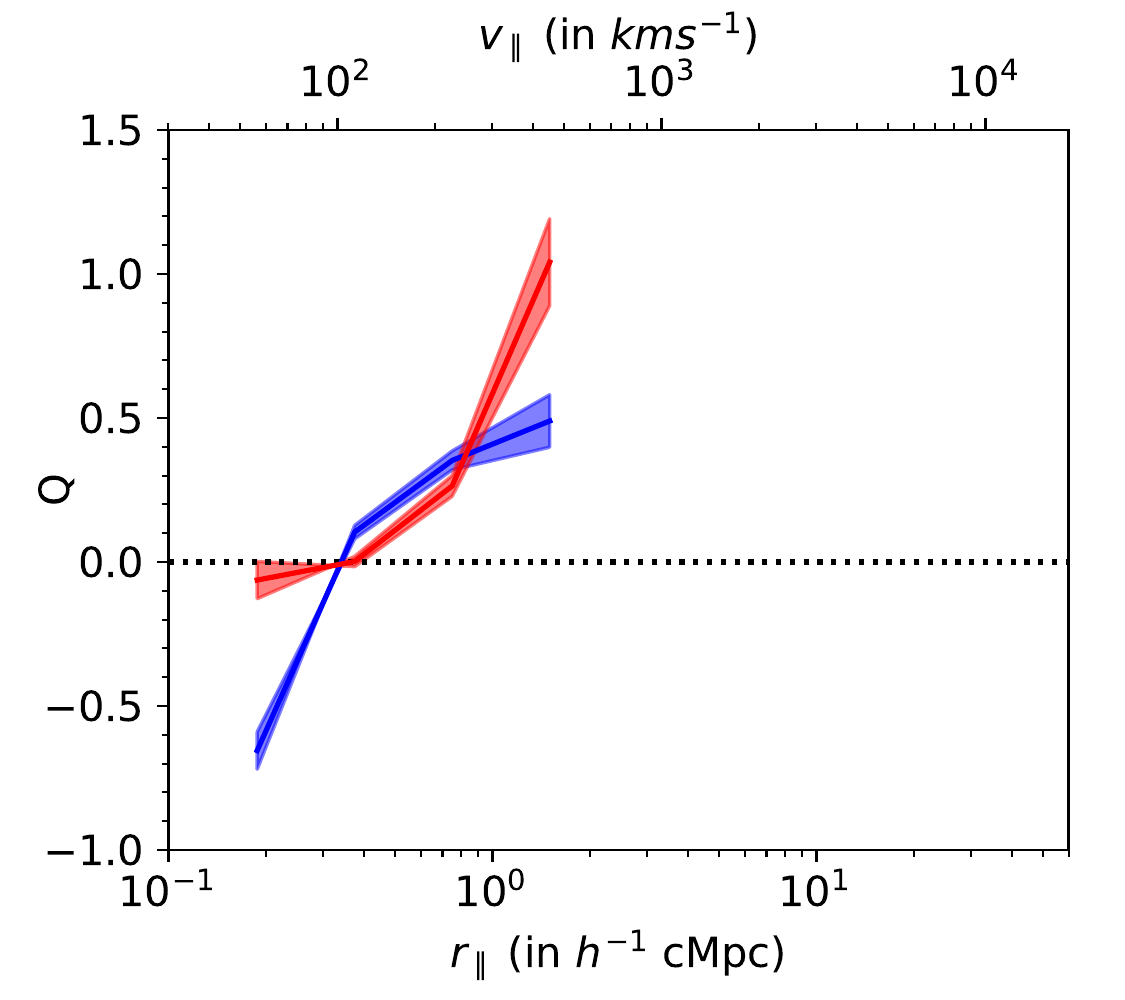}%

    \caption{ { Simulated} two-, three- and reduced three-point correlation (left to right) of \lya\ absorbers having $N_{\rm HI}>10^{13.5}$cm$^{-2}$ as a function of longitudinal separation $r_{\parallel}$ for two-different $\Gamma_{\rm HI}$ values: one corresponding to the upper limit of the errorbar of $\Gamma_{\rm HI}$ reported at $z=2$ in \citet{bolton2007} ($\Gamma_{12}=2.1$; { red curves}) and the other corresponding to the lower limit of the errorbar of ($\Gamma_{12}=0.8$; { blue curves}). It is evident from this figure that the clustering properties of absorbers above a given \NHI\ threshold depend strongly on $\Gamma_{\rm HI}$.
    }
    \label{Corr_Gamma_HI}
\end{figure*}

By changing the $\Gamma_{\rm HI}$ in the simulation, one effectively changes the ionization state of the IGM; a higher $\Gamma_{\rm HI}$ results in the \HI\ population being more photo-ionized and vice-versa. This alters the kind of baryonic overdensities that one probes for a given $N_{\rm HI}$ value. So, an absorber with a given $N_{\rm HI}$ will probe higher baryonic overdensities if the IGM is more ionized. This, in turn, would mean that for absorbers with a certain minimum $N_{\rm HI}$ threshold, the clustering signal would go up if one amplifies the $\Gamma_{\rm HI}$, since correlations would be coming from larger overdensities. With this in mind, we simulate transmitted flux with two $\Gamma_{\rm HI}$ values: one corresponding to the upper limit of the errorbar of $\Gamma_{\rm HI}$ reported at $z=2$ in \citet{bolton2007} ($\Gamma_{12}=2.1$) and the other corresponding to the lower limit of the errorbar of ($\Gamma_{12}=0.8$). 
First we compare the $N_{\rm HI}$ distribution between these two $\Gamma_{\rm HI}$ values.
in the middle panel of Fig.~\ref{NHI_dis_effects}. 
As expected, the one with larger $\Gamma_{\rm HI}$ value is more ionized and has a lower $N_{\rm HI}$ distribution. 

We plot $\xi$, $\zeta$ and Q as a function of longitudinal separation $r_{\parallel}$ for these two different $\Gamma_{\rm HI}$ values in Fig.~\ref{Corr_Gamma_HI}. We find that the correlation amplitudes are very sensitive to $\Gamma_{\rm HI}$. As expected, we find that absorbers corresponding to higher $\Gamma_{\rm HI}$ value, probing higher densities, have greater $\xi$ and $\zeta$ values at all scales. The effect is more pronounced in the case of $\zeta$. Interestingly, this large $\Gamma_{\rm HI}$ effect on $\xi$ and $\zeta$ almost nullify each other, when one considers Q.  In comparison to $\xi$ and $\zeta$, the change in Q values do not show any coeherent trend between the two $\Gamma_{\rm HI}$ values. \textit{{ In summary}, for a given set of cosmological parameters both two- and three-point correlation are highly sensitive to the background photo-ionization rates and increase with increase in $\Gamma_{\rm HI}$}.
However, in reality a well measured \NHI\ distribution will provide an independent constraints on $\Gamma_{\rm HI}$. Thus simultaneous usage of clustering (two-point, three-point correlation and Q) and
\NHI\ distribution will provide stringent constraints on $\Gamma_{\rm HI}$ and possibly on parameters of the underlying cosmology (as clustering is also sensitive to the background cosmology).

\begin{figure*}
    \centering
     
	\includegraphics[viewport=0 40 315 295,width=6cm, clip=true]{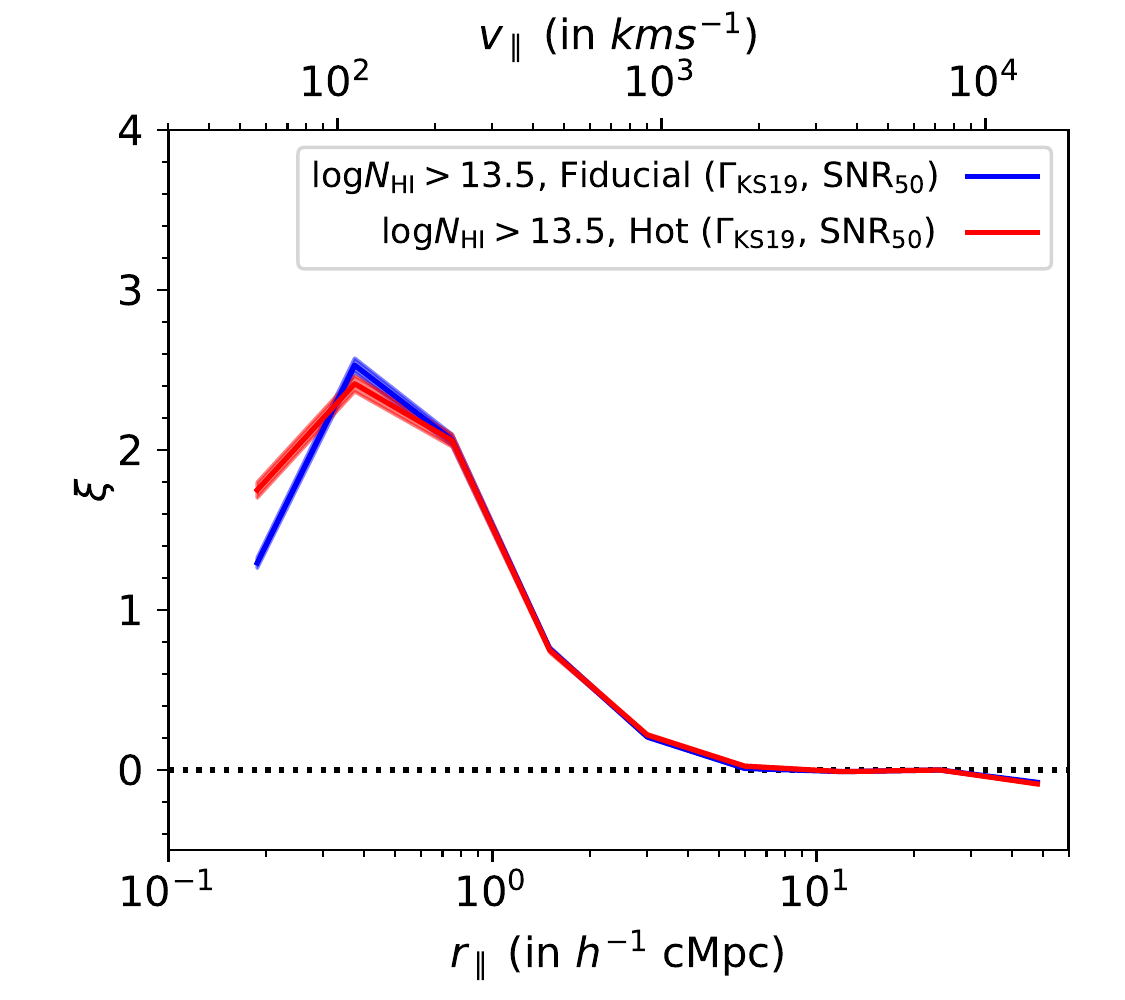}%
	\includegraphics[viewport=0 40 315 295,width=6cm, clip=true]{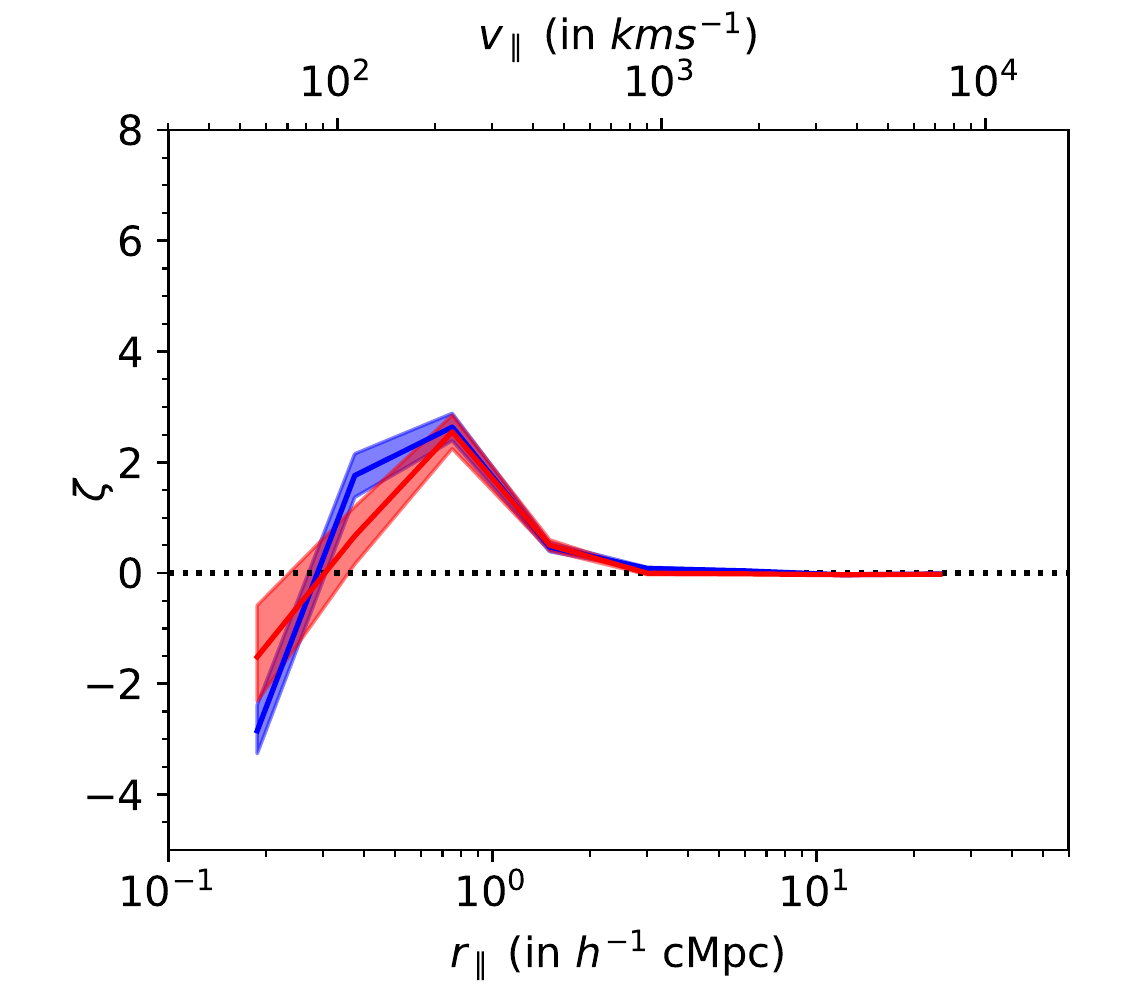}%
	\includegraphics[viewport=0 40 315 295,width=6cm, clip=true]{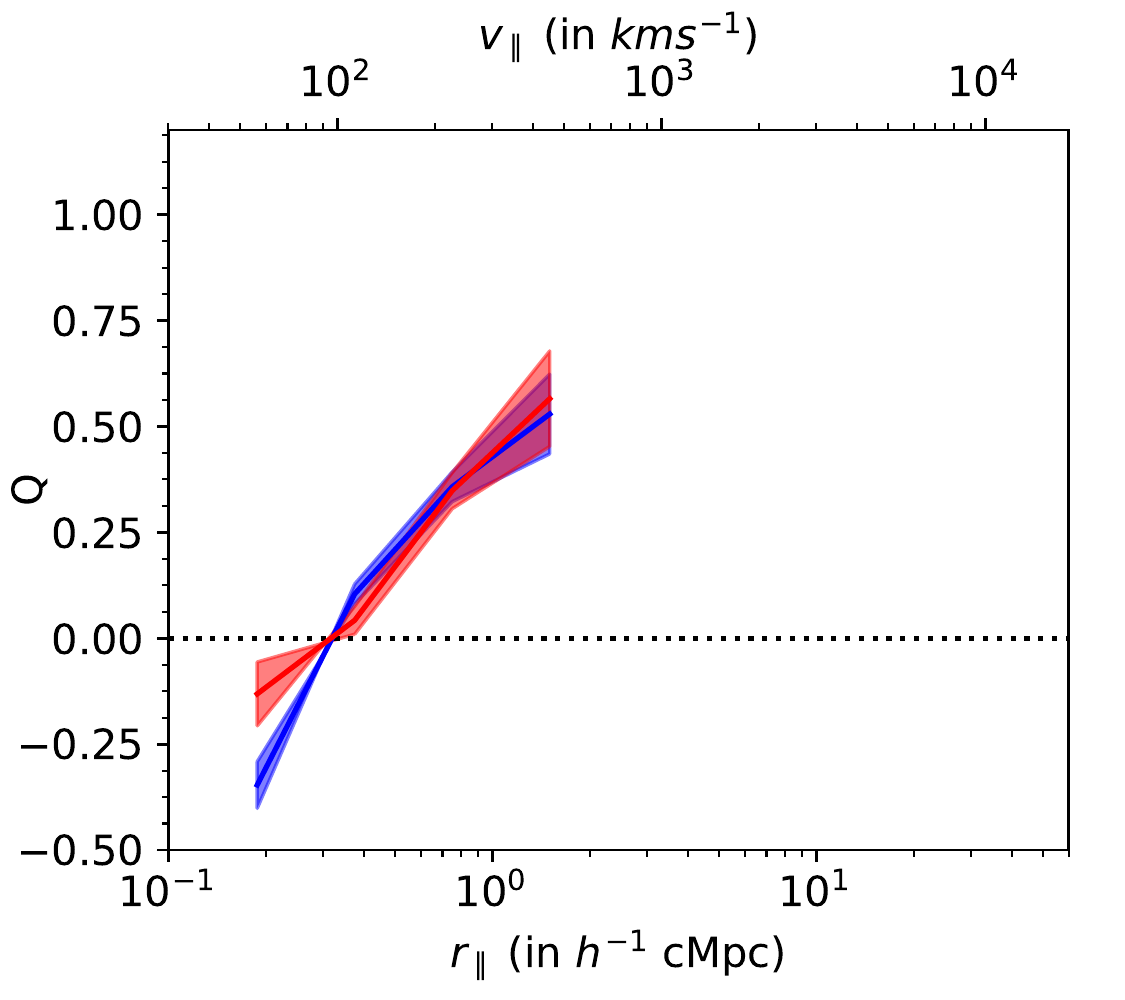}%
	
	\includegraphics[viewport=0 0 315 258,width=6cm, clip=true]{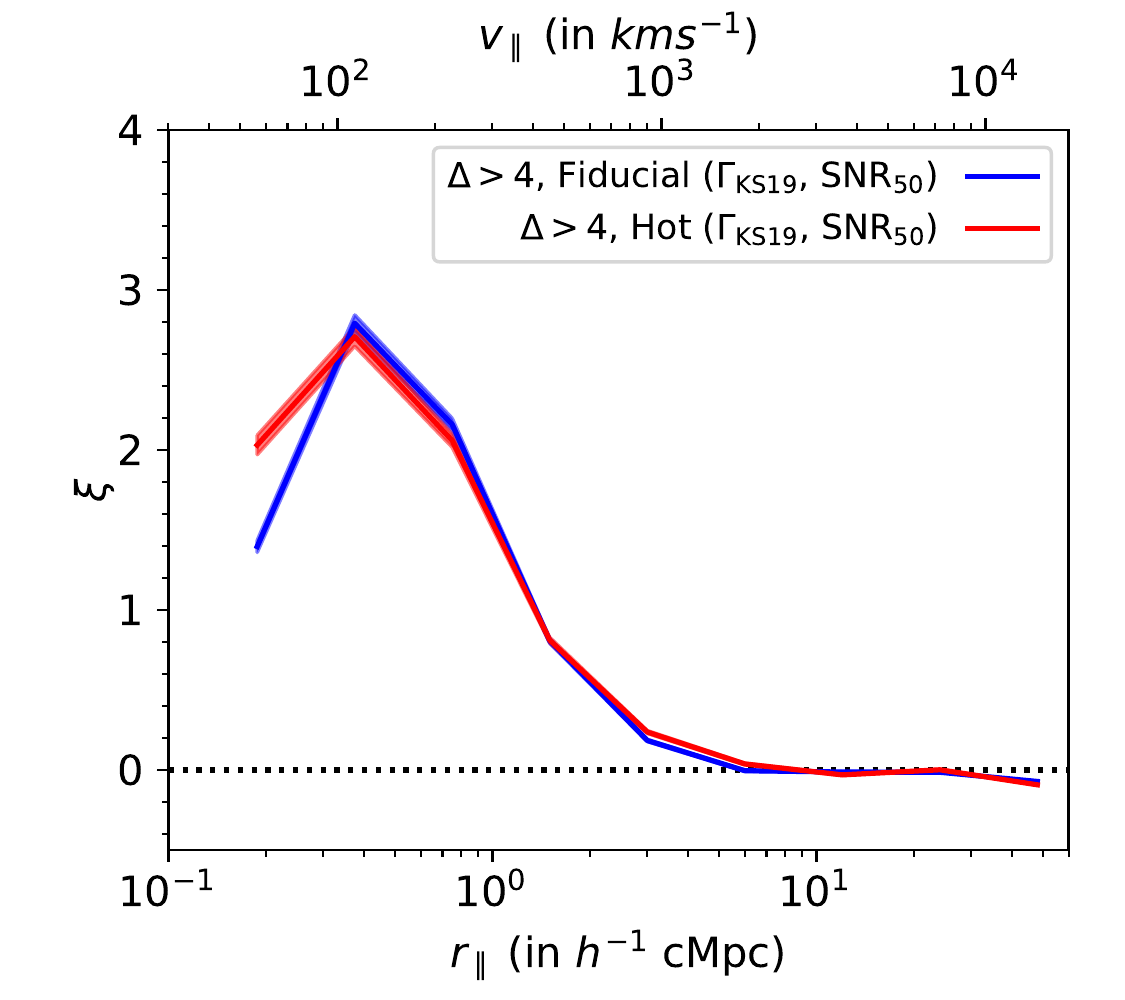}%
	\includegraphics[viewport=0 0 315 258,width=6cm, clip=true]{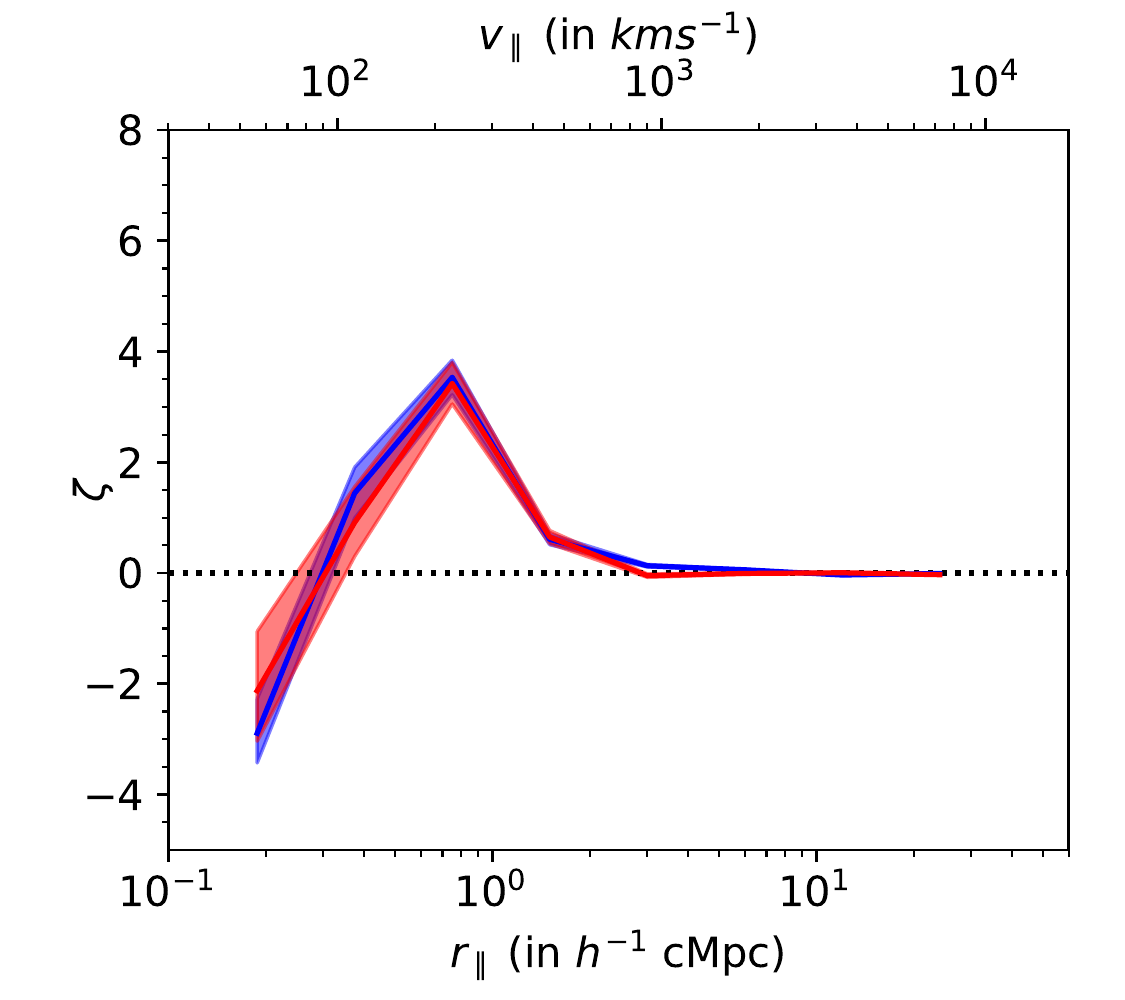}%
	\includegraphics[viewport=0 0 315 258,width=6cm, clip=true]{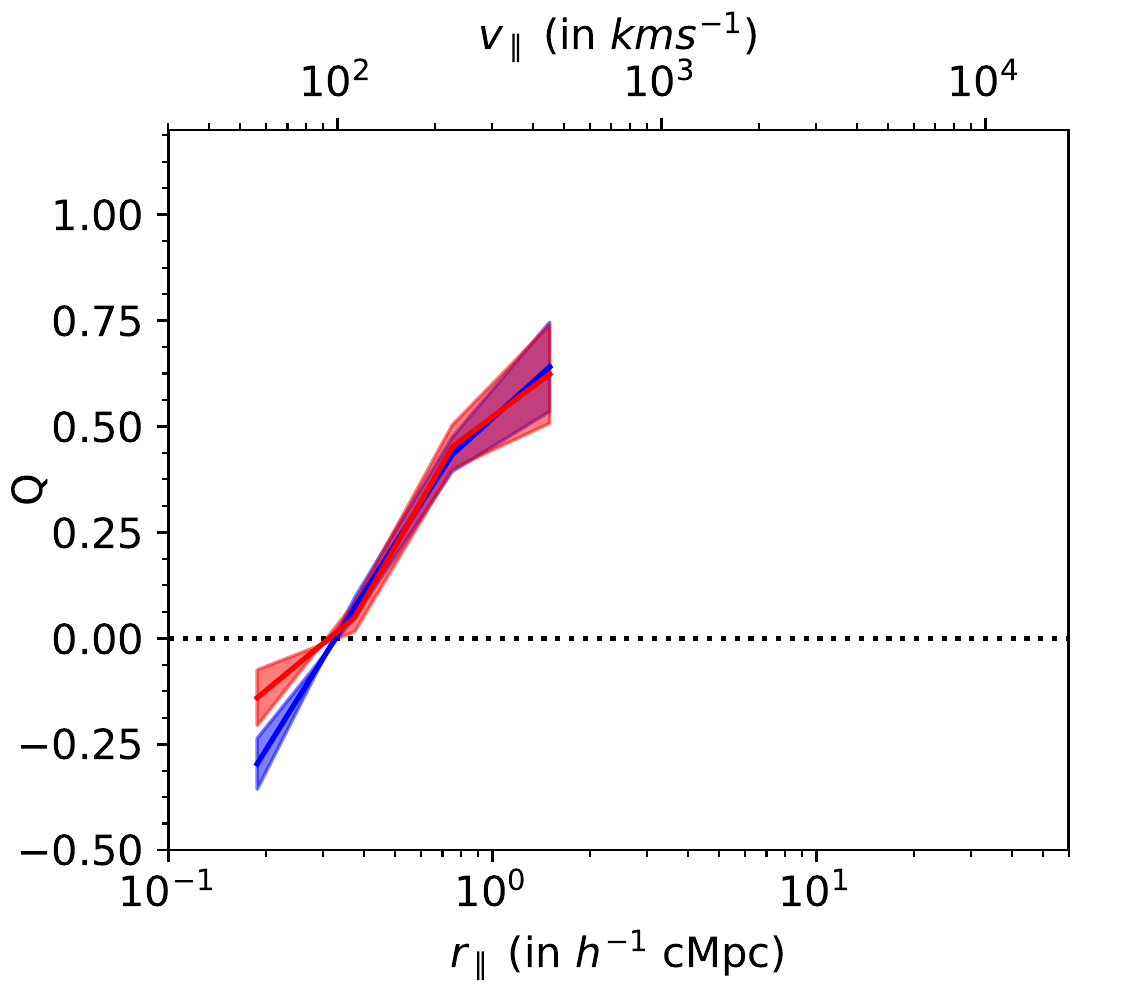}%
	
    \caption{Dependence of { simulated} clustering on thermal parameters of the IGM. \textit{Top}: Two-, three- and reduced three-point correlation (left to right) of \lya\ absorbers having $N_{\rm HI}>10^{13.5}$cm$^{-2}$ as a function of longitudinal separation $r_{\parallel}$ for two simulations, one with "Fiducial" KS19 UV background and another with Enhanced "Hot" KS19 UV background. The two simulations represent the joint effect of thermal broadening and pressure broadening (from different thermal histories) on clustering.
    \textit{Bottom}: Same correlations for the two simulations,
    but with a constant overdensity threshold of $\Delta>4$. { Using a} constant $\Delta$ threshold nullifies any differences originating from different local thermal parameters and { captures} the effect of pressure broadening only from different thermal histories. 
    }
    \label{Corr_astro}
\end{figure*}

\subsection{Dependence on thermal and pressure broadening}

Using the clustering analysis of simulated \lya\ forest, \citet{peeples2010a,peeples2010b} had reported that correlations in redshift space are dominated by thermal broadening, while transverse correlations using projected quasar pair spectra are mainly dominated by pressure broadening effects. However, this analysis was based on pixel based statistics using \lya\ forest transmitted flux. Our absorber-based approach may have different dependencies on the thermal state of the IGM which we would like to explore.

As mentioned before, we base our explorations on 
the "fiducial" and
"hot" models discussed in Section~\ref{Simulation}. We shoot random sightlines and calculate the correlations for the \lya\ absorbers separately for the two boxes. While the thermal and pressure effects are different between the two boxes, their photo-ionization rates are similar. So, the correlation functions should only show the compound effect of thermal and pressure broadening between the two simulations. Firstly, we compare the $N_{\rm HI}$ distribution between these two boxes in the right panel of Fig.~\ref{NHI_dis_effects}. We find that there is a small difference in the distributions between the two simulations in comparison to the effect of $\Gamma_{\rm HI}$. The difference is more pronounced in the higher $N_{\rm HI}$ end due to logarithmic scaling. This difference in the $N_{\rm HI}$ distribution between the two simulations is primarily due to difference in temperature-dependent recombination rates.
There can also be an additional effect of the absorbers being more thermally broadened thereby making the multiple component fitting of line-blended absorption systems more difficult (especially when we do not use higher order Lyman lines as in {\sc viper}).

Next, we plot two-point, three-point and reduced three-point correlation as a function of $r_{\parallel}$ in the top panels of Fig.~\ref{Corr_astro} for these two boxes. In the case of two-point correlation, there are negligible differences between the two boxes based on the observed correlation functions at scales greater than $r_{\parallel}=0.25h^{-1}$cMpc. In the $r_{\parallel}=0.125-0.25h^{-1}$cMpc bin, the amplitude of the two-point correlation is smaller for the "fiducial" simulation.  
The three-point and reduced three-point correlation are similar for the two boxes at all scales within the errorbars. 
This similarity in the correlation functions between the "fiducial" and "hot" simulation is contrary to what \citet{peeples2010a,peeples2010b} had found based on a pixel based statistics using transmitted flux.
The main difference between our absorber-based approach and a flux-based approach is that in absorber based approach, the \lya\ absorbers contains information of two physical quantities: $N_{\rm HI}$ and the $b$-parameter. The $b$-parameter is responsible for the thermal broadening seen in \lya\ absorption. On the other hand, thermal broadening does not affect the measured $N_{\rm HI}$ value of an absorber. So, for a fixed $N_{\rm HI}$ threshold, our correlation statistics is relatively independent of the thermal broadening effect. This is not the case for flux-based approach. Since thermal broadening smooths the transmitted flux field, the flux-deviation is altered in each pixel thereby altering the correlation function also. The only thermal effect we see for our absorber-based approach is at scales close to the $b$-parameter where the inability to distinguish line-blending in thermally broadened structures will slightly change the correlation amplitudes. We see this effect at scales $<0.25h^{-1}$cMpc for the two-point correlation profile, wherein both $\xi$ and $\zeta$ values are slightly lower for the "hot" box having larger $b$-parameter.

\begin{figure}
    \centering
    \includegraphics[width=8cm]{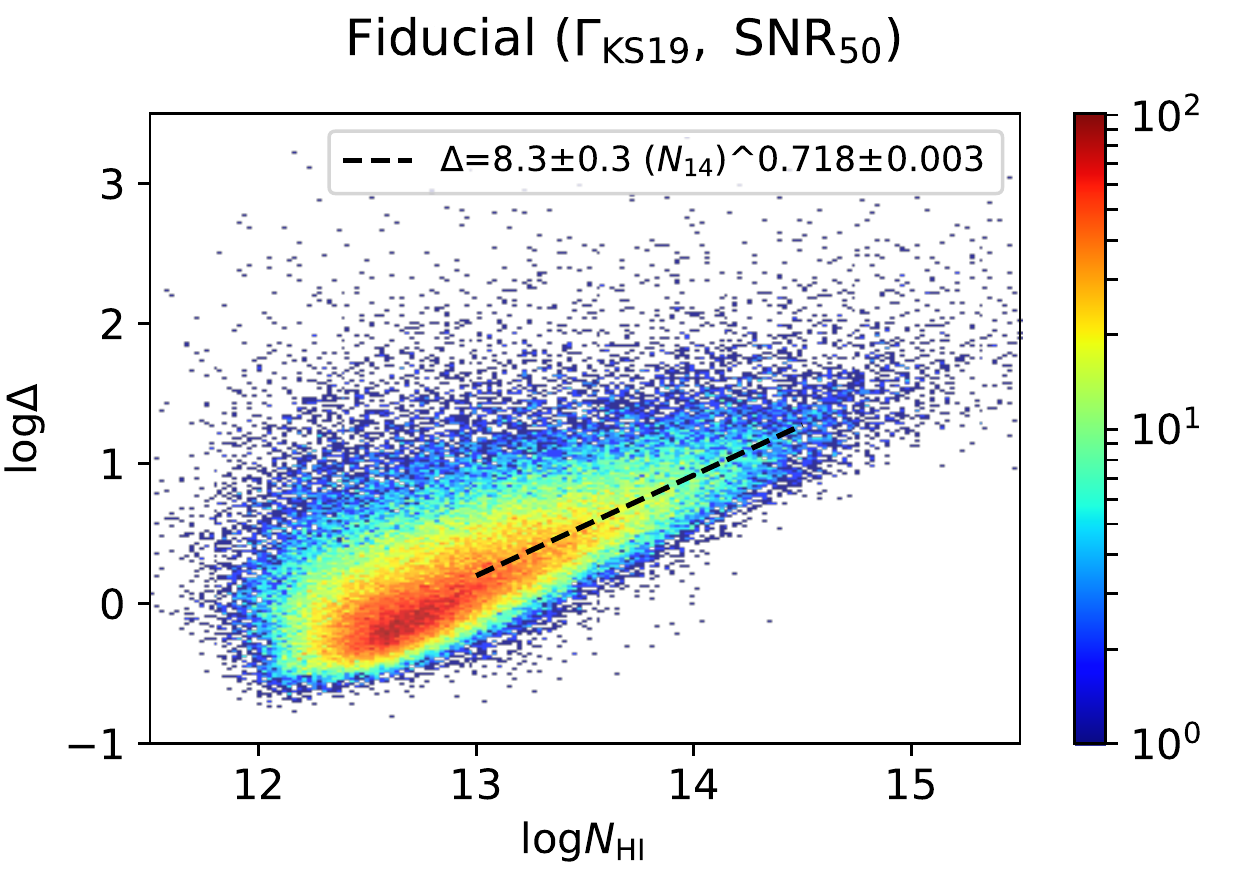}
    \includegraphics[width=8cm]{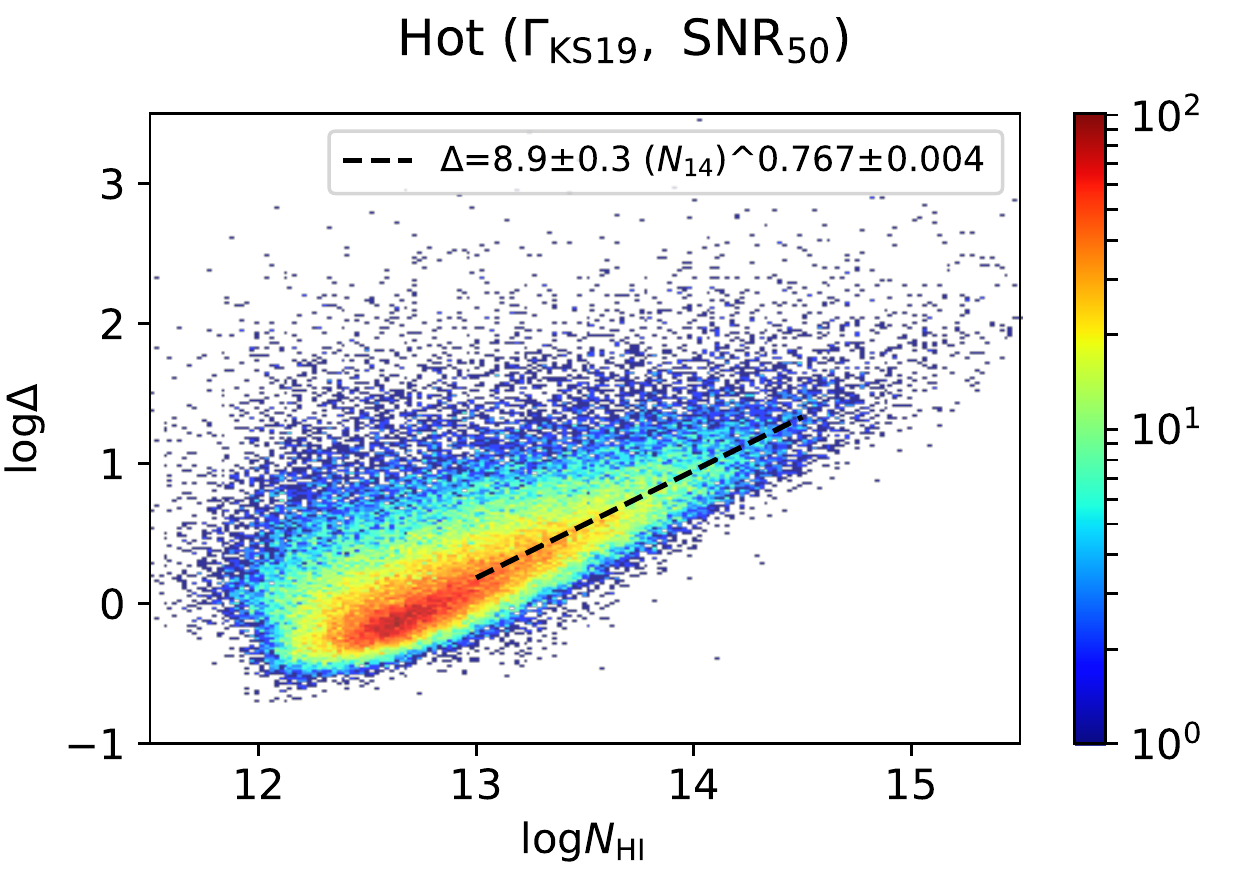}
    \caption{Neutral hydrogen column density ($N_{\rm HI}$) vs $\tau$ weighted baryon overdensity ($\Delta$) plots at $z=2$ for $\Gamma_{\rm HI}=\Gamma_{\rm KS19}$ and SNR=50 for our "fiducial" and "hot" simulations. The color represents the number of points in a certain grid. The black dashed line represents the best fit relationship (best fit values in legend) followed by the median $\Delta$ in the range of log($N_{\rm HI}$)=(13,14.5).  }
    \label{N_vs_Delta}
\end{figure}

Next, we try to identify the source of the small scale discrepancy seen between the two simulations in two-point correlation. This difference can come simply from the difference in line-blending in thermally broadened structures, or from the effect of pressure broadening on the small scale structures. In order to distinguish between the two effects, we try to probe the clustering of \lya\ absorbers as a function of the underlying baryonic density fields \citep[see Sec 3.5 of][for example]{maitra2020}. While clustering based on a uniform $N_{\rm HI}$ threshold will contain information of both thermal and pressure broadening, clustering analysis based on uniform baryonic overdensity is affected by pressure broadening only. Statistically, there exists a power law relationship between the optical depth ($\tau$) weighted overdensity $\Delta$ and $N_{\rm HI}$ of \lya\ absorbers given by,
\begin{equation}\label{delta-N}
    \Delta=\Delta_0 N_{14} ^{\eta}\ ,
\end{equation}
where, $N_{14}$ is the $N_{\rm HI}$ given in units of $10^{14}$cm$^{-2}$. Based on our simulated spectra, we can compute the best fit $\Delta_0$ and $\eta$ values.
We assign optical depth ($\tau$) weighted baryonic overdensity to these $N_{\rm HI}$ absorbers  \citep[see][]{dave1999,schaye1999}.  We associate the overdensities of all the pixels $i$ in real space that contribute to the optical depth at the position of a given absorber in velocity space identified by index $j$. This association comes as a weight factor $\tau_{ij}$ which gives the optical depth contribution of the overdensity at pixel $i$ to the total optical depth at the position of absorber $j$. The $\tau$ weighted overdensity of absorber $j$ is then given as,
\begin{equation}
   \bar{\Delta}_j=\frac{\sum\limits_i \tau_{ij} \Delta_i}{\sum\limits_i \tau_{ij}}
\end{equation}
where the summation is over all the pixels $i$ in the spectra \citep[see][]{gaikwad2017a}. We show the $N_{\rm HI}$ vs $\tau$ weighted $\Delta$ density plots for our two simulations in Fig.~\ref{N_vs_Delta}. We find the median $\Delta$ values in { different log~\NHI\ bins of width 0.1 over the log~\NHI\ range of 13.0-14.5}.
We then find the best fit parameters for the power-law relation between $N_{\rm HI}$ and $\Delta$. The best fit parameter values for the Fiducial and Hot simulations are $\Delta_0=8.3\pm 0.3$, $\eta=0.718\pm 0.003$ and $\Delta_0=8.9\pm 0.3$, $\eta=0.767\pm 0.004$, respectively.{ The $\Delta_0$ value for our fiducial simulation appears to be larger than $\Delta_0=6.6\pm0.1$ given in \citet{dave2010} at $z=2$ for their no-wind simulation. Also, our $\eta$ value is slightly larger than their value of $0.741\pm 0.003$. This difference may source from the fact that their simulations use \citet{haardt2001} ionizing background radiation while our simulations use that from \citet{khaire2019}.}

Using this, we find the $N_{\rm HI}$ thresholds of the absorbers for an overdensity threshold of $\Delta>4$ (corresponding to log$N_{\rm HI}>13.56$ and 13.55 for the "fiducial" and "hot" simulations, respectively). 
In the bottom panels of Fig.~\ref{Corr_astro}, we plot the two-point, three-point and reduced three-point correlation as a function of $r_{\parallel}$ for a constant overdensity threshold of $\Delta>4$. Overall, the correlations are all consistent with each other between the two simulations (as expected based on nearly identical log$N_{\rm HI}$ cut-offs found for the two models). { But the  discrepancy in two-point correlation between the "fiducial" and "hot" simulations at small scales still remains.} This suggests that for scales below $r_{\parallel}=0.25h^{-1}$cMpc { the clustering} is affected by the pressure broadening of the baryonic gas. So, probing clustering in \lya\ absorbers at this scale will provide valuable information regarding the thermal history of the IGM. { However, in order to do so, one should  have a good handle on the systematics introduced by the spectral SNR 
as well as the multiple Voigt profile component decomposition of the blended
absorption profile .}

\subsection{Dependence on feedback processes}

Here we study the impact of Wind and AGN feedback on the clustering properties of the \lya\ absorbers at $z\sim 2$ using two Sherwood simulations: one without any feedback and the other with Wind+AGN feedback (for details see Sec.~\ref{Simulation}). As evident from the left panel in Fig.~\ref{NHI_dis_effects}, the two simulations are consistent in producing the number of low-$N_{\rm HI}$ absorbers with each other. However, the one with feedback produce larger number of high-$N_{\rm HI}$ absorbers. 
This effect was also pointed out by \citet{Nasir2017} in their analysis of low-$z$ simulations \citep[see also][]{viel2016,gurvich2017,maitra2020b}. Therefore, accurately measured \NHI\ distribution at the high \NHI\ end can in principle constrain the effect of various feedback processes. However, one has to use simultaneous fitting of \lya\ and higher series Lyman lines to have a robust deblending of the Voigt profile components at higher \NHI\ ranges.

\begin{figure*}
    \centering
     \includegraphics[viewport=0 0 315 290,width=6cm, clip=true]{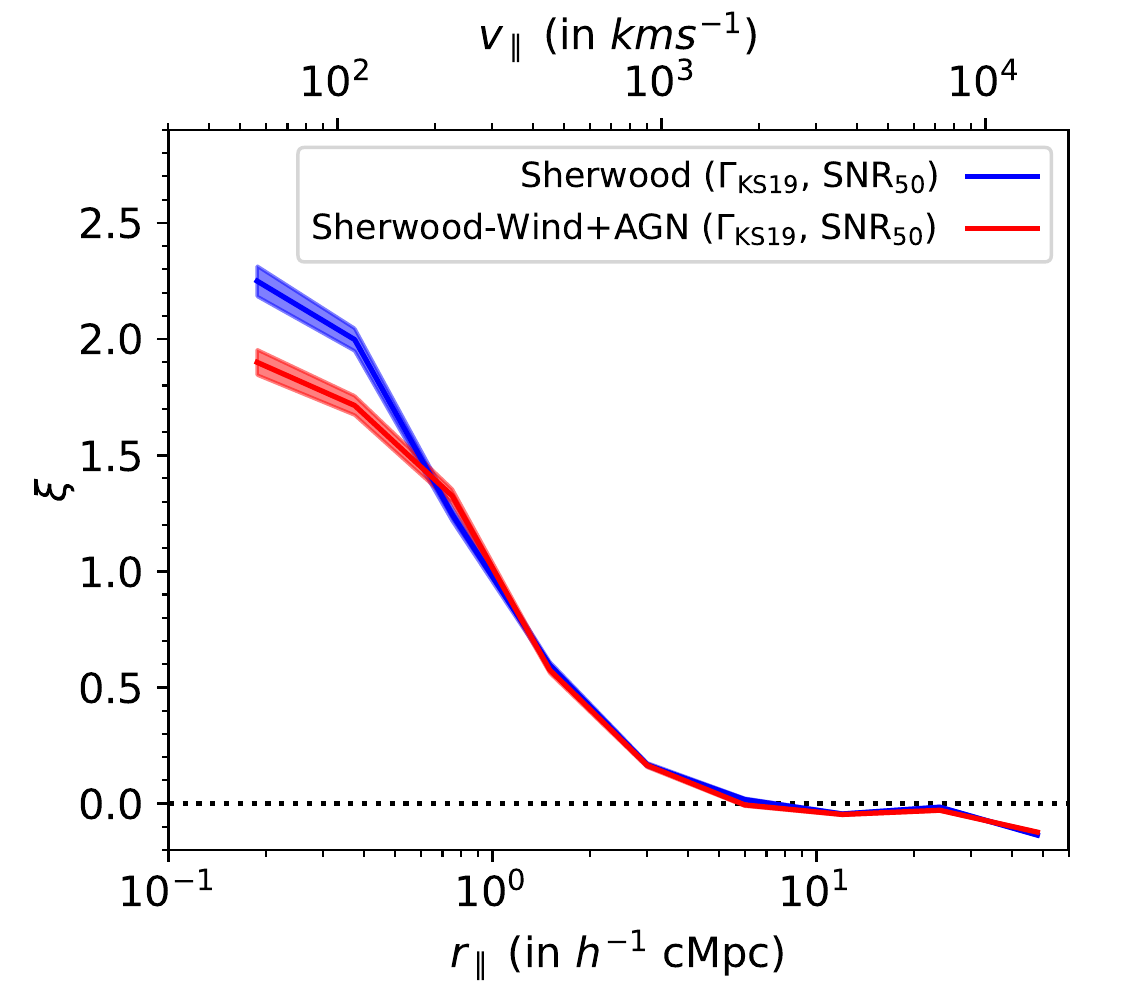}%
	\includegraphics[viewport=0 0 315 290,width=6cm, clip=true]{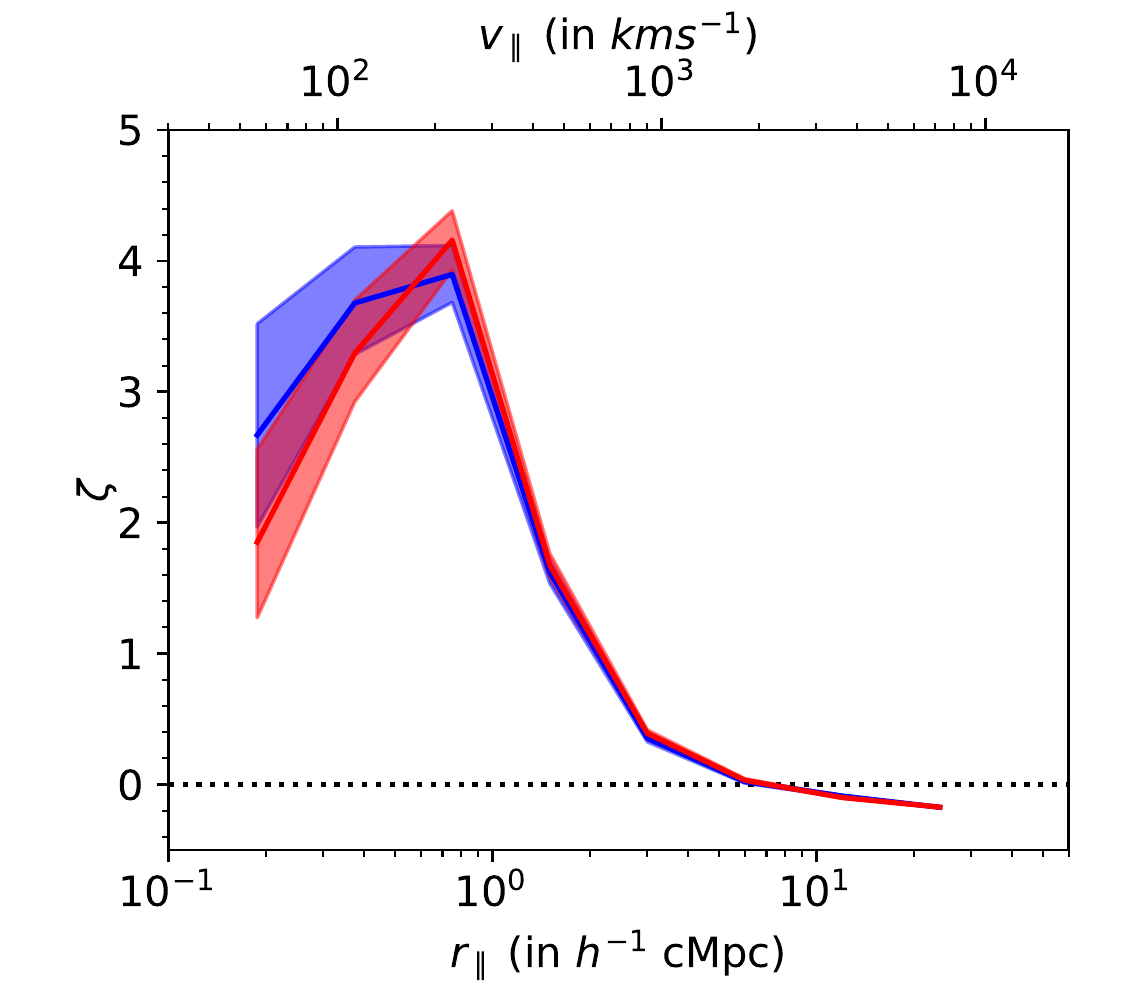}%
	\includegraphics[viewport=0 0 315 290,width=6cm, clip=true]{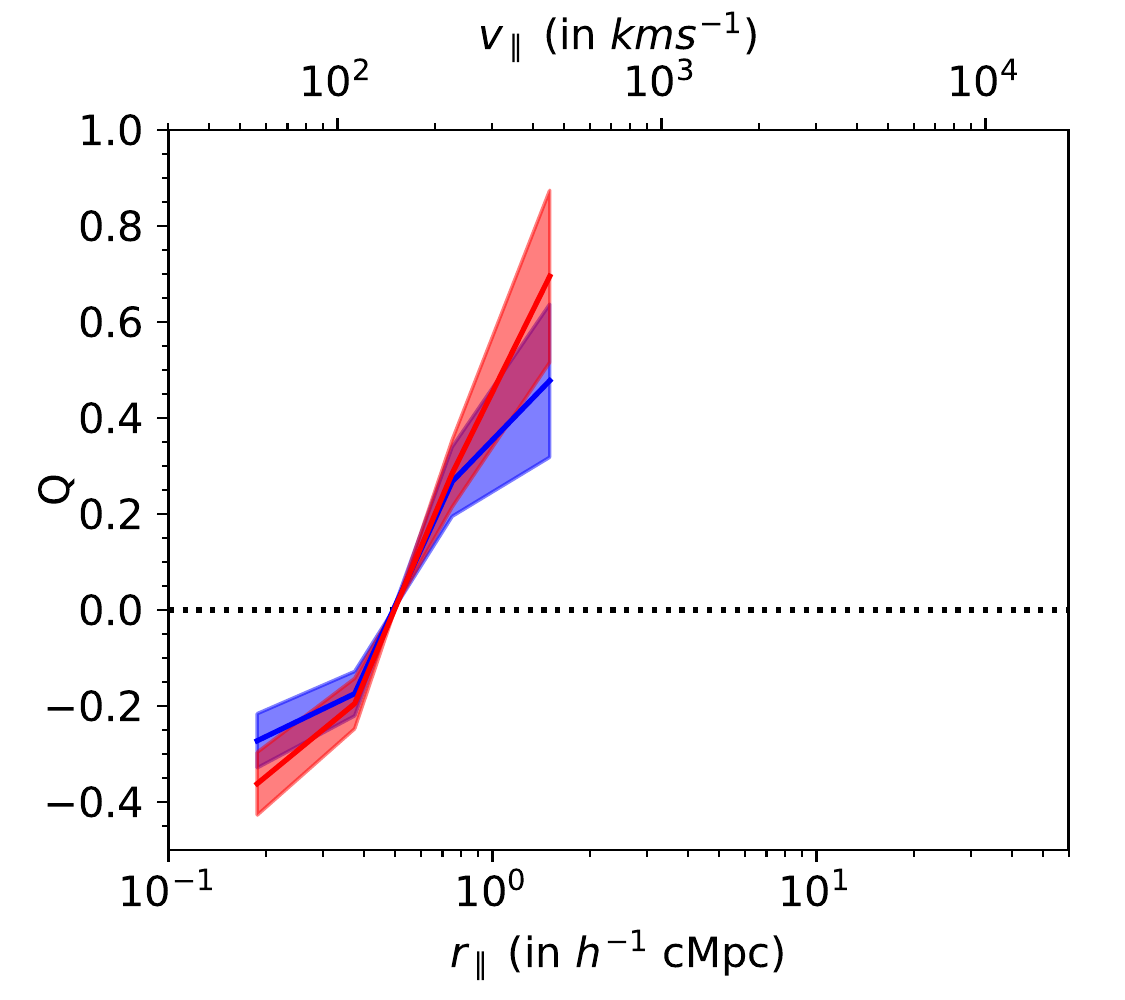}%

    \caption{ Two-point, three-point and reduced three-point correlation (left to right) of \lya\ absorbers having $N_{\rm HI}>10^{13.5}$cm$^{-2}$ as a function of longitudinal separation $r_{\parallel}$ { in Sherwood simulations} with { (red curves)} and without Wind+AGN { (blue curves)} feedback.
    }
    \label{Corr_feedback}
\end{figure*}

In Fig.~\ref{Corr_feedback}, we plot the two-, three- and reduced three-point correlations as a function of longitudinal separation $r_{\parallel}$ for the Sherwood simulation with and without Wind+AGN feedback. It is seen that both two- and three-point correlations are similar between the two simulations at scales greater than $0.5h^{-1}$cMpc. However, at scales below $0.5h^{-1}$cMpc, the correlation amplitudes are suppressed for the simulation with feedback. The effect is more clearly resolved in the case of two-point correlation where the errorbars are much less. Interestingly, the reduced three-point correlation remains relatively unaffected at all scales. These effects are consistent with what was found for the low-$z$ \lya\ absorbers by \citet{maitra2020b}.

\textit{In Summary, we find that both two- and three-point correlations are affected by peculiar velocities at all scales and hence, getting the correct spatial distribution requires a proper modelling of the redshift space distortion effects. We also find that clustering in \lya\ absorbers is affected by feedback processes and thermal and pressure effects at scales below $0.5h^{-1}$cMpc. It is, however, sensitive to \HI\ photo-ionization rates at all scales. So, at scales above $0.5h^{-1}$cMpc, \lya\ clustering can be a sensitive probe of photo-ionization rates for a given set of cosmological parameters. }

\section{Redshift evolution of clustering and its implications to evolution in the physical conditions of the IGM}
\label{Sec:evolution}

In this section, we compare the measured clustering at different redshifts with the predictions of our fiducial simulation (see Section~\ref{Simulation}). Assuming the cosmological parameters used here are correct, we try to see what kind of evolution in physical state of the IGM (in particular $\Gamma_{\rm HI}$) is required to explain the observed trend.
As shown above, the absorber-based correlation is relatively independent of thermal and pressure broadening or any feedback process at scales greater than 0.5$h^{-1}$cMpc and is largely sensitive to
$\Gamma_{\rm HI}$ 
(see Section \ref{Astro_effects}) for a given set of cosmological parameter. Hence our absorber-based correlation statistics should act as an independent robust probe for estimating the ionization state of the IGM.

\begin{figure*}
	\includegraphics[viewport=0 38 800 295,width=18cm, clip=true]{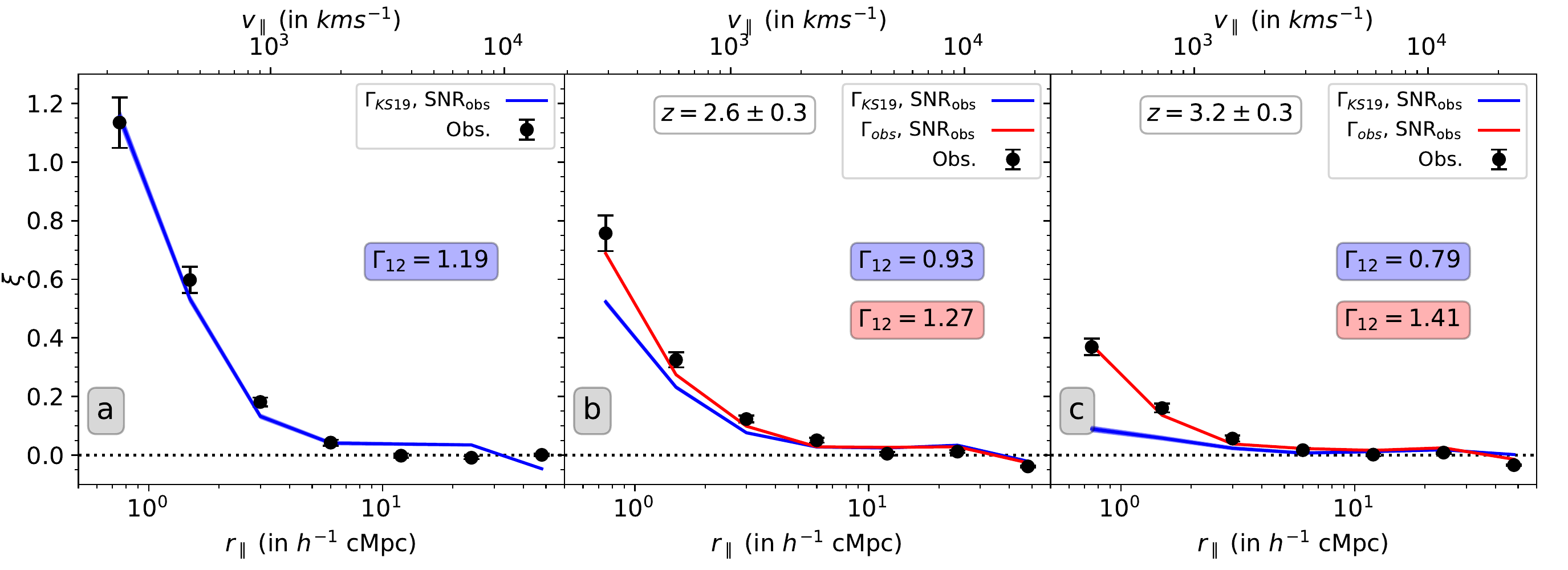}
	
	\includegraphics[viewport=0 0 800 254,width=18cm, clip=true]{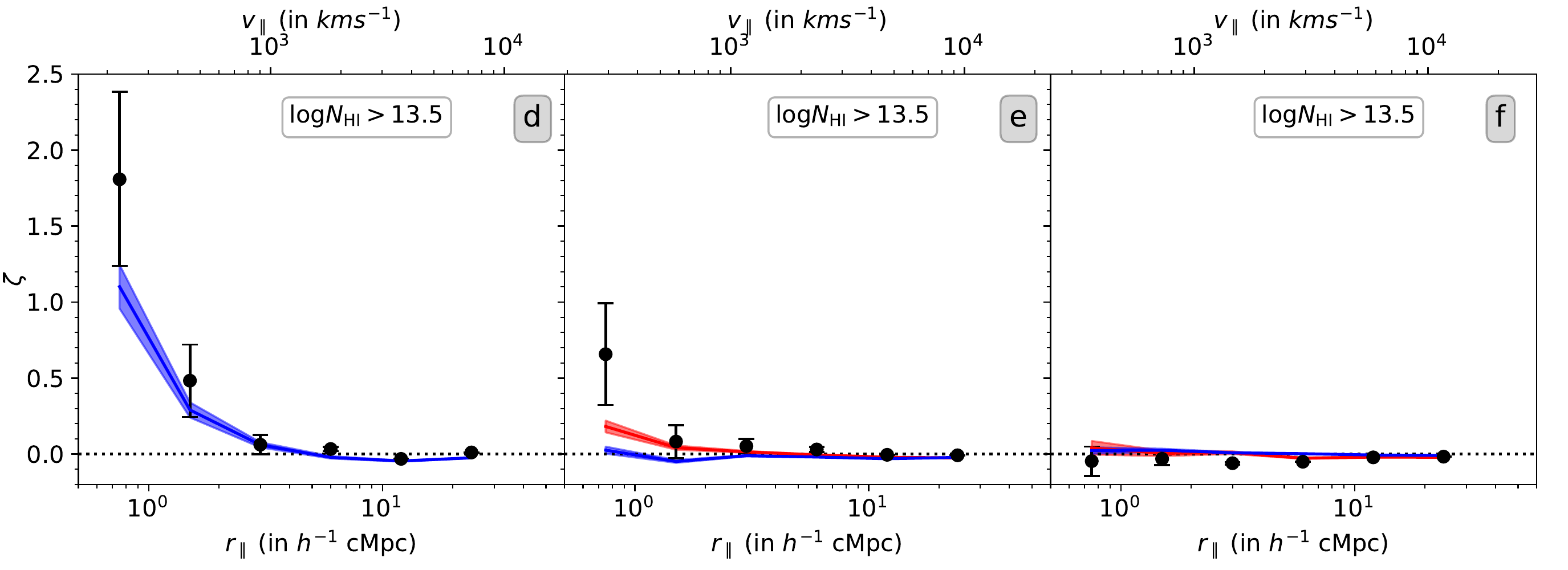}
	
	\caption{Absorber-based longitudinal two-point (\textit{top panel}) and three-point correlation (\textit{bottom panel}) of \lya\ absorbers ($N_{\rm HI}>10^{13.5}$cm$^{-2}$) as a function of longitudinal scale for different redshift intervals. The errorbar shown corresponds to the larger of the two errors: one-sided poissonian uncertainty in the number of absorber pairs (or triplets) corresponding to $\pm 1\sigma$ or bootstrapping error. The black points represent the observed correlation function obtained from joint {\sc KODIAQ+SQUAD} sample. The blue curve represents correlation function obtained from {\sc GADGET-3} simulation using $\Gamma_{\rm HI}$ from \citet{khaire2019}. The red curve represent correlation function obtained from simulation by matching the observed mean transmitted flux.   }
\label{Corr_Gammacorr}
\end{figure*}

\subsection{Observed clustering and \HI\ photo-ionization rates}\label{Gamma_est}

For this exercise, we use simulated spectra from our fiducial box with Gaussian noise based on the SNR distribution from the observations ($\rm SNR_{\rm obs}$, see Section~\ref{Forward_model}). We take 4000 simulated sightlines each for $z=1.7- 2.3$ and $2.3-2.9$ bins and 1000 for $z=2.9-3.5$. We consider less realisations for the high-$z$ bin as decomposing the \lya\ forest with Voigt profiles takes lot more time for the high-$z$ case due to heavy blending of lines.

First, we use the $\Gamma_{\rm HI}$ values 
from \citet[][denoted by $\Gamma_{\rm KS19}$]{khaire2019} for generating the simulated sightlines. These values mainly trace the mean $\Gamma_{\rm HI}$ measurements of \citet{becker2013}.
 In Fig.~\ref{Corr_Gammacorr}, we compare the two- and three-point correlations obtained from these simulations (blue curves) with the observed statistics (black points). Note that we only plot { results for} scales greater than 0.5$h^{-1}$cMpc as these are mostly unaffected by
thermal and pressure smoothing effects, feedback processes or systematics involving multiple component fitting of \lya\ absorption profiles. In the lowest redshift bin (i.e $1.7\le z\le 2.3$), we find that the two- and three-point correlation from the simulations obtained with $\Gamma_{\rm KS19}$ agree well with the observations. This is not surprising as the ionization by the uniform UV background is valid at these redshifts and thermal parameters predicted for the \citet{khaire2019} UV background model matches well with the recently measured values by \citet[][see their figure 18]{gaikwad2021}. Also we are using consistent cosmological parameters from the CMB experiments.

However, for $z=2.3-2.9$ bin, the predicted two- and three-point correlation amplitudes are less in comparison to the observed correlations. This deviation is even more pronounced for the $z=2.9-3.5$ bin. Thus it  appears that we need to produce the same range in \HI\ column densities from relatively higher overdense regions at high redshifts. Just to explore this possibility,
we next simulate spectra using  $\Gamma_{\rm HI}$ values that we get by matching the simulated mean transmitted flux with the observed mean transmitted flux (called $\Gamma_{\rm obs}$). In this case we need to increase $\Gamma_{\rm HI}$ by 36\% and 100\% for the redshift ranges 
$z=2.3-2.9$ and $z=2.9-3.5$ respectively. The resulting two-point correlations are also shown in Fig.~\ref{Corr_Gammacorr}. It is clear that if we adjust $\Gamma_{\rm HI}$ to match the observed mean transmitted flux in our simulations the predicted clustering results have better agreement with the observations. Three-point correlation functions predicted by our models are also shown in the bottom panel of Fig.~\ref{Corr_Gammacorr}. Our models reproduce these measurements reasonably well.

\citet{becker2013} measured $\Gamma_{\rm HI}$ in the ranges (0.78-1.39)$\times10^{-12}$ s$^{-1}$, (0.65-1.16)$\times10^{-12}$ s$^{-1}$  and  (0.60-1.11)$\times10^{-12}$ s$^{-1}$ for redshifts 2.4, 2.8 and 3.2 respectively. Thus our requirement that $\Gamma_{\rm HI} = 1.27\times10^{-12}$ s$^{-1}$  is not abnormal for the redshift bin $2.3\le z \le 2.9$. However the required $\Gamma_{\rm HI}$ of  $1.41\times10^{-12}$ s$^{-1}$ for the redshift interval $2.9\le z\le 3.5$ is 2.3$\sigma$ higher than the measurements of \citet{becker2013}.
Note here we are assuming the thermal state of the gas in our simulations are correct. However, \citet{gaikwad2021} have shown that $T_0$ ($\gamma$) measured in the redshift range $2.5\le z \le 3.2$ are systematially higher (lower) than those predicted by the background assumed here. Also the assumption of  ionization equilibrium evolution of the IGM may not be valid due to the He~{\sc ii} reionization at these redshifts. Especially for the highest redshift bin where one expects inhomogeneous thermal fluctuations from the end stages of He~{\sc ii} reionization.

{\it In summary, we find that our clustering measurements over the redshift range $2.3\le z\le 3.5$ are not reproduced by our fiducial model. To match our observations, an absorber with a given \NHI\ should come from regions of higher overdensity compared what has been predicted by our simulations. One possibility is to have higher $\Gamma_{\rm HI}(z)$ compared to what has been used in our simulations. While our exercise demonstrates the need to revisit $\Gamma_{\rm HI}(z)$ measurements, we reiterate that it has to be done by simultaneously matching various observables. As we use Voigt profile decomposed components for our clustering analysis, performing joint constraints on various parameters 
\citep[like the one performed by][for constraining thermal parameters]{gaikwad2021}, is computationally intensive and time consuming. While that is beyond the scope of this paper, which focuses mainly on presenting the observed clustering measurements, we hope to engage in such an exercise in a near future.}

\subsection{Evolution of clustering in baryonic density fields}

\begin{figure}
    \centering
    \includegraphics[width=8cm]{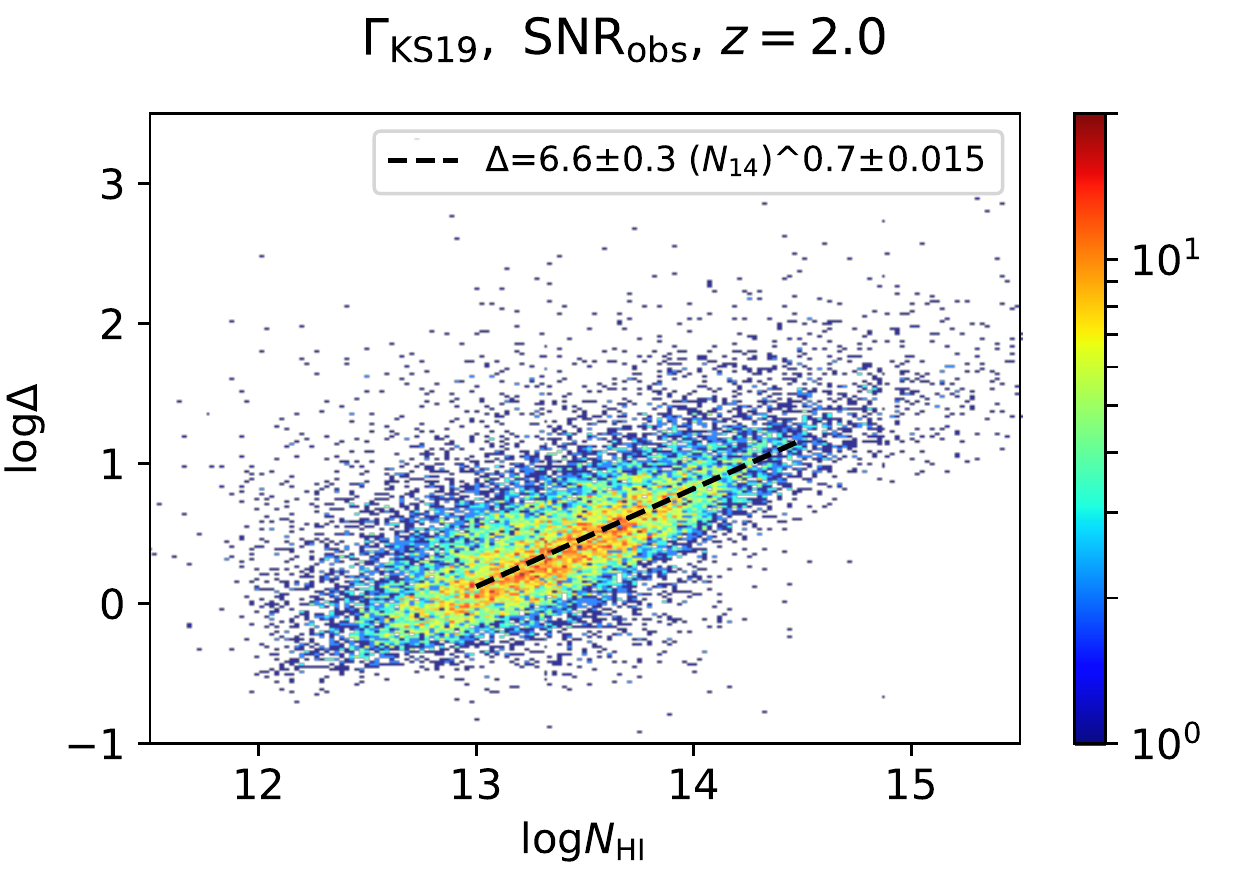}%
    
    \includegraphics[width=8cm]{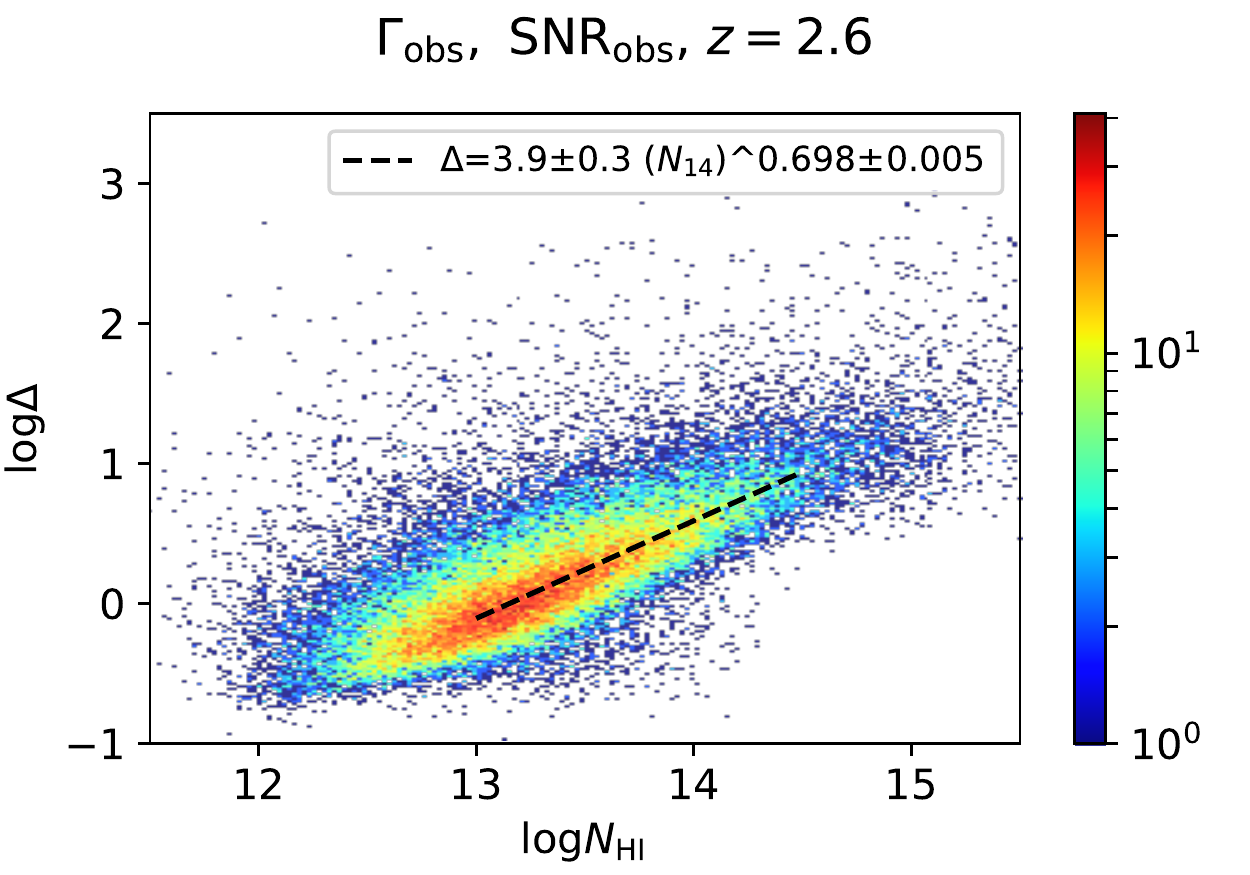}%
    
    \includegraphics[width=8cm]{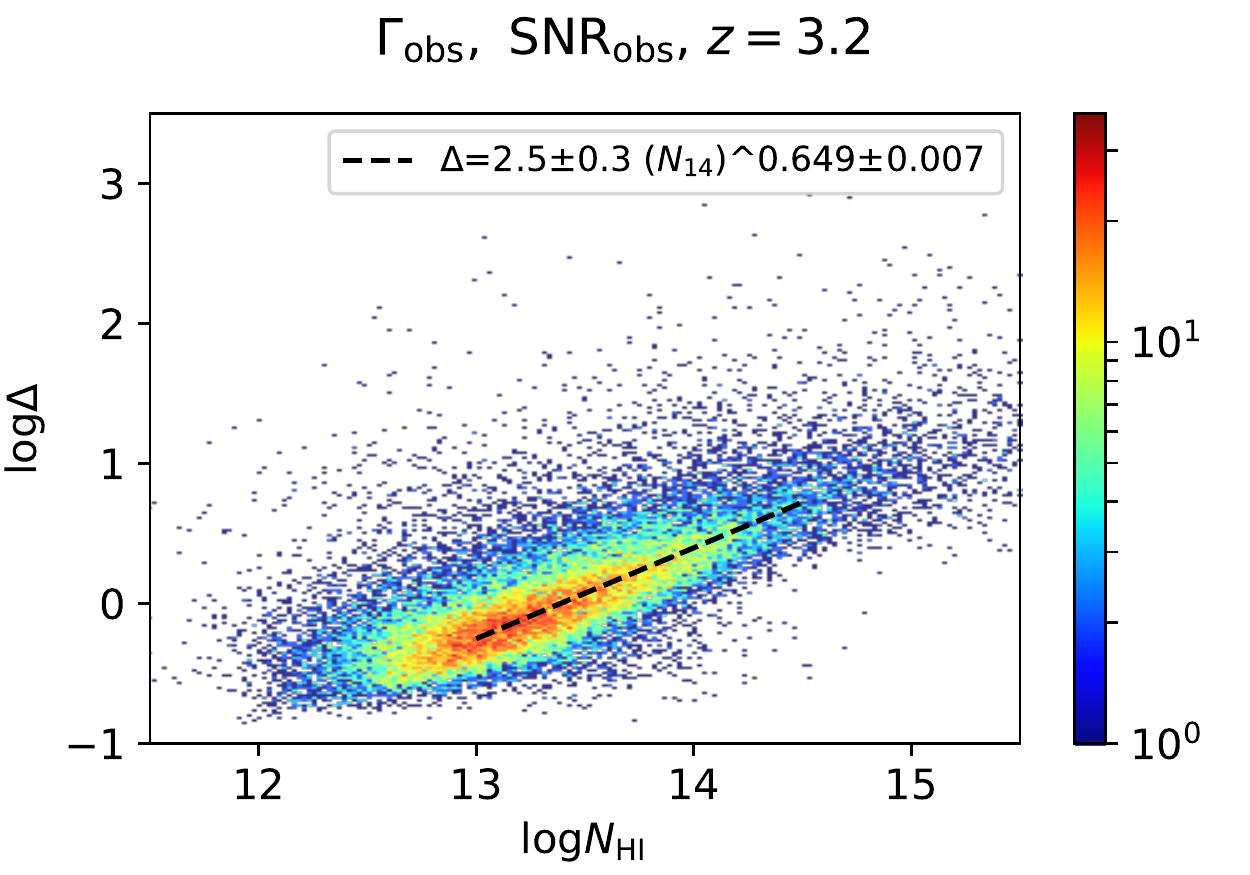}
    \caption{Neutral hydrogen column density ($N_{\rm HI}$) vs $\tau$ weighted baryon overdensity ($\Delta$) plots at $z=2$ for $\Gamma_{\rm HI}=\Gamma_{\rm KS19}$ and at $z=2.6,\ 3.2$ with $\Gamma_{\rm obs}$ with $\rm SNR_{obs}$ for our Fiducial simulation. The color represents the number of points in a certain grid. The black dashed line represents the best fit relationship (best fit values are provided in legend) followed by the median $\Delta$ in the range of log~$N_{\rm HI}$ = (13,14.5). 
      }
    \label{N_vs_Delta_obs}
\end{figure}

\begin{figure*}

\includegraphics[viewport=0 38 730 250,width=18cm, clip=true]{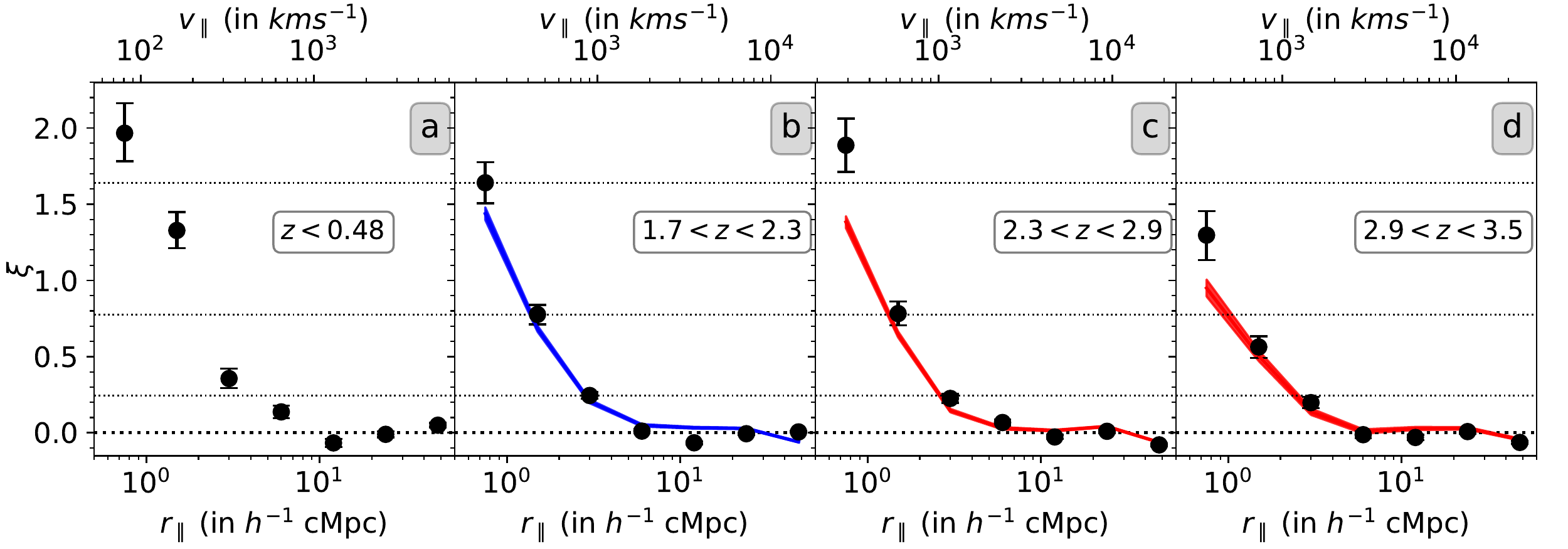}
	
	\includegraphics[viewport=0 38 730 220,width=18cm, clip=true]{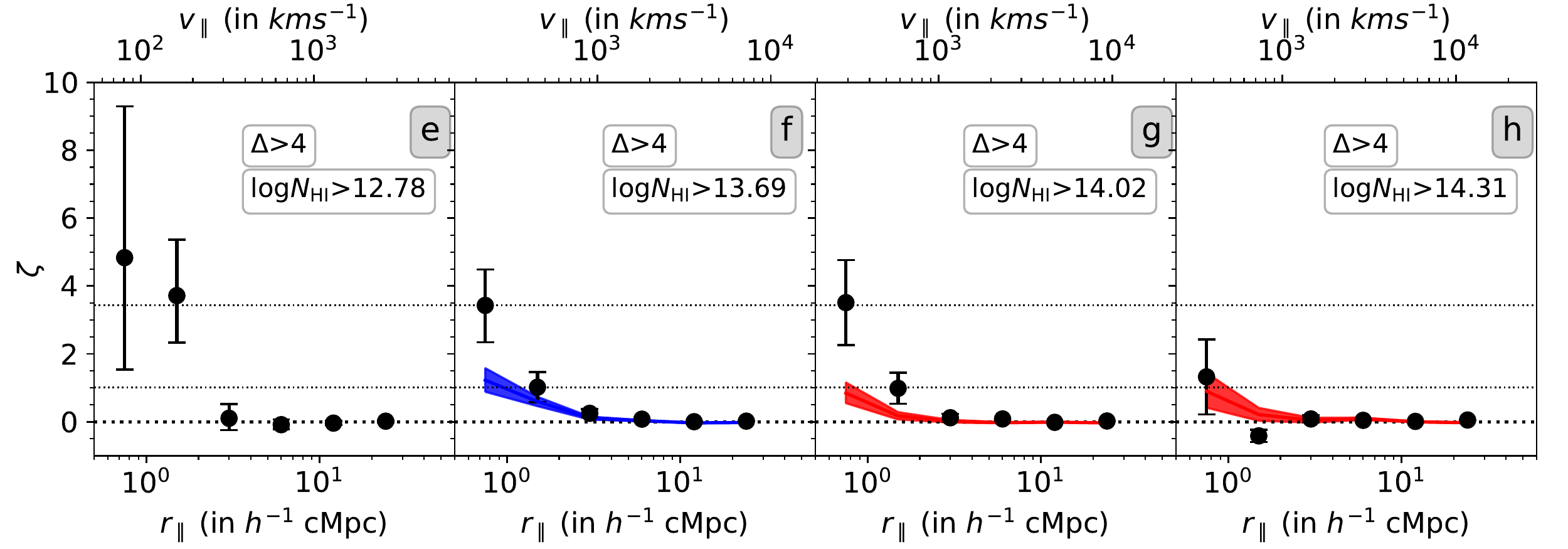}
	
	\includegraphics[viewport=0 0 730 220,width=18cm, clip=true]{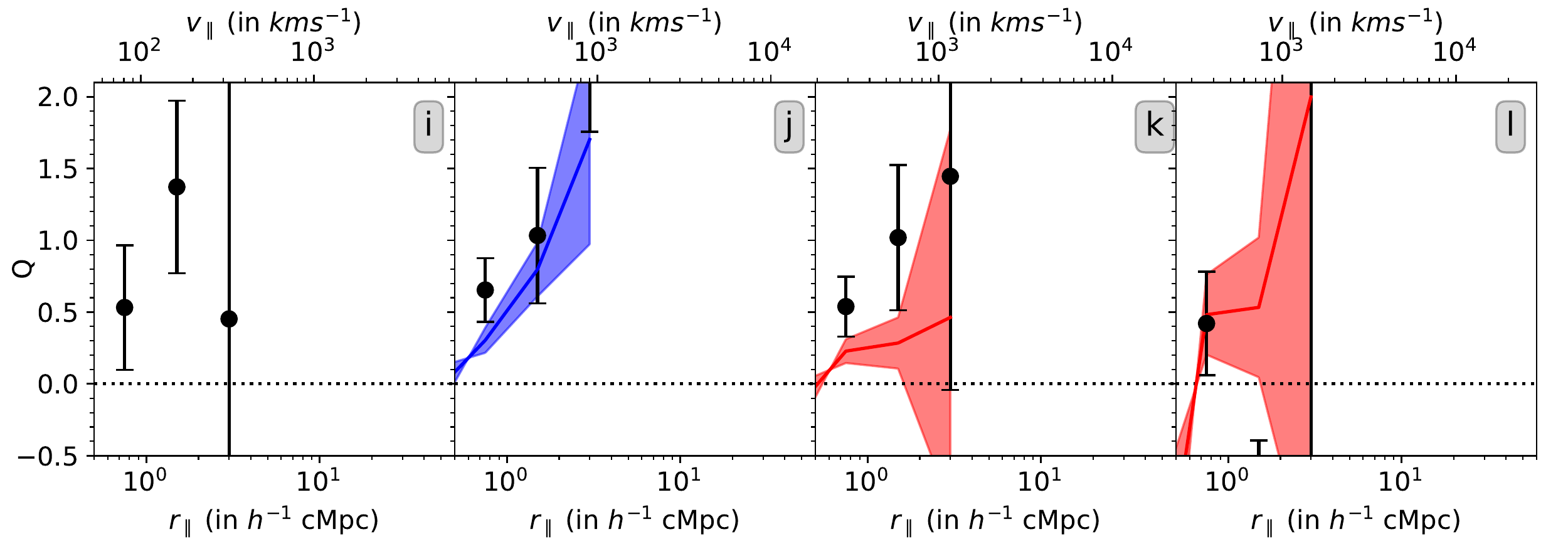}
	
	\caption{Absorber-based longitudinal two-, three- and reduced three-point correlation (top to bottom) of \lya\ absorbers for $\Delta>4$ (corresponding log$N_{\rm HI}$ thresholds given) as a function of longitudinal scale for different redshift intervals. The errorbar shown corresponds to the larger of the two errors: one-sided poissonian uncertainty in the number of absorber pairs (or triplets) corresponding to $\pm 1\sigma$ or bootstrapping error. The left plot represents correlation function obtained at $z<0.48$ using HST-COS sample \citet{danforth2016}. The black points represent the observed correlation function obtained from HST-COS or joint {\sc KODIAQ+SQUAD} sample. The blue and red curves represent correlation function obtained from simulation using $\Gamma_{\rm KS19}$ and $\Gamma_{\rm obs}$, respectively. The black dotted horizontal lines are used as a perspective for comparing correlation amplitudes at different redshifts with respect to the $1.7<z<2.3$ bin.}
\label{Corr_Delta}
\end{figure*}

In Section~\ref{Sec_Corr}, we studied the observed redshift evolution of clustering of \lya\ absorbers for a fixed \NHI\ threshold. This redshift evolution sources most probably from the evolution of thermal and ionization state of the IGM as well as the evolution in clustering of baryonic density field \citep[see dicussions in][]{maitra2020}. Now, instead of having a constant \NHI\ threshold at each redshift, we vary the \NHI\ threshold such that we will have a constant baryonic overdensity $\Delta$ threshold at all redshifts. In doing so, we would be able to study the evolution in the clustering properties of the baryonic field. This will allow us to disentagle the effects of thermal and ionization evolution of the IGM and the clustering of the baryonic density field. 

For this study, we also include clustering measurements by \citet{maitra2020b} of low-$z$ ($z<0.48$) \lya\ absorbers using quasar sample of \citet{danforth2016} observed with HST-COS. This allows us to probe the evolution in clustering of the baryonic density field measured for the Voigt profile components over a large redshift range (i.e $0\le z\le 3.5$). Firstly, in order to get a constant overdensity threshold at each redshift, we compute the $N_{\rm HI}$ vs. $\Delta$ relation from the simulations at $z=2.0,\ 2.6$ and 3.2. In the case of $z=2.0$, we will use $\Gamma_{\rm HI}$ from \citet{khaire2019} since it reproduces the observed clustering signals nicely. For $z=2.6$ and 3.2, we use the $\Gamma_{\rm HI}$ obtained by matching the observed mean flux instead, since have a better agreement with the observed clustering signals. In Fig.~\ref{N_vs_Delta_obs}, we plot the $N_{\rm HI}$ vs. $\tau$ weighted $\Delta$ relationship plot at $z=2.0\,\ 2.6$ and $3.2$. The best fitted relationship are also given in each panel.
For the low-$z$ \lya\ absorbers, we take the $N_{\rm HI}$ vs. $\Delta$ relation from Figure~8 in \citet{gaikwad2017a}, i.e, $\Delta=34.8\pm 5.9\times N_{14}^{0.770\pm 0.022}$.

In Fig.~\ref{Corr_Delta}, we plot the two- and three-point correlations as a function of $r_{\parallel}$ for a constant overdensity threshold of $\Delta>4$ at $z<0.48$, $z=1.7-2.3$, $z=2.3-2.9$ and $z=2.9-3.5$ (corresponding to log$N_{\rm HI}>12.77$, 13.69, 14.02 and 14.31 in the four redshift bins, respectively). We also plot simulated correlation curves for $z=2.0,\ 2.6$ and 3.2. We provide black dotted horizontal lines for two- and three-point correlations as a perspective for the comparison of the correlation amplitudes at different redshift with respect to $z=1.7-2.3$. { The observed and simulated correlation functions agree reasonably well with each other, except in the $r_{\parallel}=0.5-1h^{-1}$cMpc bin, where the simulations produce slightly reduced correlation amplitudes. Comparing this with Fig.~\ref{Corr_Gammacorr} (where the correlation amplitudes were calculated for a smaller $N_{\rm HI}$ threshold of $10^{13.5}$cm$^{-2}$) indicates a somewhat weaker $N_{\rm HI}$ dependence of \lya\ absorbers clustering at this scale in simulations.} 

Interestingly, we also find that for a fixed $\Delta$ threshold, the redshift evolution of the correlation amplitudes are much weaker in comparison to a fixed \NHI\ threshold. This is true for simulations too. In the case of two-point correlations, we see a weak evolution in the amplitudes at scales less than 2$h^{-1}$cMpc in going from the highest to the lowest redshift bin. In case of three-point correlation, a slightly stronger evolution is seen in the $r_{\parallel}=1-2h^{-1}$cMpc bin. The significance of this is, however, less because of larger errorbars at $z<0.48$ and the correlations being noisy at $z-2.9-3.5$. The reduced three-point correlation Q remains more or less constant within the given errorbars.

This weak evolution of two-point and three-point correlation seen for a fixed overdensity threshold confirms the fact that the large redshift evolution seen in the correlations for a fixed $N_{\rm HI}$ threshold sources primarily from the redshift evolution of $N_{\rm HI}$ vs. $\Delta$ relation and does not reflect the evolution of clustering in the baryonic field itself, or the underlying dark matter probed by the \lya\ absorption. This is in agreement to our findings in \citep[][]{maitra2020b} about the redshift evolution of transverse clustering in \lya\ absorbers.

\section{Summary}
\label{Sec:summary}

 We present measurements of line of sight (i.e redshift space) two- and three-point correlation function of 
 \lya\ absorbers at $1.7<z<3.5$ using the largest sample till date (i.e., 292 quasar sightlines) of high resolution ($\sim 6 $\kms) spectra from the publicly available {\sc KODIAQ} and {\sc SQUAD} samples. We report the measurements of redshift space two-point correlation (in three redshift bins) with high significance and first detections of redshift space three-point correlation for high-$z$ \lya\ absorbers. We study their spatial profile, redshift evolution and $N_{\rm HI}$ dependence. We use hydrodynamical simulations to investigate the effects of peculiar velocities, feedback processes, the ionization state of the IGM as well as thermal and pressure smoothing effects. We also compare our measurements with the predictions of simulations and draw some conclusions related to the physical state of the gas.
 The main results are summarized below.

\vskip 0.1in
\par \noindent{\bf Clustering properties of \lya\ absorbers:}
We studied redshift space two- and three-point correlation for \lya\ absorbers having $N_{\rm HI}>10^{13.5}$cm$^{-2}$ in { three} redshift bins: $z=1.7-2.3,\ 2.3-2.9$ and $2.9-3.5$. We detect positive two-point correlation up to a length scale of $r_{\parallel}=8h^{-1}$cMpc in all the redshift bins studied with the strongest detection ($>10\sigma$) coming from the length scale $r_{\parallel}=0.25-0.5h^{-1}$cMpc. We also find a strong evolution of two-point correlations with redshift, which is stronger than the redshift evolution expected purely on the basis of dark matter evolution \citep{cristiani1997}. We show that the radial profile of the two-point correlation can be well approximated with a power-law for scale above 0.5$h^{-1}$cMpc for $z=2.3-2.9$ and $2.9-3.5$, and above 1$h^{-1}$cMpc for $z=1.7-2.3$. The power-law index typically varies between $-1.9$ to $-1.4$ with an error of $\sim 0.2$. 

We also study redshift space three-point correlation using triplets having equal arm length configurations. We detect positive three-point correlation upto a length scale of 2$h^{-1}$cMpc at $z=1.7-2.3$ and upto 1$h^{-1}$cMpc at $z=2.3-2.9$. We do not detect any three-point correlation at $z=2.9-3.5$ for $N_{\rm HI}>10^{13.5}$cm$^{-2}$ absorbers (corresponding to mean overdensity at such redshifts). The strongest detection in three-point correlation is seen at $1.7\le z \le 2.3$ for scales  $\sim 1-2h^{-1}$cMpc with an amplitude of $1.81\pm 0.59$ ( $\sim 3.1\sigma$ significance). The corresponding reduced three-point correlation is found to be Q=$0.68\pm 0.23$. Unlike three-point correlation, the reduced three-point correlation is found to be relatively scale independent.

Both two- and three-point correlation amplitudes are found to be suppressed at scales below 0.25$h^{-1}$cMpc ($\sim$75 \kms\ at $z=2.0$, $\sim$97\kms\ at $z=2.6$ and $\sim$122\kms\ at $z=3.2$) in all the redshift bins. This small scale suppression can be attributed to the difficulty in identification of multiple Voigt profile components for thermally broadened absorption structure as well as instrumental resolution and other systematic effects.

\vskip 0.1in
\par \noindent{\bf Dependence of $N_{\rm HI}$ on clustering:}
Based on the assumption of a simple $N_{\rm HI}$ vs $\Delta$ relation, one expects stronger $N_{\rm HI}$ absorbers to arise from more overdense regions. It is also expected that clustering strength will scale with overdensity.
With this motivation, we studied the $N_{\rm HI}$ dependence of both two-point and three-point correlation at scales $r_{\parallel}=0.5-1,\ 1-2,\ 2-4 h^{-1}$cMpc in all the redshift bins. We find that both two-point and three-point correlation increases with increasing $N_{\rm HI}$ threshold at all scales and in all the redshift bins. Their $N_{\rm HI}$ threshold dependence can be well approximated with a power-law. While the two-point correlation increase with $N_{\rm HI}$ threshold with a typical power-law index of $\sim 0.7$, the three-point correlation has a stronger dependence with an index of $\sim 1.4$. Interestingly, we find that while two-point and three-point correlation increase with increasing $N_{\rm HI}$ thresholds, the reduced three-point correlation Q remains relatively unaffected by it. 

\vskip 0.1in
\par \noindent{\bf Effect of redshift space distortion:}
Using hydrodynamical simulations at $z=2$, we find that peculiar velocity amplifies the real space two-point correlation at all scales. The effect is strongest at the scale of 0.5-1$h^{-1}$ cMpc with an amplification factor of $\sim 3$. The effect goes weaker with { increasing} scale with an amplification factor of only $\sim 1.2$ at $r_{\parallel}=2-4h^{-1}$cMpc. So, the peculiar velocities makes the real space two-point correlation profile steeper in redshift space.
The effect of peculiar velocity is even stronger in the case of three-point correlation with an amplification factor of $\sim 5$ at the scale of 0.5-1$h^{-1}$cMpc. The amplification seen in two-point and three-point correlations in redshift space suggests that the \lya\ absorbers are part of converging flows.

\vskip 0.1in
\par \noindent{\bf Effect of astrophysical parameters:}
We also investigated the effect of astrophysical parameters of the IGM { governed by} its thermal history and ionization state, as well as, the effect of feedback processes on clustering of \lya\ absorbers. By comparing our "fiducial" simulation with a "hot" simulation, we found that thermal and pressure smoothing effects only affect the clustering signal at scales less than 0.25$h^{-1}$cMpc. We also consider two Sherwood simulations to study the effect of wind and AGN feedback and found it to be relevant only at scales below 0.5$h^{-1}$cMpc. { The \HI\ photo-ionization rate affects the clustering signals at all scales with  clustering amplitude increasing with increasing $\Gamma_{\rm HI}$ values.} Hence, one can in principle use the two-point and three-point correlation of \lya\ absorbers at scale above $0.5h^{-1}$cMpc as a sensitive probe of $\Gamma_{\rm HI}$.

We also investigated whether simulations correctly reproduce the observed clustering statistics at scales above $0.5h^{-1}$cMpc using $\Gamma_{\rm HI}$ values from \citet{khaire2019} ($\Gamma_{\rm KS19}$). At the lowest redshift bin of $z=1.7-2.3$, we find that the observed clustering statistics of \lya\ forest is reproduced reasonably well by our models. However these models under-predict the clustering amplitudes at higher redshifts. We show that better matching can be found by increasing the $\Gamma_{\rm HI}$ that will also make our simulations reproduce the observed mean transmitted flux. While this exercise demonstrates the need to revisit the $\Gamma_{\rm HI}$ measurements, we reiterate that it has to be done by simultaneously matching various observables. We leave such a comprehensive parameter extraction to a future project.

\vskip 0.1in
\par \noindent{\bf Redshift evolution of clustering:} 
As mentioned before, we found a strong evolution of two- and three-point correlation of \lya\ absorbers having $N_{\rm HI}>10^{13.5}$cm$^{-2}$. In order to understand whether this corresponds to an actual evolution of clustering in the baryonic overdensities or is just a by-product of how $N_{\rm HI}$ vs $\Delta$ relation evolves with redshift, we calculated the correlation with a fixed overdensity threshold of $\Delta>4$. We found that both two- and three-point correlations remain relatively constant in all the redshift bins. We also compared our results with the correlation statistics in $z<0.48$ \lya\ absorbers \citep{maitra2020} and found only a slight enhancement in two- and three-point correlation signals. This confirms the fact that a strong evolution of two- and three-point correlation for a fixed $N_{\rm HI}$ threshold sources from the evolution of $N_{\rm HI}$ vs $\Delta$ relation.

\vskip 0.1in
\par \noindent{\bf Future perspective:}
We present all our clustering measurements in the tabulated form in the appendix to enable future parameter estimation studies by other groups.
As discussed before, our simulations suggested the need for a slightly elevated \HI\ photo-ionization rates in comparison to the presently available measurements to match the clustering signals at higher redshifts. This necessitates a more thorough study of the entire parameter space involving thermal and ionization parameters of IGM as well as cosmological parameters to properly constrain the \HI\ photo-ionization rate. This, however, requires simultaneous Voigt profile fitting over sightlines covering the entire parameter space which is computationally expensive. This is beyond the scope of this work and is hence left as a future exercise.

{The study of redshift-space clustering in \lya\ absorbers at small scales (i.e $\le$ 0.5 cMpc) will benefit greatly from the spectra obtained using very high-resolution (i.e R$\sim$100,000) spectrographs like EELT-HIRES \citep[][]{marconi2016} and ESPRESSO \citep[][]{pepe2021}. 
Both two- and three-point correlation signals at larger scales can be determined with better significance with the help of large number of high-resolution spectra from future surveys like the  Maunakea Spectroscopic Explorer \citep[MSE;][]{McConnachie2016}.
Such large surveys will also pave way for the measurement of transverse three-point correlation of \lya\ absorbers using projected quasar triplets.
However, one needs a large number of triplet sightlines \citep[i.e $\sim$ 100 triplet sightlines are needed for a $\sim 5\sigma$ detection at scales $\sim 4 h^{-1}$ cMpc as discussed in section~5 of][]{maitra2020b} 
in order to obtain significant transverse three-point correlation signals.
Such triplet sightlines are very rare in nature.
However, with the help of future large surveys like MSE, Dark Energy Spectroscopic Instrument DESI \citep{desi2016} and Legacy Survey of Space and Time (LSST), one expects a significant improvement in the number of known quasar triplets.}

 In this work, we also explored the effect of metal contamination in the clustering of \lya\ forest and found its effect to be negligible. It will, however, be interesting to investigate the clustering of these metal line systems and its redshift evolution by making a list of metal line system for the entire sample.
Such an exercise will also allow us to perform cross-correlation studies to associate these metal line absorptions with \HI\ absorption. Our measurement will also allow us to infer linear bias and possibly the first non-linear bias terms for the IGM at high-$z$. We hope to present a detailed self-consistent discussion on the bias in our upcoming work.

\section*{Acknowledgments}
{
We acknowledge the use of High performance computing facilities PERSEUS and PEGASUS at IUCAA. 
{ We thank the referee for useful comments},
J. Bolton for the Sherwood simulations and K. Subramanian, T. R. Choudhury, P. Petitjean and A. Pranjape for useful discussions. PG acknowledges the support of the UK Science and Technology Facilities Council (STFC). The Sherwood simulations were performed using the Curie supercomputer at the Tre Grand Centre de Calcul (TGCC), and the DiRAC Data Analytic system at the
University of Cambridge, operated by the University of Cambridge High
Performance Computing Service on behalf of the STFC DiRAC HPC Facility
(www.dirac.ac.uk) . This was funded by BIS National E-infrastructure
capital grant (ST/K001590/1), STFC capital grants ST/H008861/1 and
ST/H00887X/1, and STFC DiRAC Operations grant ST/K00333X/1. DiRAC is part of
the National E-Infrastructure.
}

\section*{Data Availability}
The data underlying this article are available in the article and in its online supplementary material.




\bibliographystyle{mnras}
\bibliography{main} 




\appendix

\section{Appendix}

\begin{table*}
	\caption{\HI\ column density distribution $f= log (d^2N/dN_{\rm HI}dX)$ for the combined {\sc KODIAQ+SQUAD} sample in three redshift bins.}
	\centering
	\begin{filecontents*}{NHI_dis_Datalist.txt}
12.875 -10.894 0.051 0.058 -11.065 0.039 0.043 -11.158 0.038 0.042
13.125 -11.298 0.040 0.044 -11.405 0.031 0.033 -11.399 0.038 0.041
13.375 -11.696 0.032 0.035 -11.701 0.028 0.030 -11.658 0.034 0.037
13.625 -12.056 0.030 0.032 -12.052 0.028 0.030 -11.989 0.033 0.036
13.875 -12.449 0.031 0.034 -12.408 0.028 0.030 -12.369 0.033 0.035
14.125 -12.901 0.031 0.034 -12.832 0.027 0.028 -12.738 0.033 0.035
14.375 -13.374 0.031 0.034 -13.228 0.031 0.031 -13.133 0.036 0.036
14.625 -13.858 0.036 0.036 -13.765 0.045 0.045 -13.553 0.046 0.047
14.875 -14.283 0.049 0.050 -14.264 0.058 0.058 -14.062 0.058 0.058
15.125 -14.720 0.058 0.058 -14.687 0.077 0.078 -14.536 0.069 0.069
15.375 -15.256 0.080 0.081 -15.217 0.084 0.086 -15.020 0.097 0.099
15.625 -15.590 0.092 0.094 -15.517 0.106 0.109 -15.466 0.129 0.133
15.875 -15.909 0.087 0.089 -15.924 0.120 0.123 -15.888 0.174 0.185
\end{filecontents*}
\pgfplotstabletypeset[col sep = space,
	header=false,
	every head row/.append style={
		before row=\toprule,
		after row/.add={}{ &  \multicolumn{3}{c}{($1.7\le z \le 2.3$)}  & \multicolumn{3}{c}{($2.3\le z \le 2.9$)} & \multicolumn{3}{c}{($2.9\le z\le 3.5$)}  \\ \midrule
		}
	},
	every last row/.style={after row=\bottomrule},
	display columns/0/.style={string type,column type={l},column name=log$N_{\rm HI}$},
	display columns/1/.style={string type,column type={l},column name=$f$},
	display columns/2/.style={string type,column type={l},column name=$+\Delta f$},
	display columns/3/.style={string type,column type={l},column name=$-\Delta f$},
	display columns/4/.style={string type,column type={l},column name=$f$},
	display columns/5/.style={string type,column type={l},column name=$+\Delta f$},
	display columns/6/.style={string type,column type={l},column name=$-\Delta f$},
	display columns/7/.style={string type,column type={l},column name=$f$},
	display columns/8/.style={string type,column type={l},column name=$+\Delta f$},
	display columns/9/.style={string type,column type={l},column name=$-\Delta f$},
	]{NHI_dis_Datalist.txt}
	\label{tab:NHI_dis}
\end{table*}

\begin{table*}
	\caption{Observed values of longitudinal two-point ($\xi$), three-point ($\zeta$) and reduced three-point (Q) correlations for the combined {\sc KODIAQ+SQUAD} sample in three redshift bins.}
	\begin{tabular}{cccccccc}
		\toprule
		$r_{\parallel}$  & \multicolumn{3}{c}{$1.7\le z\le 2.3$} & \multicolumn{3}{c}{$2.3\le z\le 2.9$} & \multicolumn{1}{c}{$2.9\le z \le 3.5$} \\
		in $h^{-1}$cMpc & $\xi$ & $\zeta$ & Q & $\xi$ & $\zeta$ & Q & $\xi$  \\
		\midrule
		\midrule
		0.125-0.25 & $0.924\pm 0.086$ & $1.28\pm 1.15$ & $0.37\pm 0.34$ & $0.64\pm 0.06$ & $-1.0\pm 0.35$ & $-0.56\pm 0.21$ & $0.07\pm 0.04$ \\
		0.25-0.5 & $1.39\pm 0.11$ & $2.83\pm 1.02$ & $0.56\pm 0.21$ & $1.08\pm 0.09$ & $1.19\pm 0.59$ & $0.43\pm 0.22$ & $0.41\pm 0.04$ \\
		0.5-1    & $1.13\pm 0.09$ & $1.81\pm 0.59$ & $0.68\pm 0.23$ & $0.76\pm 0.06$ & $0.66\pm 0.35$ & $0.62\pm 0.33$ & $0.37\pm 0.03$ \\
		1-2     & $0.60\pm 0.04$ & $0.48\pm 0.25$ & $0.84\pm 0.44$ & $0.33\pm 0.03$ & $0.08\pm 0.12$ & $0.44\pm 0.62$ & $0.16\pm 0.03$\\
		2-4     & $0.18\pm 0.02$ & $0.06\pm 0.07$ & -   & $0.12\pm 0.01$ & $0.05\pm 0.05$ & - & $0.06\pm 0.02$\\
		4-8     & $0.04\pm 0.01$ & $0.03\pm 0.02$    &   -   & $0.05\pm 0.01$ & $0.03\pm 0.02$ & - & $0.02\pm 0.01$ \\
		8-16     & $0.00\pm 0.01$ & $-0.03\pm 0.01$    &   -   & $0.01\pm 0.01$ & $0.00\pm 0.01$ & - & $0.00\pm 0.01$ \\
		16-32     & $-0.01\pm 0.01$ & $0.01\pm 0.01$    &   -   & $0.01\pm 0.01$ & $-0.01\pm 0.01$ & - & $0.01\pm 0.01$ \\
		
		\bottomrule
	\end{tabular}
	\label{tab:Corr}
\end{table*}

\begin{table}
	\caption{SQUAD sample}
	\centering
	\begin{filecontents*}{SQUAD_Data1.txt}
QSO_name z_em z_min z_max Median_SNR
J001210-012207 1.999 1.7 1.9 8.06
J012417-374423 2.2 1.70036 2.1 39.8
J022620-285750 2.171 1.7 2.1 13.72
J031006-192124 2.144 1.7 2.1 12.24
J031009-192207 2.122 1.7 2.09978 14.7
J042707-130253 2.159 1.7 2.1 12.64
J043037-485523 1.94 1.7 1.9 14.88
J055158-211949 2.245 1.73799 2.22062 6.49
J092129-261843 2.3 1.7844 2.27462 18.94
J110325-264515 2.145 1.7 2.1 46.56
J114608-244732 1.94 1.7 1.9 5.18
J120342+102831 1.888 1.7 1.86954 7.55
J121140+103002 2.193 1.7 2.1 14.8
J124913-055919 2.247 1.73968 2.22259 16.16
J134427-103541 2.134 1.7 2.09484 43.93
J143040+014939 2.113 1.7 2.09093 6.72
J143229-010616 2.085 1.7 2.06371 8.49
J145102-232930 2.215 1.71268 2.1 48.95
J211719-433424 2.041 1.7 2.02013 10.07
J213314-464030 2.204 1.7034 2.1 8.05
J222756-224302 1.891 1.7 1.87249 46.8
J223922-294947 1.849 1.7 1.83112 7.02
J223941-294954 2.102 1.7 2.08012 8.37
J223948-294748 2.068 1.7 2.04669 8.85
J223951-294836 2.121 1.7 2.0988 10.54
J224752-123719 1.892 1.7 1.87348 7.08
J225154-314520 1.938 1.7 1.9 11.25
J234819+005721 2.16 1.7 2.1 13.26
J002110-242248 2.256 1.9 2.23093 7.62
J004201-403039 2.37 1.9 2.3 10.04
J005102-252846 2.082 1.9 2.06045 6.53
J011143-350300 2.41 1.9 2.3 65.29
J022928-364357 2.115 1.9 2.0929 5.78
J030049-072137 2.106 1.9 2.08356 5.82
J033106-382404 2.423 1.9 2.39528 40.95
J112442-170517 2.4 1.9 2.3 106.93
J120044-185944 2.448 1.90927 2.41979 51.53
J135038-251216 2.534 1.98184 2.5 84.94
J144231+013734 2.277 1.9 2.1 5.28
\end{filecontents*}
\pgfplotstabletypeset[col sep = space,
	every head row/.style={%
		before row=\toprule,
		after row=\midrule,
		typeset cell/.code={
			\ifnum\pgfplotstablecol=\pgfplotstablecols
			\pgfkeyssetvalue{/pgfplots/table/@cell content}{\multicolumn{1}{c}{##1}\\}%
			\else
			\pgfkeyssetvalue{/pgfplots/table/@cell content}{\multicolumn{1}{c}{##1}&}%
			\fi
		}
	},
	every last row/.style={after row=\bottomrule},
	display columns/0/.style={string type,column type={l}},
	display columns/1/.style={string type,column type={l}},
	display columns/2/.style={string type,column type={l}},
	display columns/3/.style={string type,column type={l}},
	display columns/4/.style={string type,column type={l}},
	]{SQUAD_Data1.txt}
	\label{tab:squad_sample}
\end{table}

\begin{table}
	\contcaption{}
	\centering
	\begin{filecontents*}{SQUAD_Data2.txt}
QSO_name z_em z_min z_max Median_SNR
J211739-433537 2.05 1.9613 2.02898 5.71
J215155-302753 2.345 1.9 2.3 16.08
J222006-280323 2.406 1.9 2.3 76.0
J223850-295300 2.387 1.9 2.1 5.76
J232046-294406 2.401 1.9 2.1 7.42
J000448-415728 2.76 2.17253 2.7 79.44
J001130+005550 2.309 2.1 2.27 7.67
J004351-265128 2.786 2.19446 2.7 10.95
J005109-255216 2.249 2.1 2.22455 7.09
J015327-431137 2.74 2.15565 2.7 81.81
J033108-252443 2.685 2.10924 2.65187 33.83
J045214-164016 2.6 2.1 2.5 32.38
J091127+055054 2.793 2.20029 2.7 40.91
J114254+265457 2.596 2.1 2.5 65.62
J140445-013021 2.499 2.1 2.46958 10.6
J213302-464026 2.23 2.1 2.20589 9.03
J232059-295521 2.317 2.1 2.2913 7.39
J234825+002040 2.654 2.1 2.62115 6.76
J235600-682003 2.751 2.16493 2.7 14.26
J035405-272421 2.819 2.3 2.7 23.23
J235034-432559 2.885 2.3 2.84733 109.23
J004342-262209 3.053 2.5 3.01123 14.67
J025140-220026 3.197 2.54125 3.1 16.58
J033626-201940 3.125 2.5 3.0814 7.99
J094253-110426 3.054 2.5 3.01221 167.19
J212912-153841 3.268 2.60115 3.22061 119.75
J214222-441929 3.22 2.56065 3.1 7.95
J214225-442018 3.23 2.56909 3.1 26.45
J004452-264009 3.33 2.7 3.28091 16.79
J010604-254651 3.365 2.7 3.3 25.83
J033755-120404 3.442 2.74797 3.1 8.76
J005758-264314 3.655 2.92769 3.5 56.38
J122602+132114 3.533 2.9 3.47788 20.32
J124957-015928 3.637 2.91233 3.5 55.81
J132029-052335 3.7 2.96566 3.5 54.92
J194025-690756 3.154 2.9 3.1 53.28
J024401-013403 4.053 3.3 3.5 21.14
J040356-170321 4.227 3.41032 3.5 19.25
\end{filecontents*}
\pgfplotstabletypeset[col sep = space,
	every head row/.style={%
		before row=\toprule,
		after row=\midrule,
		typeset cell/.code={
			\ifnum\pgfplotstablecol=\pgfplotstablecols
			\pgfkeyssetvalue{/pgfplots/table/@cell content}{\multicolumn{1}{c}{##1}\\}%
			\else
			\pgfkeyssetvalue{/pgfplots/table/@cell content}{\multicolumn{1}{c}{##1}&}%
			\fi
		}
	},
	every last row/.style={after row=\bottomrule},
	display columns/0/.style={string type,column type={l}},
	display columns/1/.style={string type,column type={l}},
	display columns/2/.style={string type,column type={l}},
	display columns/3/.style={string type,column type={l}},
	display columns/4/.style={string type,column type={l}},
	]{SQUAD_Data2.txt}
\end{table}

\begin{table}
	\caption{KODIAQ sample}
	\centering
	\begin{filecontents*}{KODIAQ_Data1.txt}
QSO_name z_em z_min z_max Median_SNR
J000520+052410 1.887 1.70000 1.86844 12.08
J000931+021707 2.350 1.82659 2.3 11.67
J002952+020606 2.333 1.81224 2.3 7.41
J005202+010129 2.270 1.75908 2.24502 10.96
J010054+021136 1.959 1.70000 1.90000 8.83
J014944+150106 2.070 1.7 2.04853 12.34
J022853-033737 2.066 1.7 2.04459 14.13
J025127+341442 2.230 1.72533 2.20575 7.88
J031003-004645 2.114 1.7 2.09178 7.11
J042408+020424 2.044 1.7 2.02296 9.46
J075054+425219 1.896 1.70000 1.87731 11.71
J082849+085854 2.271 1.75993 2.24601 6.73
J092708+582319 1.910 1.70000 1.89109 8.30
J093857+412821 1.935 1.70000 1.90000 8.23
J095309+523029 1.875 1.70000 1.85663 10.57
J095822+014524 1.960 1.70000 1.90000 9.30
J101939+524627 2.170 1.7 2.1 11.64
J102410+060013 2.130 1.7 2.1 7.6
J104213+062853 2.035 1.7 2.01411 7.48
J105648+120826 1.923 1.70000 1.90000 6.24
J110610+640009 2.202 1.70171 2.17825 22.08
J110621+104432 1.859 1.70000 1.84087 9.27
J122824+312837 2.200 1.70002 2.17629 39.63
J123643-020420 1.864 1.70000 1.84579 7.99
J124924-023339 2.116 1.7 2.09375 8.17
J131011+460124 2.133 1.7 2.1 11.61
J131040+542449 1.929 1.70000 1.90000 5.51
J131341+144140 1.884 1.70000 1.86549 5.17
J133532+082404 1.909 1.70000 1.89011 5.61
J134544+262506 2.031 1.7 2.01017 15.97
J140501+444800 2.218 1.71521 2.19397 9.19
J141719+413237 2.023 1.7 2.0023 11.6
J141906+592312 2.320 1.80127 2.2941 20.11
J145435+094100 1.945 1.70000 1.90000 7.90
J152156+520238 2.208 1.70677 2.18415 37.7
J160547+511330 1.786 1.70000 1.76894 7.73
J162645+642655 2.320 1.80127 2.2941 28.52
J162902+091322 1.985 1.7 1.96491 7.57
J212329-005052 2.262 1.75233 2.23717 36.02
J231324+003444 2.083 1.7 2.06131 12.49
J234023-005327 2.085 1.7 2.06328 14.53
J002127-020333 2.596 2.03415 2.56459 11.96
J002830-281704 2.400 1.9 2.37257 5.91
J003501-091817 2.418 1.9 2.39022 20.79
J004358-255115 2.501 1.95399 2.47156 12.49
J004448+372114 2.410 1.9 2.38237 6.27
J005814+011530 2.494 1.94809 2.4647 10.4
J011150+140141 2.470 1.92784 2.44119 7.81
J015227-200107 2.147 1.90000 2.10000 6.62
J023145+132254 2.059 1.90000 2.03771 7.12
J023359+004938 2.522 1.97171 2.49213 7.74
J023924-090138 2.471 1.92868 2.44217 6.15
J024008-230915 2.225 1.9 2.20084 12.7
J025515+014828 2.470 1.92784 2.44119 6.32
\end{filecontents*}
\pgfplotstabletypeset[col sep = space,
	every head row/.style={%
		before row=\toprule,
		after row=\midrule,
		typeset cell/.code={
			\ifnum\pgfplotstablecol=\pgfplotstablecols
			\pgfkeyssetvalue{/pgfplots/table/@cell content}{\multicolumn{1}{c}{##1}\\}%
			\else
			\pgfkeyssetvalue{/pgfplots/table/@cell content}{\multicolumn{1}{c}{##1}&}%
			\fi
		}
	},
	every last row/.style={after row=\bottomrule},
	display columns/0/.style={string type,column type={l}},
	display columns/1/.style={string type,column type={l}},
	display columns/2/.style={string type,column type={l}},
	display columns/3/.style={string type,column type={l}},
	display columns/4/.style={string type,column type={l}},
	]{KODIAQ_Data1.txt}
	\label{tab:kodiaq_sample}
\end{table}

\begin{table}
	\contcaption{}
	\centering
	\begin{filecontents*}{KODIAQ_Data2.txt}
QSO_name z_em z_min z_max Median_SNR
J030046+022245 2.520 1.97002 2.49017 5.63
J035116+061914 2.059 1.90000 2.03771 5.12
J040241-064124 2.432 1.9 2.40394 7.71
J083102+335803 2.427 1.9 2.39904 15.68
J083712+145917 2.512 1.96327 2.48234 6.85
J101723-204658 2.545 1.99112 2.5 38.21
J115940-003203 2.035 1.90000 2.01411 7.33
J120416+022111 2.529 1.97762 2.49899 12.05
J121117+042222 2.526 1.97509 2.49605 19.24
J124610+303117 2.560 2.00377 2.5 46.83
J124610+303131 2.560 2.00377 2.5 16.58
J131215+423900 2.566 2.00884 2.5 17.08
J134004+281653 2.517 1.96749 2.48723 12.31
J160455+381201 2.551 1.99618 2.5 28.21
J160455+381214 2.551 1.99618 2.5 67.32
J162548+264433 2.535 1.98268 2.5 7.46
J162548+264658 2.517 1.96749 2.48723 17.52
J162557+264448 2.601 2.03837 2.56948 6.77
J203642-055300 2.582 2.02234 2.55089 8.97
J220852-194400 2.573 2.01474 2.5 19.82
J225409+244523 2.328 1.9 2.3 5.72
J233823+150445 2.418 1.9 2.39022 8.51
J234628+124859 2.573 2.01474 2.5 18.82
J235050+045507 2.593 2.03162 2.56165 11.3
J005700+143737 2.643 2.1 2.61059 9.89
J010311+131617 2.721 2.13962 2.68688 69.53
J010806+163550 2.652 2.1 2.61939 40.94
J014516-094517 2.729 2.14637 2.6947 15.11
J020950-000506 2.828 2.2299 2.79144 34.8
J022839-101110 2.256 2.10000 2.23128 21.62
J025644+001246 2.251 2.10000 2.22637 5.97
J045213-164012 2.684 2.1084 2.65069 31.95
J053007-250329 2.765 2.17674 2.7 11.08
J081240+320808 2.711 2.13118 2.6771 25.74
J082107+310751 2.625 2.1 2.59297 25.94
J101155+294141 2.652 2.1 2.61939 31.81
J121303+171423 2.558 2.1 2.5 8.15
J121930+494052 2.633 2.1 2.6008 21.46
J130411+295348 2.850 2.24846 2.81293 19.31
J131855+531207 2.321 2.10000 2.29508 16.16
J143500+535953 2.635 2.1 2.60276 36.54
J144453+291905 2.660 2.1 2.62722 26.04
J155152+191104 2.843 2.24256 2.80609 70.3
J155610+374039 2.658 2.1 2.62526 10.6
J155810-003120 2.827 2.22906 2.79047 32.24
J155814+405337 2.635 2.1 2.60276 20.97
J160355+573054 2.850 2.24846 2.81293 31.88
J163412+320335 2.343 2.10000 2.30000 12.56
J170100+641209 2.734 2.15059 2.69959 31.18
J204051-010538 2.783 2.19193 2.7 9.24
J220639-181846 2.728 2.14553 2.69372 12.11
J234632+124540 2.515 2.1 2.48527 8.75
J010741-263328 2.460 2.30000 2.43139 5.98
J012156+144823 2.870 2.3 2.83246 20.19
J013301-400628 3.023 2.39443 2.98174 54.06
\end{filecontents*}
\pgfplotstabletypeset[col sep = space,
	every head row/.style={%
		before row=\toprule,
		after row=\midrule,
		typeset cell/.code={
			\ifnum\pgfplotstablecol=\pgfplotstablecols
			\pgfkeyssetvalue{/pgfplots/table/@cell content}{\multicolumn{1}{c}{##1}\\}%
			\else
			\pgfkeyssetvalue{/pgfplots/table/@cell content}{\multicolumn{1}{c}{##1}&}%
			\fi
		}
	},
	every last row/.style={after row=\bottomrule},
	display columns/0/.style={string type,column type={l}},
	display columns/1/.style={string type,column type={l}},
	display columns/2/.style={string type,column type={l}},
	display columns/3/.style={string type,column type={l}},
	display columns/4/.style={string type,column type={l}},
	]{KODIAQ_Data2.txt}
\end{table}

\begin{table}
	\contcaption{}
	\centering
	\begin{filecontents*}{KODIAQ_Data3.txt}
QSO_name z_em z_min z_max Median_SNR
J015234+335033 2.431 2.30000 2.40296 14.26
J020455+364917 2.912 2.30078 2.87346 14.69
J021129+124110 2.953 2.33537 2.9 5.76
J022554+005451 2.975 2.35393 2.9 7.48
J080518+614423 3.033 2.40287 2.99149 8.37
J093643+292713 2.923 2.31006 2.8842 13.79
J093748+730158 2.525 2.30000 2.49507 11.07
J095714+544017 2.589 2.3 2.55774 12.53
J103514+544040 2.989 2.36575 2.9 10.25
J111038+483115 2.954 2.33621 2.9 12.07
J113130+604420 2.921 2.30837 2.88224 19.72
J113508+222715 2.886 2.3 2.84808 14.35
J120006+312630 2.989 2.36575 2.9 38.61
J130426+120245 2.980 2.35815 2.9 10.59
J134328+572147 3.034 2.40372 2.99246 9.08
J135317+532825 2.920 2.30753 2.88127 12.45
J143316+313126 2.940 2.3244 2.9 42.67
J143912+295448 2.992 2.36828 2.95151 13.75
J160843+071508 2.877 2.3 2.83929 11.61
J171227+575507 3.008 2.38178 2.96711 20.22
J175746+753916 3.050 2.41722 3.00806 15.82
J194455+770552 3.020 2.3919 2.97882 39.78
J212904-160249 2.900 2.3 2.86175 7.5
J221516-294423 2.706 2.3 2.67221 8.31
J222256-094636 2.926 2.31259 2.88712 15.97
J223408+000001 3.027 2.39781 2.98564 29.26
J234352+141014 2.896 2.3 2.85784 7.83
J235129-142756 2.933 2.3185 2.89395 28.11
J000150-015940 2.810 2.5 2.77386 10.05
J010925-210257 3.226 2.56572 3.17947 20.42
J030341-002321 3.175 2.52268 3.1 17.28
J030449-000813 3.295 2.62394 3.24659 15.36
J032412-320259 3.302 2.62984 3.2534 15.42
J033900-013318 3.197 2.54125 3.15125 11.24
J043038-133546 3.249 2.58512 3.20185 7.26
J064204+675835 3.177 2.52437 3.1 28.26
J074521+473436 3.220 2.56065 3.17363 35.61
J074927+415242 3.111 2.5 3.0675 18.61
J080117+521034 3.235 2.57331 3.18823 14.43
J082619+314848 3.093 2.5 3.04997 19.99
J083933+111207 2.696 2.50000 2.66243 6.25
J090033+421546 3.290 2.61972 3.24173 20.4
J094202+042244 3.275 2.60706 3.22714 40.79
J095852+120245 3.297 2.62562 3.24854 18.59
J100841+362319 3.125 2.5 3.08114 27.01
J101447+430030 3.125 2.5 3.08114 38.46
J102009+104002 3.167 2.51593 3.1 11.86
J114308+345222 3.145 2.5 3.1 32.6
J120917+113830 3.105 2.5 3.06166 18.52
J121134+090220 3.291 2.62056 3.2427 12.79
J122518+483116 3.090 2.5 3.04704 18.22
J123748+012606 3.144 2.5 3.09965 6.75
J160413+395121 3.130 2.5 3.08601 11.73
J210025-064146 3.137 2.5 3.09283 13.64
J223627+135714 3.216 2.55728 3.16974 21.58
\end{filecontents*}
\pgfplotstabletypeset[col sep = space,
	every head row/.style={%
		before row=\toprule,
		after row=\midrule,
		typeset cell/.code={
			\ifnum\pgfplotstablecol=\pgfplotstablecols
			\pgfkeyssetvalue{/pgfplots/table/@cell content}{\multicolumn{1}{c}{##1}\\}%
			\else
			\pgfkeyssetvalue{/pgfplots/table/@cell content}{\multicolumn{1}{c}{##1}&}%
			\fi
		}
	},
	every last row/.style={after row=\bottomrule},
	display columns/0/.style={string type,column type={l}},
	display columns/1/.style={string type,column type={l}},
	display columns/2/.style={string type,column type={l}},
	display columns/3/.style={string type,column type={l}},
	display columns/4/.style={string type,column type={l}},
	]{KODIAQ_Data3.txt}
\end{table}

\begin{table}
	\contcaption{}
	\centering
	\begin{filecontents*}{KODIAQ_Data4.txt}
QSO_name z_em z_min z_max Median_SNR
J224145+122557 2.631 2.50000 2.59884 8.60
J233446-090812 3.316 2.64165 3.26701 16.61
J234451+343348 3.010 2.5 2.96906 7.5
J234531+090906 2.784 2.50000 2.70000 6.14
J234856-104131 3.172 2.52015 3.1 16.69
J001708+813508 3.366 2.7 3.3 12.11
J004530-261709 3.440 2.74628 3.38749 16.59
J025905+001121 3.365 2.7 3.3 14.77
J045142-132033 3.093 2.7 3.04997 6.82
J064632+445116 3.408 2.71928 3.35641 7.96
J092914+282529 3.399 2.71169 3.3 34.33
J093337+284532 3.425 2.73362 3.37293 13.68
J102156+300140 3.129 2.7 3.08504 13.5
J103456+035859 3.369 2.7 3.3 9.16
J104018+572448 3.408 2.71928 3.35641 16.81
J113418+574204 3.521 2.81462 3.46611 26.1
J120147+120630 3.509 2.8045 3.45447 23.45
J134002+110630 2.913 2.70000 2.87444 14.62
J144516+095836 3.529 2.82137 3.47387 18.84
J151224+465233 3.359 2.7 3.3 10.47
J162453+375806 3.380 2.7 3.3 25.56
J173352+540030 3.424 2.73278 3.37195 29.18
J230301-093930 3.454 2.75809 3.40109 45.27
J231543+145606 3.390 2.70409 3.3 12.09
J235057-005209 3.025 2.7 2.98369 8.68
J015741-010629 3.564 2.9 3.5 12.51
J020346+113445 3.610 2.9 3.5 9.77
J034943-381031 3.222 2.9 3.17558 17.09
J073149+285448 3.675 2.94456 3.5 19.83
J101336+561536 3.632 2.90828 3.5 9.25
J142438+225600 3.620 2.9 3.5 43.96
J145408+511443 3.635 2.91081 3.5 10.35
J161009+472444 3.216 2.9 3.16974 6.87
J193957-100241 3.787 3.03906 3.5 41.09
J200324-325145 3.783 3.03569 3.5 14.18
J223619+132620 3.295 2.9 3.24659 12.07
J004434-261121 3.289 3.10000 3.24076 5.12
J081435+502946 3.883 3.12006 3.5 16.0
J082540+354414 3.846 3.1 3.5 17.88
J083141+524517 3.911 3.14369 3.5 36.45
J111113-080402 3.922 3.15297 3.5 10.55
J115538+053050 3.475 3.1 3.42147 18.23
J155556+480015 3.298 3.10000 3.24951 6.40
J221527-161132 3.990 3.21035 3.5 6.76
J013340+040059 4.150 3.34535 3.50000 17.10
J020944+051714 4.180 3.37066 3.50000 32.66
J105756+455553 4.137 3.33438 3.50000 24.34
J164656+551446 4.050 3.30000 3.50000 21.95
J215502+135826 4.260 3.43816 3.50000 9.86
\end{filecontents*}
\pgfplotstabletypeset[col sep = space,
	every head row/.style={%
		before row=\toprule,
		after row=\midrule,
		typeset cell/.code={
			\ifnum\pgfplotstablecol=\pgfplotstablecols
			\pgfkeyssetvalue{/pgfplots/table/@cell content}{\multicolumn{1}{c}{##1}\\}%
			\else
			\pgfkeyssetvalue{/pgfplots/table/@cell content}{\multicolumn{1}{c}{##1}&}%
			\fi
		}
	},
	every last row/.style={after row=\bottomrule},
	display columns/0/.style={string type,column type={l}},
	display columns/1/.style={string type,column type={l}},
	display columns/2/.style={string type,column type={l}},
	display columns/3/.style={string type,column type={l}},
	display columns/4/.style={string type,column type={l}},
	]{KODIAQ_Data4.txt}
\end{table}

\bsp	
\label{lastpage}

\end{document}